\documentclass[showpacs,aps,prd]{revtex4}
\input epsf

\begin{document}

\title{$\{Q\bar{q}\}\{\bar{Q}^{(')}q\}$ molecular states}
\author{Jian-Rong Zhang and Ming-Qiu Huang}
\affiliation{Department of Physics, National University of Defense
Technology, Hunan 410073, China}
\date{\today}

%%%%%%%%%%%%%%%%%%%%%%%%%%%%%%%%%%%%%%%%%%%%%%%%%%%%%%%%%%%%%%%%%%%%%
\begin{abstract}
Masses for $\{Q\bar{q}\}\{\bar{Q}^{(')}q\}$ molecular states are
systematically studied in QCD sum rules. The interpolating currents
representing the related molecular states are proposed. Technically,
contributions of the operators up to dimension six are included in
operator product expansion (OPE). Mass spectra for molecular states
with $\{Q\bar{q}\}\{\bar{Q}^{(')}q\}$ configurations are obtained.
\end{abstract}
\pacs {11.55.Hx, 12.38.Lg, 12.39.Mk}\maketitle
%%%%%%%%%%%%%%%%%%%%%%%%%%%%%%%%%%%%%%%%%%%%%%%%%%%%%%%%%%%%%%%%%%%%%

\newpage
\section{Introduction}\label{sec1}
The field of heavy hadron spectroscopy has attracted much attention
nowadays. Experimentally, plentiful hadronic resonances have been
observed, such as the so-called X, Y, and Z states
\cite{X3872,Y3930,Y4260,Z3930,X3940,Z4430,Z4050,Y4140} (for
experimental reviews, e.g., see \cite{Swanson,PDG}). In theory,
different from conventional charmonium states, some of these
resonances may not reconcile with the quark model picture, hence it
is not easy to find appropriate positions for them in mesonic
spectroscopy. Many authors
\cite{theory,theory-X3872,theory-Z4430,theory-Y3930,Liu,theory-Y4140}
tend to interpret some of the hadrons as possible mesonic molecule
candidates for their masses are very close to the meson-meson
thresholds. For example, it has been proposed that $X(3872)$ could
be a $D^{*}\bar{D}$ molecular state \cite{theory-X3872},
$Z^{+}(4430)$ be a $D^{*}\bar{D}_{1}$ molecular state
\cite{theory-Z4430}, $Y(3930)$ be a $D^{*}\bar{D}^{*}$ molecular
state \cite{theory-Y3930,Liu}, $Y(4140)$ be a
$D_{s}^{*}\bar{D}_{s}^{*}$ molecular state \cite{Liu,theory-Y4140}
etc.. As a matter of fact, the charmed molecular states were put
forward long ago in \cite{long ago}, and it has also predicted that
the molecular states involving hidden $c\bar{c}$ pair do exist and
have a rich spectroscopy in \cite{cc}. What is also very important,
the existence of molecular states is not excluded by QCD itself.
Indubitably, investigations of molecular states could deepen one's
understanding of quark-gluon interaction. Thus, the quantitative
descriptions of molecular states' properties like masses are quite
needed for well comprehending their structures. In a word, it is
interesting and significative to study mass spectra for the
molecular states.

Motivated by the above reasons, we devote to obtain the spectra for
molecular states with $\{Q\bar{q}\}\{\bar{Q}^{(')}q\}$
configurations in this work. However, it is a great challenge to
extract information on the spectrum from first principles. While QCD
has been widely accepted as the correct theory describing strong
interaction, it is still far from clear how to gain hadron masses
from the simple Lagrangian. That's because QCD is highly
noperturbative in the low energy region where futile to attempt
perturbative calculations. Therefore, one has to deal with a
genuinely strong field in nonperturbative methods. Before thoroughly
grasping the essence of QCD confinement, for the moment, one could
make use of QCD sum rule \cite{svzsum} (for reviews see
\cite{overview,overview1,overview2,overview3} and references
therein), which is a comprehensive and reliable working tool for
evaluating the nonperturbative effects. The basic idea of this
approach is bridging the gap between the perturbative and
nonperturbative sectors by employing the language of dispersion
relations. Already, there have been several works testing some
molecular states from QCD sum rules up to now
\cite{SR-study,zgwang,zhang}. At present, we'd like to gain the
spectra for $\{Q\bar{q}\}\{\bar{Q}^{(')}q\}$ molecular states with
QCD sum rules, which could serve as a extension of our previous work
on $\{Q\bar{s}\}\{\bar{Q}^{(')}s\}$ states \cite{zhang}.

The paper's framework is as follows. In Sec. \ref{sec2}, QCD sum
rules for the molecular states are introduced, and both the
phenomenological representation and QCD side are derived, followed
by the numerical analysis to extract the hadronic masses and a brief
summary in Sec. \ref{sec3}.
%%%%%%%%%%%%%%%%%%%%%%%%%%%%%%%%%%%%%%%%%%%%%%%%%%%%%%%%%%%%%%%%%%%
\section{molecular state QCD sum rules}\label{sec2}
The QCD sum rule attempts to link the hadron phenomenology with the
interactions of quarks and gluons, the elementary point of which is
the choice of interpolating current. Following the standard scheme
\cite{PDG}, the $Q\bar{q}$ mesons with $J^{P}=0^{-},~1^{-},~ 0^{+},~
\mbox{and}~1^{+}$ are named $D$, $D^{*}$, $D_{0}^{*}$, and $D_{1}$
for charmed mesons, with $B$, $B^{*}$, $B_{0}^{*}$, and $B_{1}$ for
bottom mesons, respectively. Here the configurations for these
mesons are represented as $(Q\bar{q})$, $(Q\bar{q})^{*}$,
$(Q\bar{q})_{0}^{*}$, and $(Q\bar{q})_{1}$. In full theory, the
interpolating currents for these mesons can be found in Refs.
\cite{reinders,reinders1}. Consequently, the interpolating currents
for the related molecular states are constructed in following forms,
with
\begin{eqnarray}
j_{(Q\bar{q})(\bar{Q'}q)}&=&(\bar{q}_{a}i\gamma_{5}Q_{a})(\bar{Q'}_{b}i\gamma_{5}q_{b}),\nonumber\\
j_{(Q\bar{q})^{*}(\bar{Q'}q)^{*}}&=&(\bar{q}_{a}\gamma_{\mu}Q_{a})(\bar{Q'}_{b}\gamma^{\mu}q_{b}),\nonumber\\
j_{(Q\bar{q})_{0}^{*}(\bar{Q'}q)_{0}^{*}}&=&(\bar{q}_{a}Q_{a})(\bar{Q'}_{b}q_{b}),\nonumber\\
j_{(Q\bar{q})_{1}(\bar{Q'}q)_{1}}&=&(\bar{q}_{a}\gamma_{\mu}\gamma_{5}Q_{a})(\bar{Q'}_{b}\gamma^{\mu}\gamma_{5}q_{b}),\nonumber\\
j_{(Q\bar{q})(\bar{Q'}q)_{0}^{*}}&=&(\bar{q}_{a}i\gamma_{5}Q_{a})(\bar{Q'}_{b}q_{b}),\nonumber\\
j_{(Q\bar{q})^{*}(\bar{Q'}q)_{1}}&=&(\bar{q}_{a}\gamma_{\mu}Q_{a})(\bar{Q'}_{b}\gamma^{\mu}\gamma_{5}q_{b}),\nonumber
\end{eqnarray}
for one type of hadrons, and
\begin{eqnarray}
j^{\mu}_{(Q\bar{q})^{*}(\bar{Q'}q)}&=&(\bar{q}_{a}\gamma^{\mu}Q_{a})(\bar{Q'}_{b}i\gamma_{5}q_{b}),\nonumber\\
j^{\mu}_{(Q\bar{q})_{1}(\bar{Q'}q)}&=&(\bar{q}_{a}\gamma^{\mu}\gamma_{5}Q_{a})(\bar{Q'}_{b}i\gamma_{5}q_{b}),\nonumber\\
j^{\mu}_{(Q\bar{q})^{*}(\bar{Q'}q)_{0}^{*}}&=&(\bar{q}_{a}\gamma^{\mu}Q_{a})(\bar{Q'}_{b}q_{b}),\nonumber\\
j^{\mu}_{(Q\bar{q})_{1}(\bar{Q'}q)_{0}^{*}}&=&(\bar{q}_{a}\gamma^{\mu}\gamma_{5}Q_{a})(\bar{Q'}_{b}q_{b}),\nonumber
\end{eqnarray}
for another type, where $q$ indicates the light quark $u$ or $d$,
$Q$ and $Q'$ denote heavy quarks ($Q=Q'$ or $Q\neq Q'$), with $a$
and $b$ are color indices.

For the former case, the starting point is the two-point correlator
\begin{eqnarray}\label{correlator}
\Pi(q^{2})=i\int
d^{4}x\mbox{e}^{iq.x}\langle0|T[j(x)j^{+}(0)]|0\rangle.
\end{eqnarray}
The correlator can be phenomenologically expressed as a dispersion
integral over a physical spectral function
\begin{eqnarray}
\Pi(q^{2})=\frac{\lambda^{2}_H}{M_{H}^{2}-q^{2}}+\frac{1}{\pi}\int_{s_{0}}
^{\infty}ds\frac{\mbox{Im}\Pi^{\mbox{phen}}(s)}{s-q^{2}}+\mbox{subtractions},
\end{eqnarray}
where $M_{H}$ denotes the mass of the hadronic resonance, and
$\lambda_{H}$ gives the coupling of the current to the hadron
$\langle0|j|H\rangle=\lambda_{H}$. In the OPE side, the correlator
can be written in terms of a dispersion relation as
\begin{eqnarray}
\Pi(q^{2})=\int_{(m_{Q}+m_{Q'})^{2}}^{\infty}ds\frac{\rho^{\mbox{OPE}}(s)}{s-q^{2}}~~(m_{Q}=m_{Q'}~
\mbox{or}~m_{Q}\neq m_{Q'}),
\end{eqnarray}
where the spectral density is given by the imaginary part of the
correlator
\begin{eqnarray}
\rho^{\mbox{OPE}}(s)=\frac{1}{\pi}\mbox{Im}\Pi^{\mbox{OPE}}(s).
\end{eqnarray}
After equating the two sides, assuming quark-hadron duality, and
making a Borel transform, the sum rule can be written as
\begin{eqnarray}
\lambda_{H}^{2}e^{-M_{H}^{2}/M^{2}}&=&\int_{(m_{Q}+m_{Q'})^{2}}^{s_{0}}ds\rho^{\mbox{OPE}}(s)e^{-s/M^{2}},
\end{eqnarray}
where $M^2$ indicates Borel parameter. To eliminate the hadronic
coupling constant $\lambda_H$, one reckons the ratio of derivative
of the sum rule and itself, and then yields
\begin{eqnarray}\label{sum rule}
M_{H}^{2}&=&\int_{(m_{Q}+m_{Q'})^{2}}^{s_{0}}ds\rho^{\mbox{OPE}}s
e^{-s/M^{2}}/
\int_{(m_{Q}+m_{Q'})^{2}}^{s_{0}}ds\rho^{\mbox{OPE}}e^{-s/M^{2}}.
\end{eqnarray}

For the latter case, one starts from the two-point correlator
\begin{eqnarray}
\Pi^{\mu\nu}(q^{2})=i\int
d^{4}x\mbox{e}^{iq.x}\langle0|T[j^{\mu}(x)j^{\nu+}(0)]|0\rangle.
\end{eqnarray}
Lorentz covariance implies that the two-point correlation function
can be generally parameterized as
\begin{eqnarray}
\Pi^{\mu\nu}(q^{2})=(\frac{q^{\mu}q^{\nu}}{q^{2}}-g^{\mu\nu})\Pi^{(1)}(q^{2})+\frac{q^{\mu}q^{\nu}}{q^{2}}\Pi^{(0)}(q^{2}).
\end{eqnarray}
The part of the correlator proportional to $g_{\mu\nu}$ will be
chosen to extract the mass sum rule here. In phenomenology,
$\Pi^{(1)}(q^{2})$ can be expressed as a dispersion integral over a
physical spectral function
\begin{eqnarray}
\Pi^{(1)}(q^{2})=\frac{[\lambda^{(1)}]^{2}}{M_{H}^{2}-q^{2}}+\frac{1}{\pi}\int_{s_{0}}
^{\infty}ds\frac{\mbox{Im}\Pi^{(1)\mbox{phen}}(s)}{s-q^{2}}+\mbox{subtractions},
\end{eqnarray}
In the OPE side, $\Pi^{(1)}(q^{2})$ can be written in terms of a
dispersion relation as
\begin{eqnarray}
\Pi^{(1)}(q^{2})=\int_{(m_{Q}+m_{Q'})^{2}}^{\infty}ds\frac{\rho^{\mbox{OPE}}(s)}{s-q^{2}}~~(m_{Q}=m_{Q'}~
\mbox{or}~m_{Q}\neq m_{Q'}),
\end{eqnarray}
where the spectral density is given by
\begin{eqnarray}
\rho^{\mbox{OPE}}(s)=\frac{1}{\pi}\mbox{Im}\Pi^{\mbox{(1)}}(s).
\end{eqnarray}
After equating the two sides, assuming quark-hadron duality, and
making a Borel transform, the sum rule can be written as
\begin{eqnarray}
[\lambda^{(1)}]^{2}e^{-M_{H}^{2}/M^{2}}&=&\int_{(m_{Q}+m_{Q'})^{2}}^{s_{0}}ds\rho^{\mbox{OPE}}(s)e^{-s/M^{2}}.
\end{eqnarray}
To eliminate the hadronic coupling constant $\lambda^{(1)}$, one
reckons the ratio of derivative of the sum rule and itself, and then
yields
\begin{eqnarray}\label{sum rule 1}
M_{H}^{2}&=&\int_{(m_{Q}+m_{Q'})^{2}}^{s_{0}}ds\rho^{\mbox{OPE}}s
e^{-s/M^{2}}/
\int_{(m_{Q}+m_{Q'})^{2}}^{s_{0}}ds\rho^{\mbox{OPE}}e^{-s/M^{2}}.
\end{eqnarray}

Calculating the OPE side, one works at leading order in $\alpha_{s}$
and considers condensates up to dimension six with the similar
techniques in Refs. \cite{technique,technique1}. To keep the
heavy-quark mass finite, one uses the momentum-space expression for
the heavy-quark propagator. One calculates the light-quark part of
the correlation function in the coordinate space, which is then
Fourier-transformed to the momentum space in $D$ dimension. The
resulting light-quark part is combined with the heavy-quark part
before it is dimensionally regularized at $D=4$. For the heavy-quark
propagator with two and three gluons attached, the momentum-space
expressions given in Ref. \cite{reinders} are used. After some
lengthy OPE calculations, the concrete spectral densities can be
acquired, which are collected in the Appendix.

%%%%%%%%%%%%%%%%%%%%%%%%%%%%%%%%%%%%%%%%%%%%%%%%%%%%%%%%%%%%%%%%%%%
\section{Numerical analysis}\label{sec3}
In this section, the sum rules (\ref{sum rule}) and (\ref{sum rule
1}) will be numerically analyzed. The input values are taken as
$m_{c}=1.23~\mbox{GeV}$, $m_{b}=4.20~\mbox{GeV}$ \cite{PDG} with
$\langle\bar{q}q\rangle=-(0.23)^{3}~\mbox{GeV}^{3}$, $\langle
g\bar{q}\sigma\cdot G q\rangle=m_{0}^{2}~\langle\bar{q}q\rangle$,
$m_{0}^{2}=0.8~\mbox{GeV}^{2}$, $\langle
g^{2}G^{2}\rangle=0.88~\mbox{GeV}^{4}$, and $\langle
g^{3}G^{3}\rangle=0.045~\mbox{GeV}^{6}$
\cite{overview2,SR-study,technique}. Complying with the standard
procedure of sum rule analysis, the threshold $s_{0}$ and Borel
parameter $M^{2}$ are varied to find the optimal stability window,
in which the perturbative contribution should be larger than the
condensate contributions while the pole contribution larger than
continuum contribution. Thus, the regions of thresholds are taken as
those values presented in the related figure captions, with
$M^{2}=3.5\sim4.5~\mbox{GeV}^{2}$ for $D\bar{D}$, $D^{*}\bar{D}$,
$D^{*}\bar{D}^{*}$, $D_{0}^{*}\bar{D}_{0}^{*}$,
$D_{1}\bar{D}_{0}^{*}$, $D_{1}\bar{D}_{1}$, $D\bar{D}_{0}^{*}$,
$D_{1}\bar{D}$, $D^{*}\bar{D}_{0}^{*}$, and $D^{*}\bar{D}_{1}$,
$M^{2}=7.5\sim9.0~\mbox{GeV}^{2}$ for $B\bar{D}$, $B^{*}\bar{D}$,
$B^{*}\bar{D}^{*}$, $B_{0}^{*}\bar{D}_{0}^{*}$,
$B_{1}\bar{D}_{0}^{*}$, $B_{1}\bar{D}_{1}$, $B\bar{D}_{0}^{*}$,
$B_{1}\bar{D}$, $B^{*}\bar{D}_{0}^{*}$, $B^{*}\bar{D}_{1}$,
$D^{*}\bar{B}$, $D_{1}\bar{B}_{0}^{*}$, $D\bar{B}_{0}^{*}$,
$D_{1}\bar{B}$, $D^{*}\bar{B}_{0}^{*}$, and $D^{*}\bar{B}_{1}$, and
$M^{2}=9.5\sim11.0~\mbox{GeV}^{2}$ for $B\bar{B}$, $B^{*}\bar{B}$,
$B^{*}\bar{B}^{*}$, $B_{0}^{*}\bar{B}_{0}^{*}$,
$B_{1}\bar{B}_{0}^{*}$, $B_{1}\bar{B}_{1}$, $B\bar{B}_{0}^{*}$,
$B_{1}\bar{B}$, $B^{*}\bar{B}_{0}^{*}$, and $B^{*}\bar{B}_{1}$,
respectively. The corresponding Borel curves are exhibited in Figs.
1-18, and the numerical results are listed in Tables I-II. It is
worth noting that the numerical errors reflect the uncertainty due
to sum rule windows (variation of the threshold $s_{0}$ and Borel
parameter $M^{2}$) only; the uncertainty resulted from the variation
of the quark masses and QCD parameters is not included. After a
comparison, one can find that the numerical results for
$D^{*}\bar{D}$, $D^{*}\bar{D}_{1}$, and $D^{*}\bar{D}^{*}$ are in
good agreement with the experimental data for $X(3872)$,
$Z^{+}(4430)$, and $Y(3930)$, respectively, which could support the
molecular interpretations for these hadrons.

\begin{table}\caption{The mass spectra of molecular states with same heavy quarks.}
 \centerline{\begin{tabular}{p{2.5cm} p{2.5cm} p{2.5cm} p{2.5cm} p{2.5cm} p{2.5cm}}  \hline\hline
Hadron                            & configuration                                &     mass (GeV)                      &  Hadron                            & configuration                                &     mass (GeV)                   \\
\hline
$D\bar{D}$                        &$(c\bar{q})(\bar{c}q)$                        & $3.76\pm0.10$                       &$B\bar{B}$                          &$(b\bar{q})(\bar{b}q)$                        & $10.58\pm0.10$                   \\
\hline
$D^{*}\bar{D}$                    &$(c\bar{q})^{*}(\bar{c}q)$                    & $3.88\pm0.10$                       &$B^{*}\bar{B}$                      &$(b\bar{q})^{*}(\bar{b}q)$                    & $10.62\pm0.10$                   \\
\hline
$D^{*}\bar{D}^{*}$                &$(c\bar{q})^{*}(\bar{c}q)^{*}$                & $3.91\pm0.11$                       &$B^{*}\bar{B}^{*}$                  &$(b\bar{q})^{*}(\bar{b}q)^{*}$                & $10.67\pm0.10$                   \\
\hline
$D_{0}^{*}\bar{D}_{0}^{*}$        &$(c\bar{q})_{0}^{*}(\bar{c}q)_{0}^{*}$        & $4.56\pm0.11$                       &$B_{0}^{*}\bar{B}_{0}^{*}$          &$(b\bar{q})_{0}^{*}(\bar{b}q)_{0}^{*}$        & $11.28\pm0.08$                   \\
\hline
$D_{1}\bar{D}_{0}^{*}$            &$(c\bar{q})_{1}(\bar{c}q)_{0}^{*}$            & $4.62\pm0.11$                       &$B_{1}\bar{B}_{0}^{*}$              &$(b\bar{q})_{1}(\bar{b}q)_{0}^{*}$            & $11.32\pm0.09$                   \\
\hline
$D_{1}\bar{D}_{1}$                &$(c\bar{q})_{1}(\bar{c}q)_{1}$                & $4.66\pm0.13$                       &$B_{1}\bar{B}_{1}$                  &$(b\bar{q})_{1}(\bar{b}q)_{1}$                & $11.33\pm0.12$                   \\
\hline
$D\bar{D}_{0}^{*}$                &$(c\bar{q})(\bar{c}q)_{0}^{*}$                & $4.21\pm0.07$                       &$B\bar{B}_{0}^{*}$                  &$(b\bar{q})(\bar{b}q)_{0}^{*}$                & $11.03\pm0.09$                   \\
\hline
$D_{1}\bar{D}$                    &$(c\bar{q})_{1}(\bar{c}q)$                    & $4.34\pm0.07$                       &$B_{1}\bar{B}$                      &$(b\bar{q})_{1}(\bar{b}q)$                    & $11.04\pm0.09$                   \\
 \hline
$D^{*}\bar{D}_{0}^{*}$            &$(c\bar{q})^{*}(\bar{c}q)_{0}^{*}$            & $4.26\pm0.07$                       &$B^{*}\bar{B}_{0}^{*}$              &$(b\bar{q})^{*}(\bar{b}q)_{0}^{*}$            & $11.02\pm0.09$                   \\
\hline
$D^{*}\bar{D}_{1}$                &$(c\bar{q})^{*}(\bar{c}q)_{1}$                & $4.44\pm0.09$                       &$B^{*}\bar{B}_{1}$                  &$(b\bar{q})^{*}(\bar{b}q)_{1}$                & $11.03\pm0.09$                   \\
\hline \hline
\end{tabular}} \label{table:1}
\end{table}

\vspace{2cm}

\begin{table}\caption{The mass spectra of molecular states with differently heavy quarks.}
 \centerline{\begin{tabular}{p{2.5cm} p{2.5cm} p{2.5cm} p{2.5cm} p{2.5cm} p{2.5cm}}  \hline\hline
Hadron                            & configuration                                &     mass (GeV)                      &  Hadron                            & configuration                                &     mass (GeV)                   \\
$B\bar{D}$                        &$(b\bar{q})(\bar{c}q)$                        & $7.12\pm0.09$                       &$B^{*}\bar{D}_{0}^{*}$              &$(b\bar{q})^{*}(\bar{c}q)_{0}^{*}$            & $7.67\pm0.06$                    \\
\hline
$B^{*}\bar{D}$                    &$(b\bar{q})^{*}(\bar{c}q)$                    & $7.28\pm0.09$                       &$B^{*}\bar{D}_{1}$                  &$(b\bar{q})^{*}(\bar{c}q)_{1}$                & $7.74\pm0.07$                    \\
\hline
$B^{*}\bar{D}^{*}$                &$(b\bar{q})^{*}(\bar{c}q)^{*}$                & $7.29\pm0.10$                       &$D^{*}\bar{B}$                      &$(c\bar{q})^{*}(\bar{b}q)$                    & $7.21\pm0.09$                    \\
\hline
$B_{0}^{*}\bar{D}_{0}^{*}$        &$(b\bar{q})_{0}^{*}(\bar{c}q)_{0}^{*}$        & $8.04\pm0.08$                       &$D_{1}\bar{B}_{0}^{*}$              &$(c\bar{q})_{1}(\bar{b}q)_{0}^{*}$            & $8.04\pm0.10$                    \\
\hline
$B_{1}\bar{D}_{0}^{*}$            &$(b\bar{q})_{1}(\bar{c}q)_{0}^{*}$            & $8.06\pm0.13$                       &$D\bar{B}_{0}^{*}$                  &$(c\bar{q})(\bar{b}q)_{0}^{*}$                & $7.70\pm0.06$                    \\
\hline
$B_{1}\bar{D}_{1}$                &$(b\bar{q})_{1}(\bar{c}q)_{1}$                & $8.07\pm0.11$                       &$D_{1}\bar{B}$                      &$(c\bar{q})_{1}(\bar{b}q)$                    & $7.74\pm0.07$                    \\
\hline
$B\bar{D}_{0}^{*}$                &$(b\bar{q})(\bar{c}q)_{0}^{*}$                & $7.68\pm0.06$                       &$D^{*}\bar{B}_{0}^{*}$              &$(c\bar{q})^{*}(\bar{b}q)_{0}^{*}$            & $7.76\pm0.06$                    \\
\hline
$B_{1}\bar{D}$                    &$(b\bar{q})_{1}(\bar{c}q)$                    & $7.77\pm0.06$                       &$D^{*}\bar{B}_{1}$                  &$(c\bar{q})^{*}(\bar{b}q)_{1}$                & $7.76\pm0.07$                    \\
\hline\hline
\end{tabular}} \label{table:2}
\end{table}

\begin{figure}
\centerline{\epsfysize=5.5truecm
\epsfbox{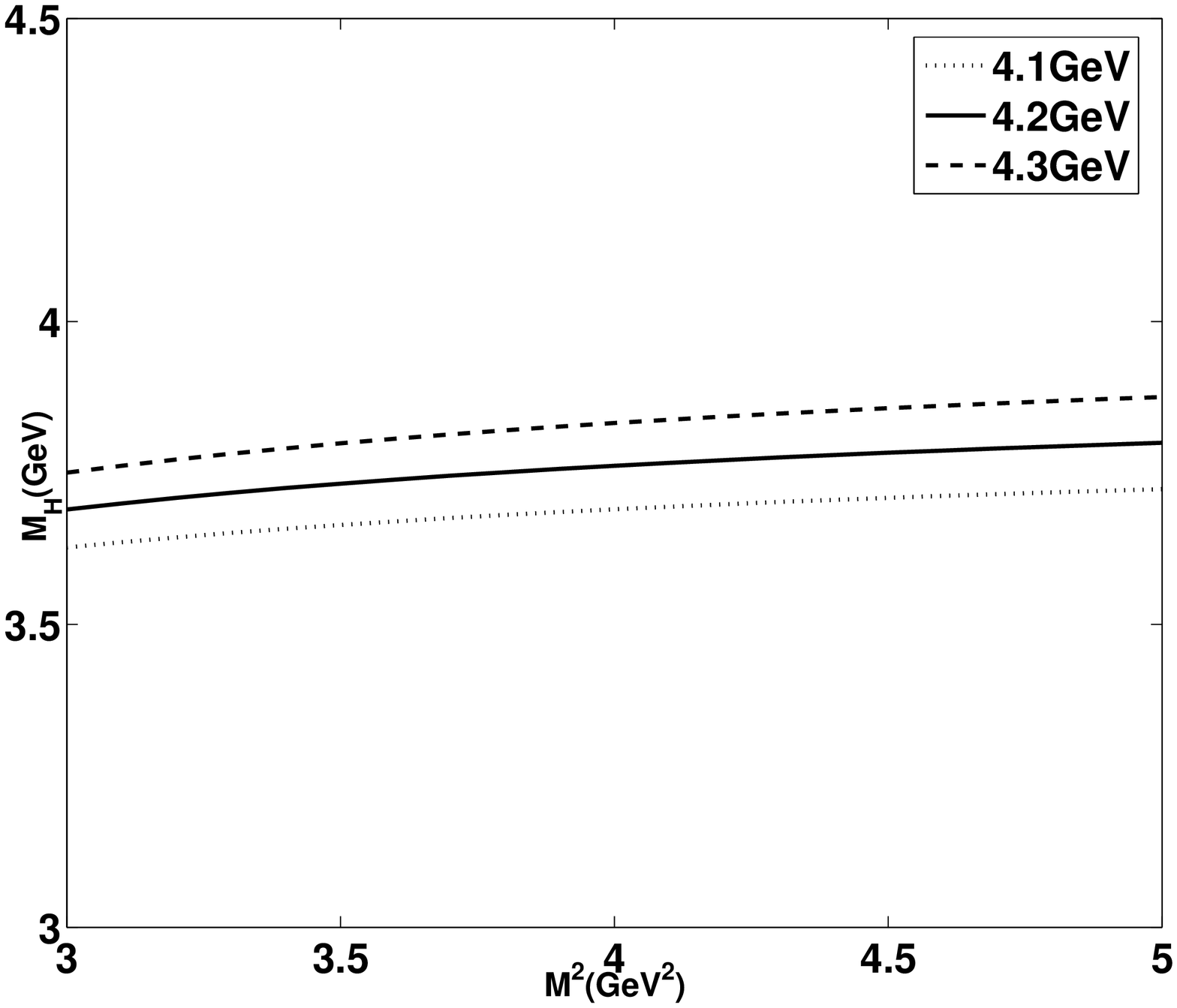}\epsfysize=5.5truecm\epsfbox{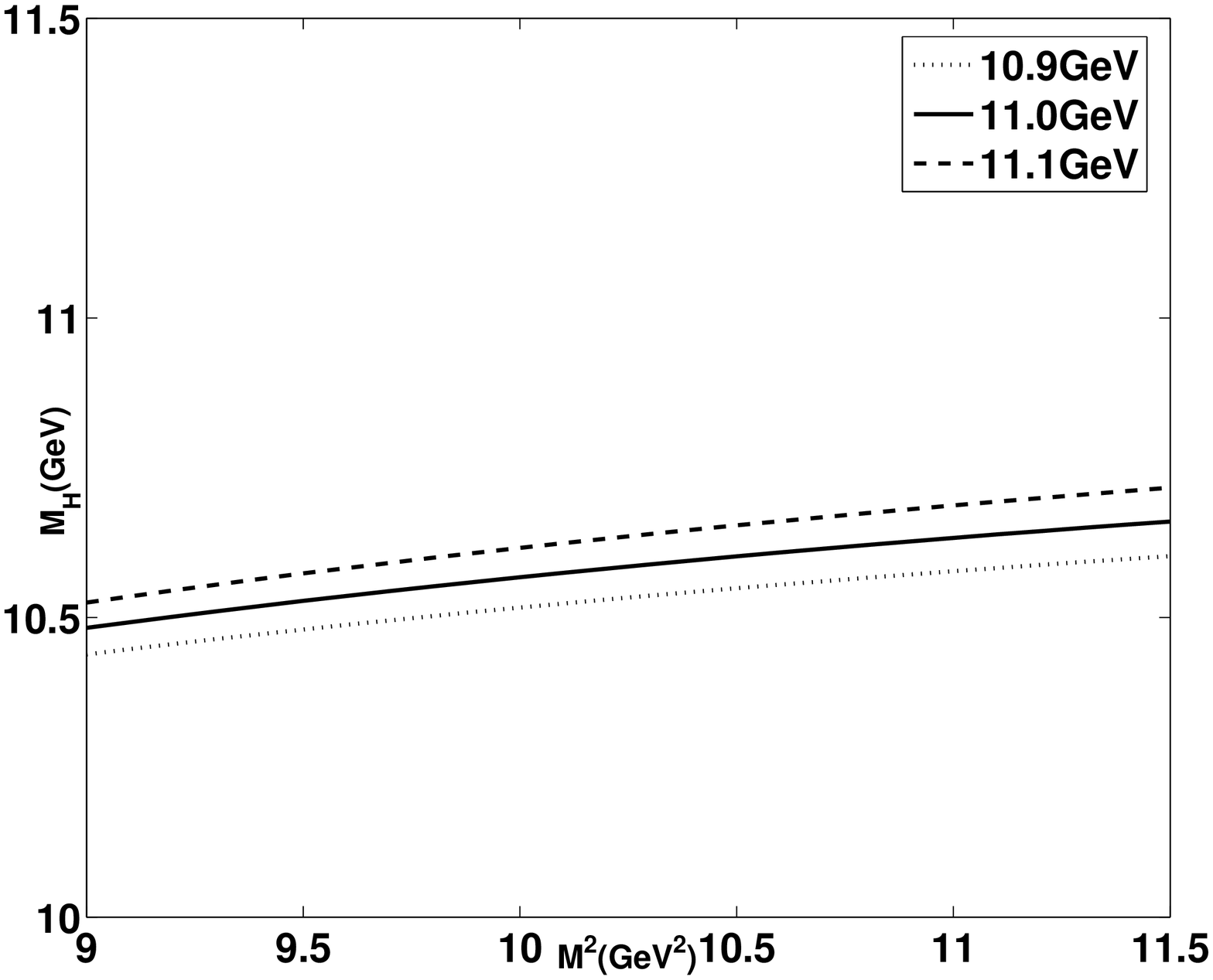}}\caption{The
dependence on $M^2$ for the masses of $D\bar{D}$ and $B\bar{B}$ from
sum rule (\ref{sum rule}). The continuum thresholds are taken as
$\sqrt{s_0}=4.1\sim4.3~\mbox{GeV}$ and
$\sqrt{s_0}=10.9\sim11.1~\mbox{GeV}$, respectively.} \label{fig:1}
\end{figure}

\begin{figure}
\centerline{\epsfysize=5.5truecm
\epsfbox{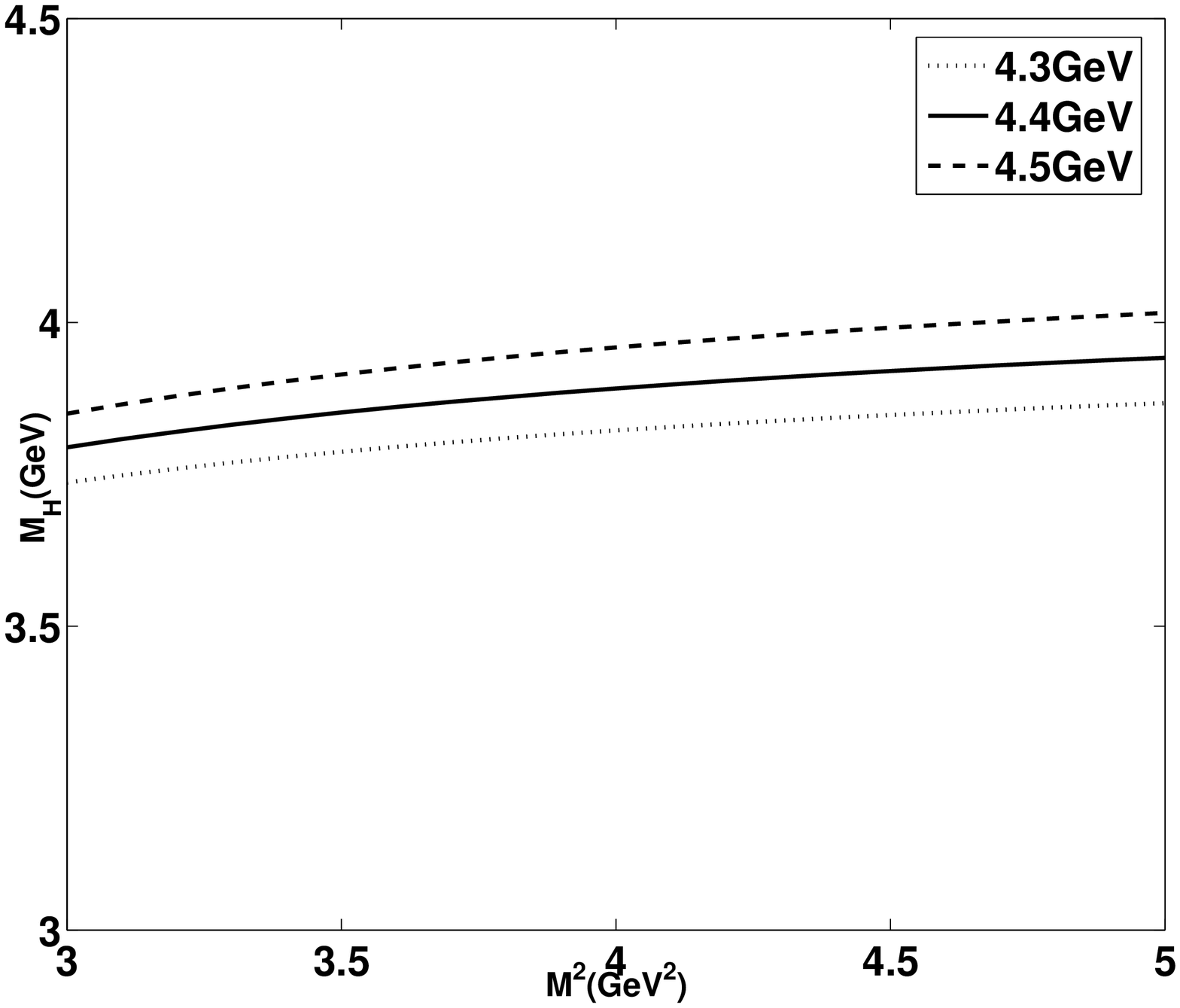}\epsfysize=5.5truecm\epsfbox{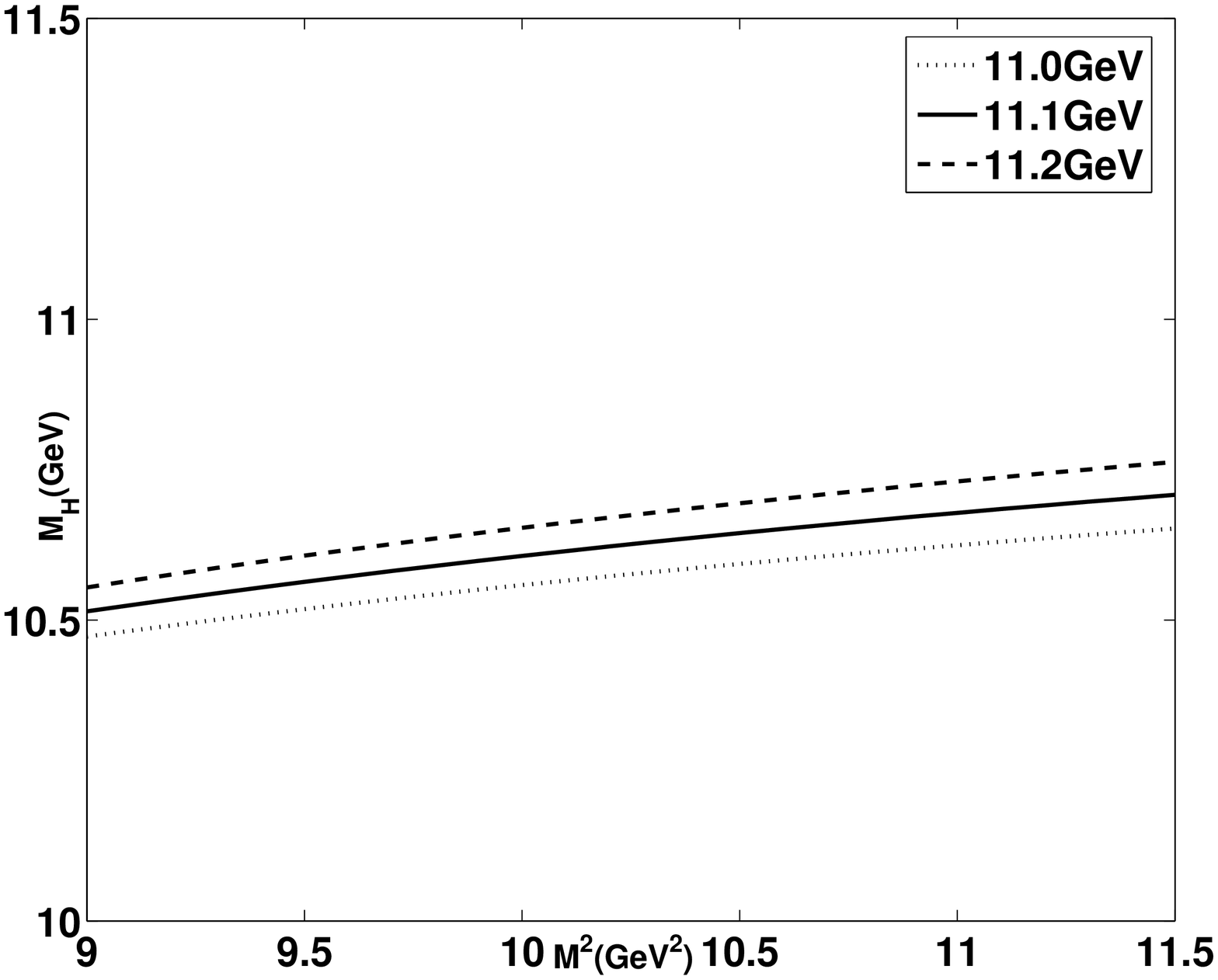}}\caption{The
dependence on $M^2$ for the masses of $D^{*}\bar{D}$ and
$B^{*}\bar{B}$ from sum rule (\ref{sum rule 1}). The continuum
thresholds are taken as $\sqrt{s_0}=4.3\sim4.5~\mbox{GeV}$ and
$\sqrt{s_0}=11.0\sim11.2~\mbox{GeV}$, respectively.} \label{fig:2}
\end{figure}

\begin{figure}
\centerline{\epsfysize=5.5truecm
\epsfbox{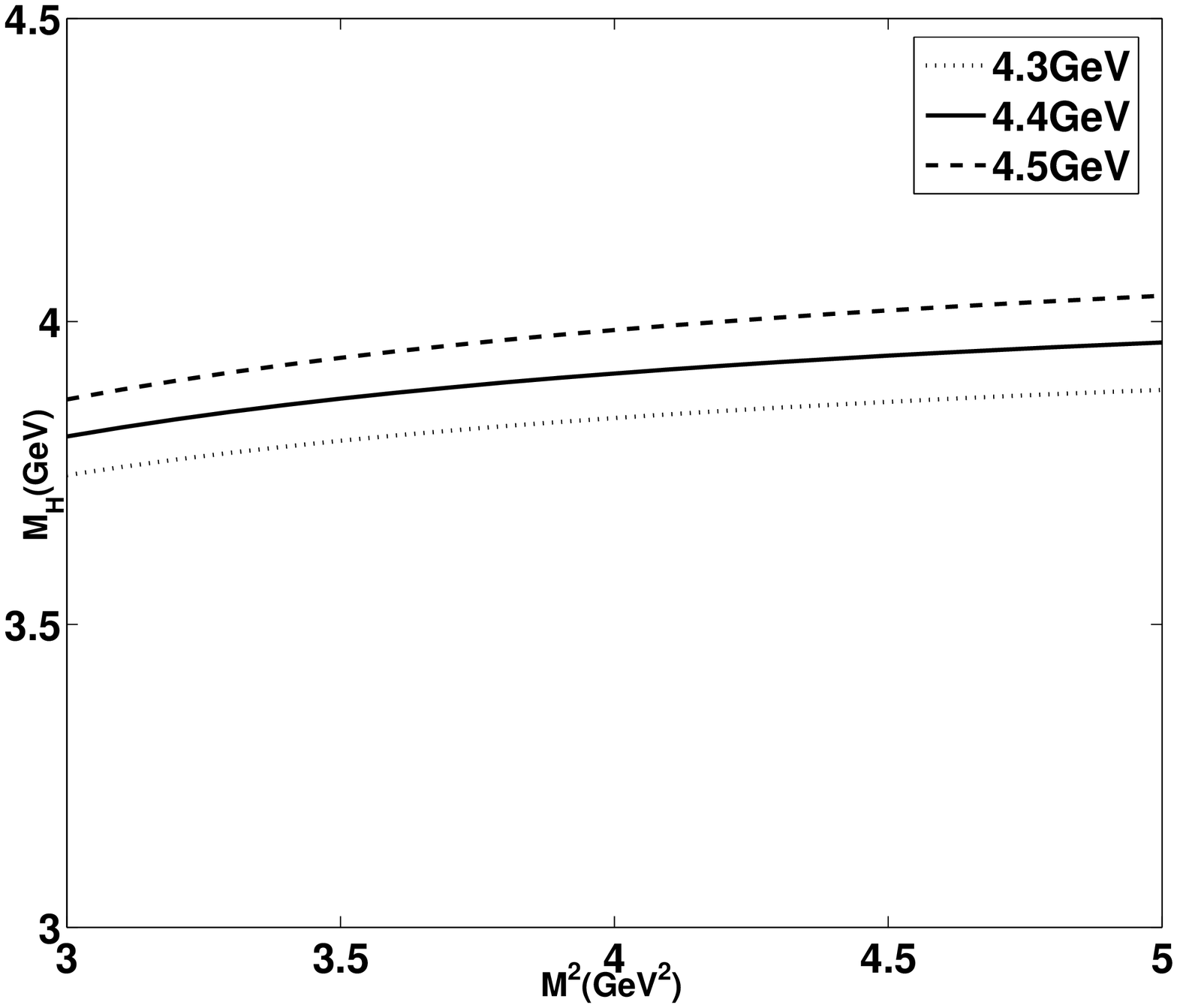}\epsfysize=5.5truecm\epsfbox{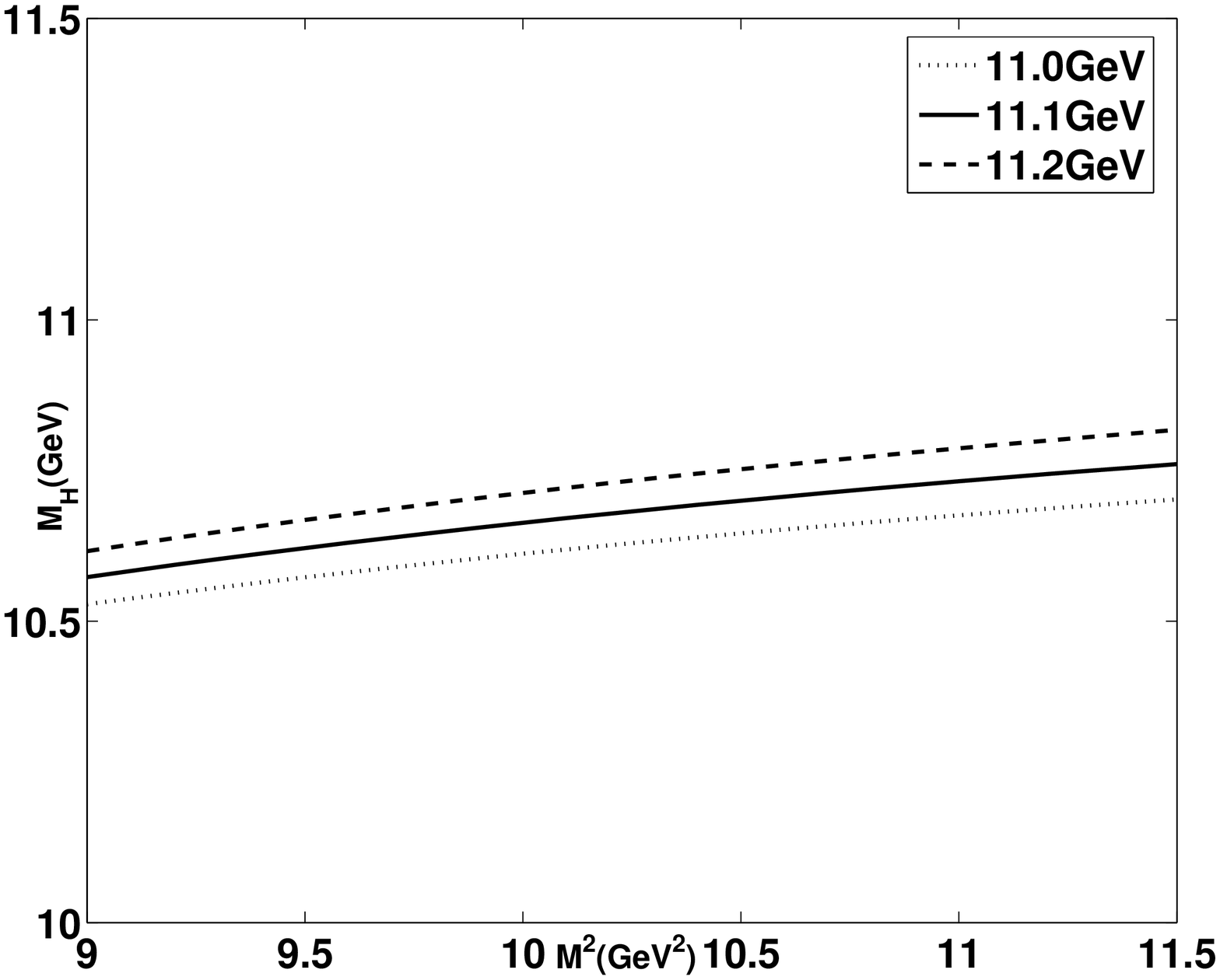}}\caption{The
dependence on $M^2$ for the masses of $D^{*}\bar{D}^{*}$ and
$B^{*}\bar{B}^{*}$ from sum rule (\ref{sum rule}). The continuum
thresholds are taken as $\sqrt{s_0}=4.3\sim4.5~\mbox{GeV}$ and
$\sqrt{s_0}=11.0\sim11.2~\mbox{GeV}$, respectively.} \label{fig:3}
\end{figure}

\begin{figure}
\centerline{\epsfysize=5.5truecm
\epsfbox{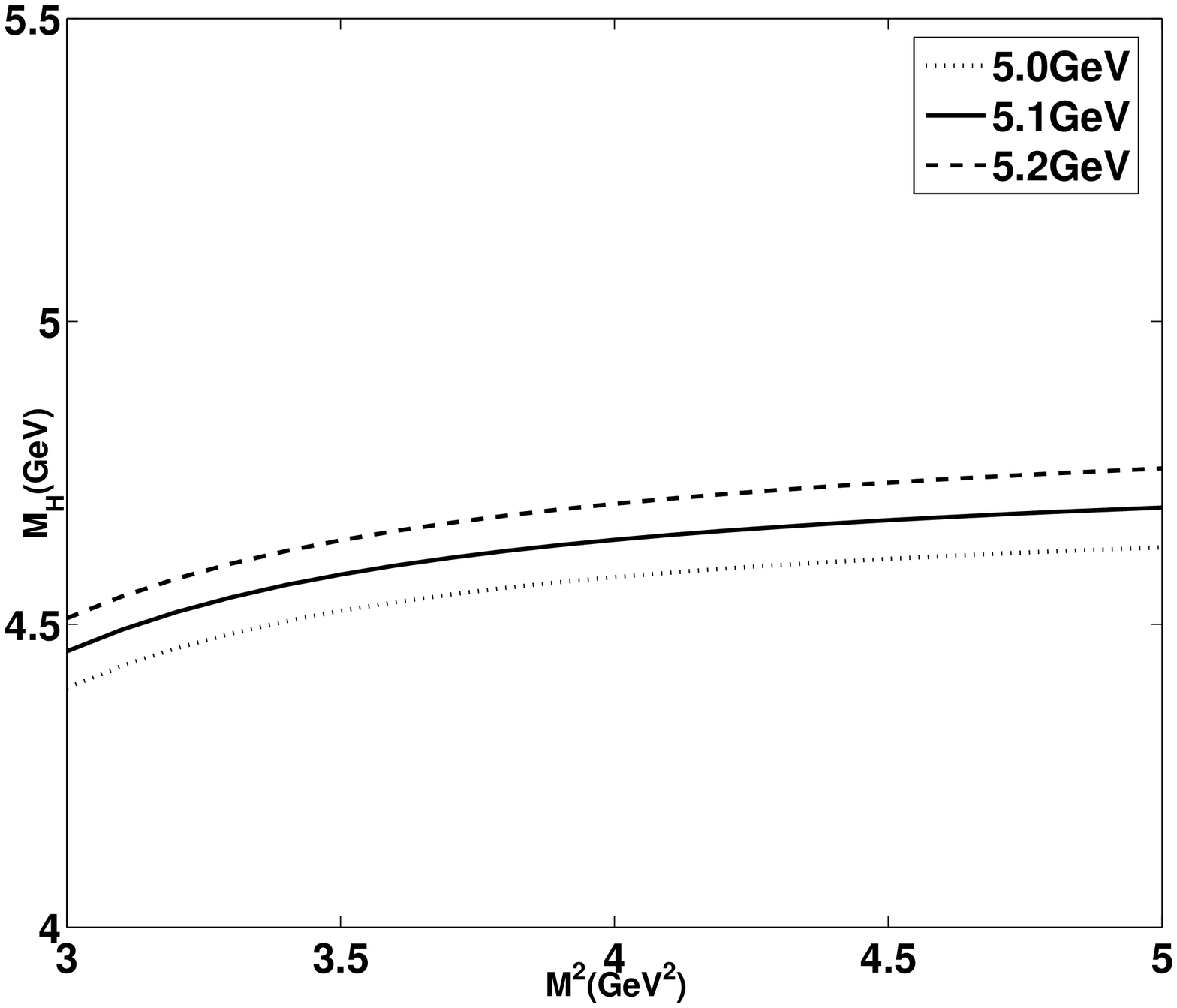}\epsfysize=5.5truecm\epsfbox{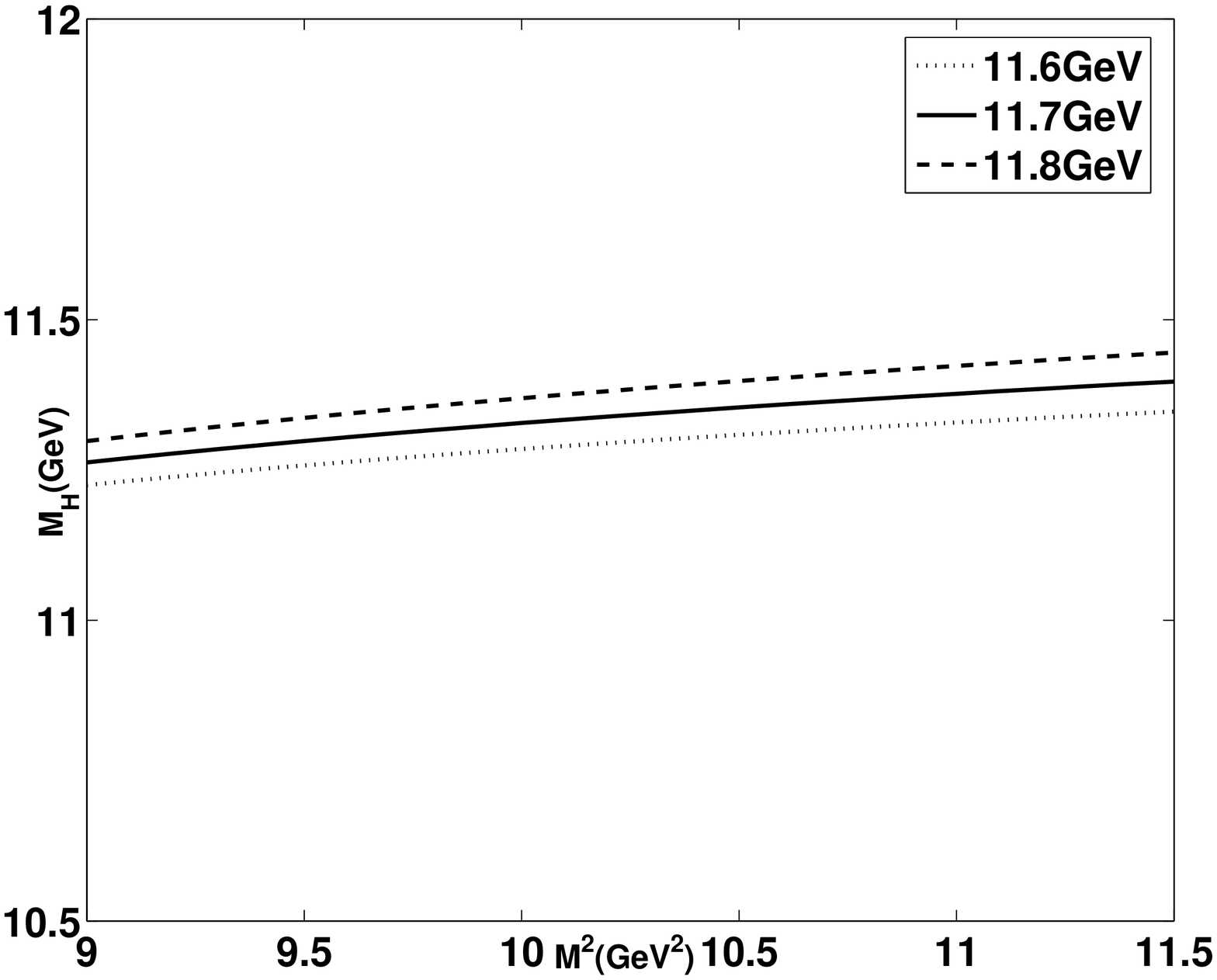}}\caption{The
dependence on $M^2$ for the masses of $D_{1}\bar{D}_{0}^{*}$ and
$B_{1}\bar{B}_{0}^{*}$ from sum rule (\ref{sum rule 1}). The
continuum thresholds are taken as $\sqrt{s_0}=5.0\sim5.2~\mbox{GeV}$
and $\sqrt{s_0}=11.6\sim11.8~\mbox{GeV}$, respectively.}
\label{fig:4}
\end{figure}

\begin{figure}
\centerline{\epsfysize=5.5truecm
\epsfbox{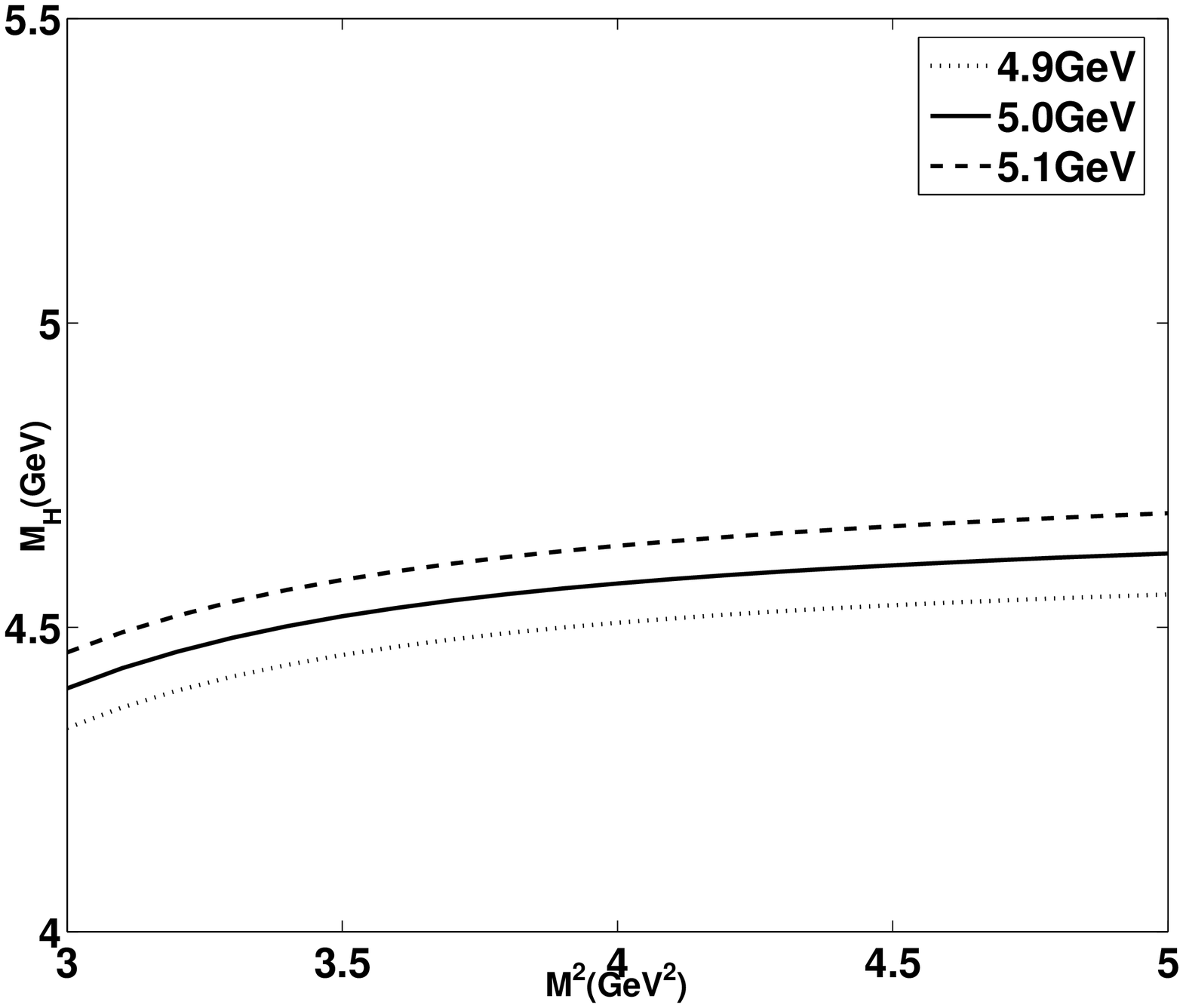}\epsfysize=5.5truecm\epsfbox{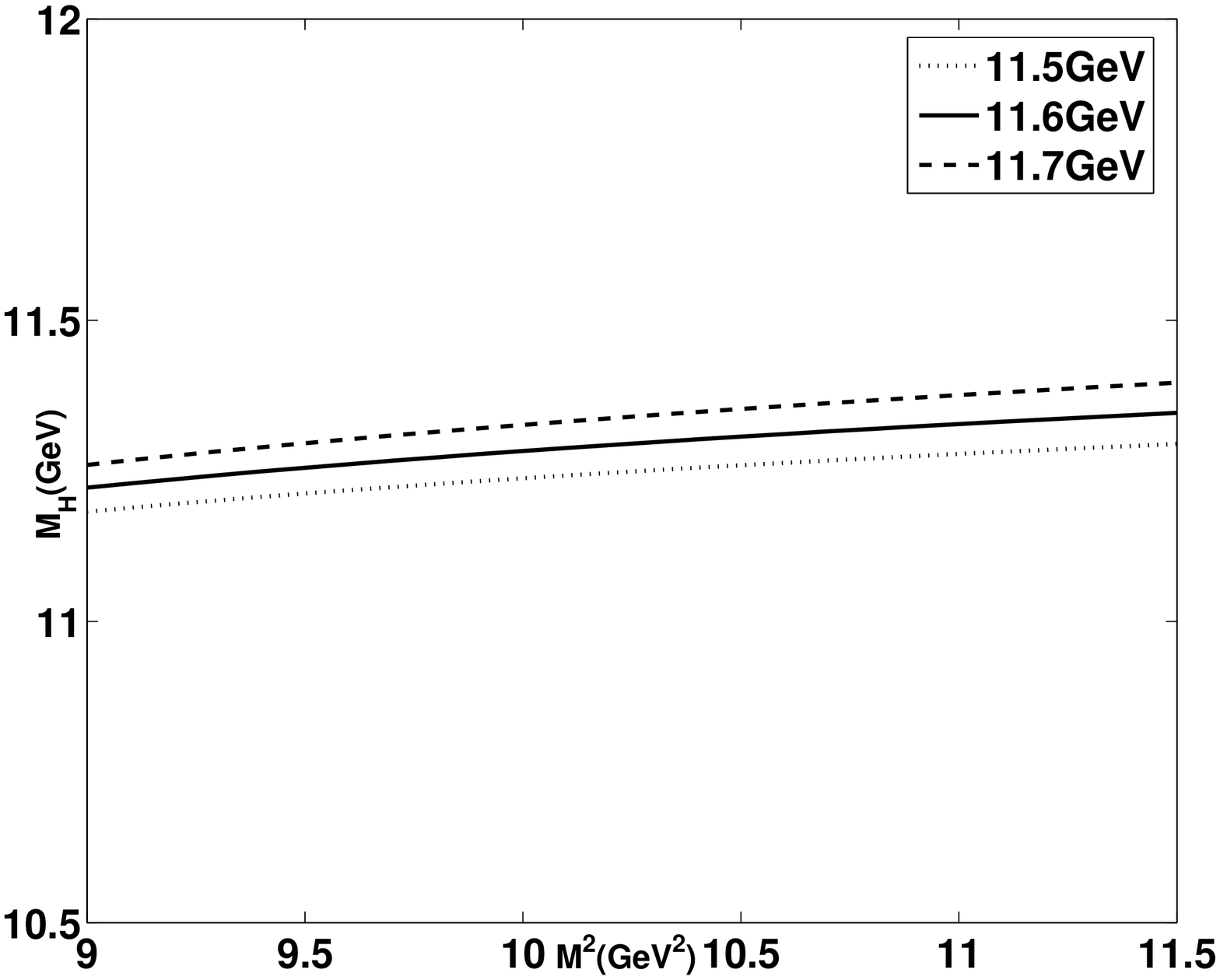}}\caption{The
dependence on $M^2$ for the masses of $D_{0}^{*}\bar{D}_{0}^{*}$ and
$B_{0}^{*}\bar{B}_{0}^{*}$ from sum rule (\ref{sum rule}). The
continuum thresholds are taken as $\sqrt{s_0}=4.9\sim5.1~\mbox{GeV}$
and $\sqrt{s_0}=11.5\sim11.7~\mbox{GeV}$, respectively.}
\label{fig:5}
\end{figure}

\begin{figure}
\centerline{\epsfysize=5.5truecm
\epsfbox{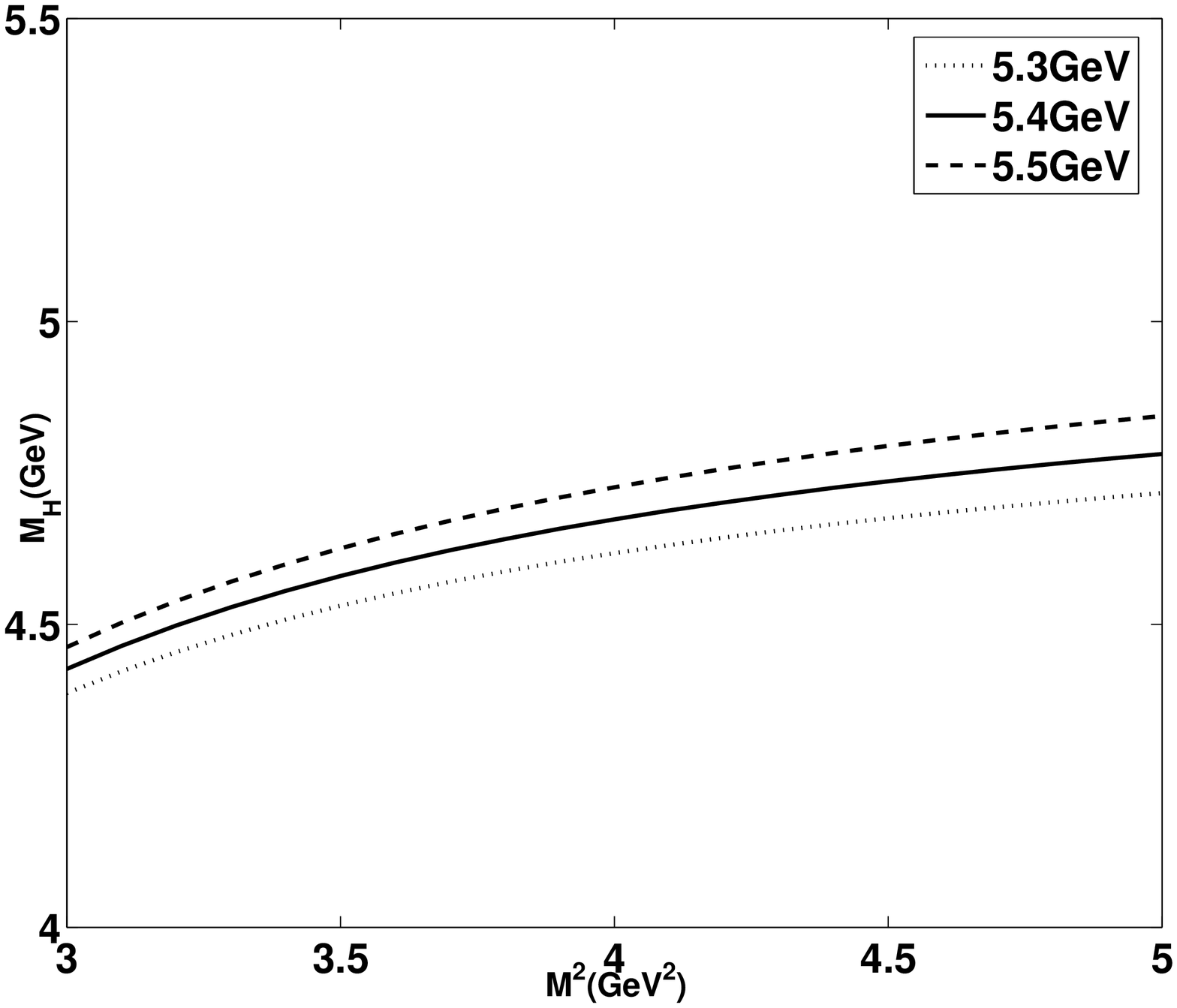}\epsfysize=5.5truecm\epsfbox{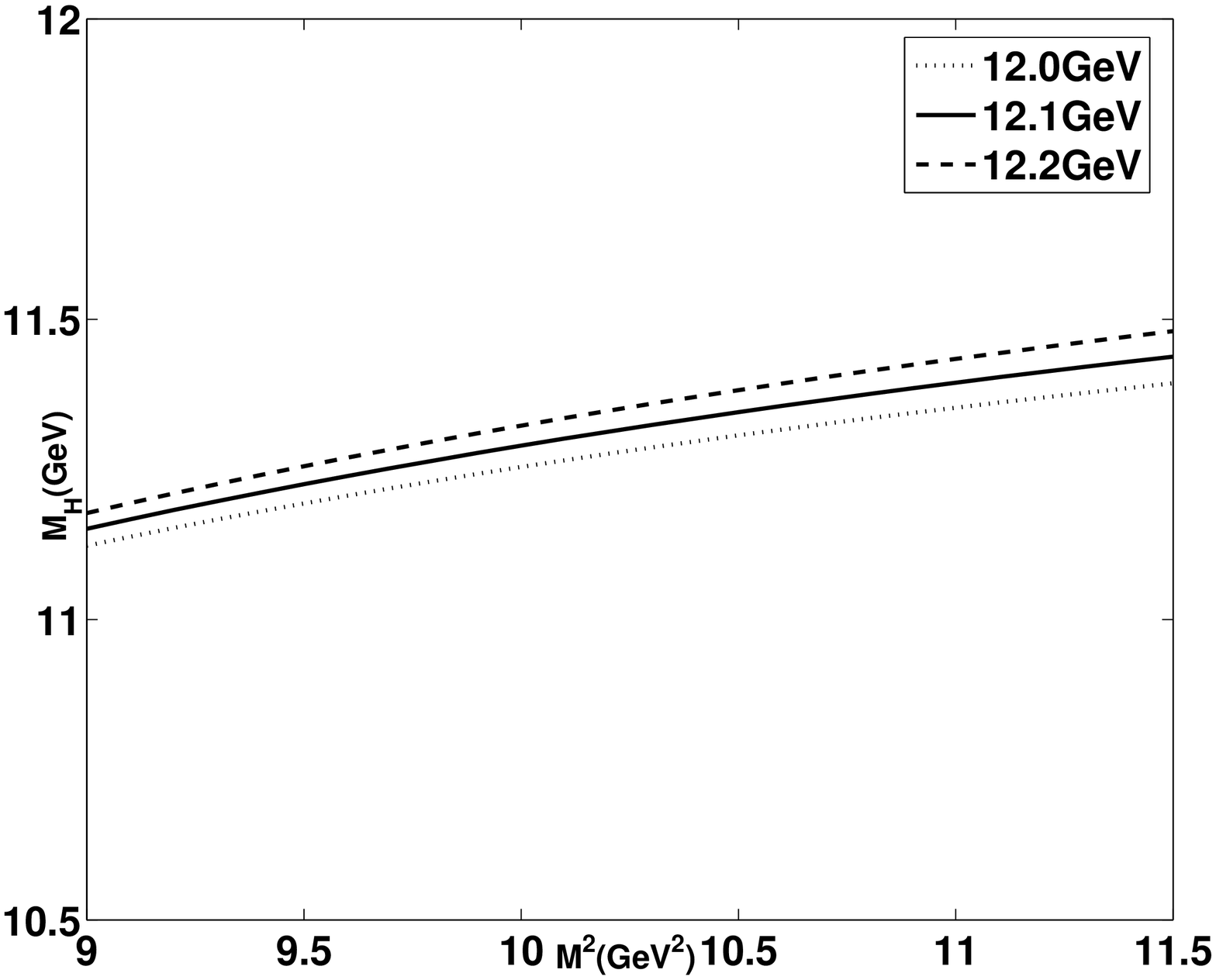}}\caption{The
dependence on $M^2$ for the masses of $D_{1}\bar{D}_{1}$ and
$B_{1}\bar{B}_{1}$ from sum rule (\ref{sum rule}). The continuum
thresholds are taken as $\sqrt{s_0}=5.3\sim5.5~\mbox{GeV}$ and
$\sqrt{s_0}=12.0\sim12.2~\mbox{GeV}$, respectively.} \label{fig:6}
\end{figure}

\begin{figure}
\centerline{\epsfysize=5.5truecm
\epsfbox{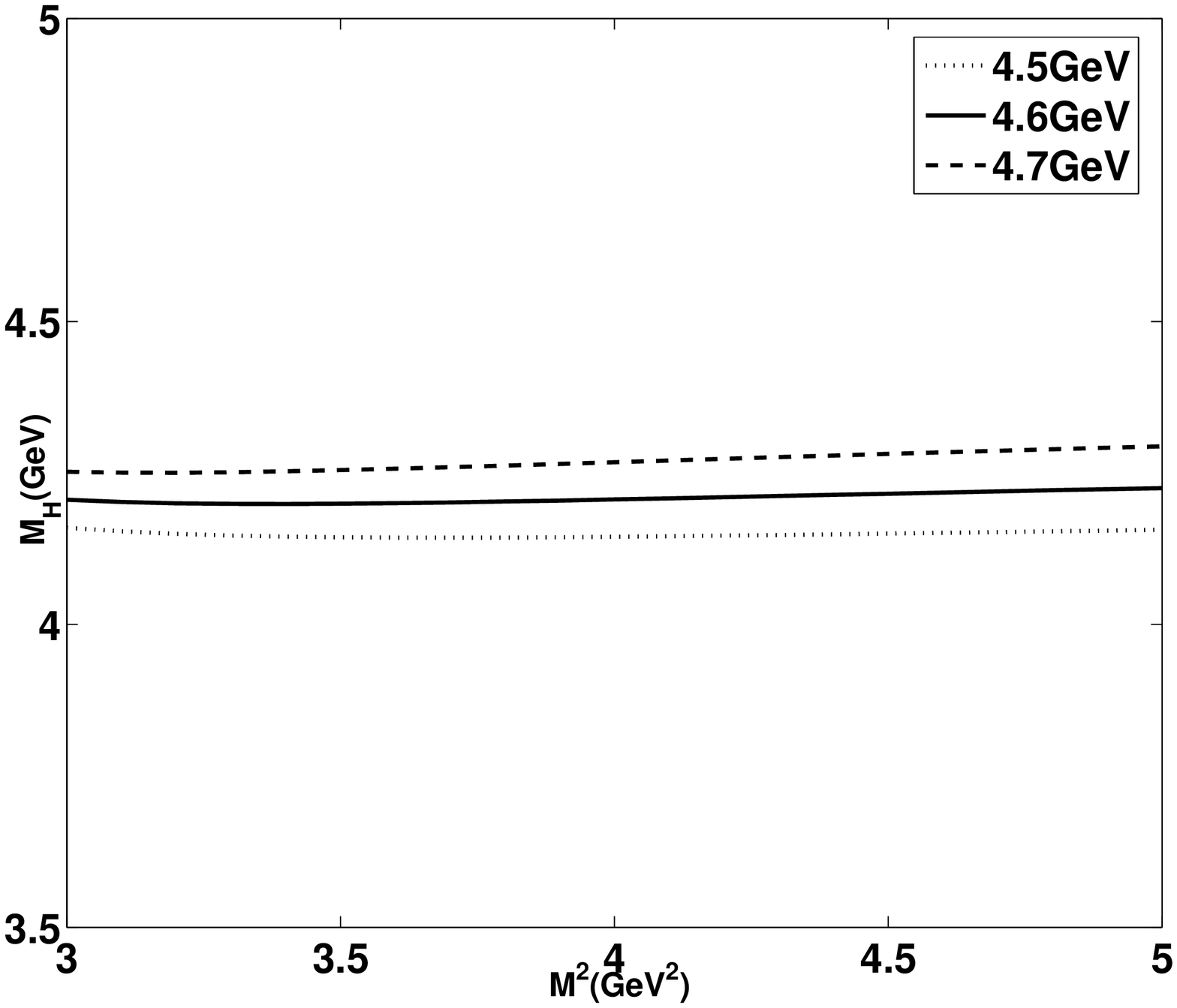}\epsfysize=5.5truecm\epsfbox{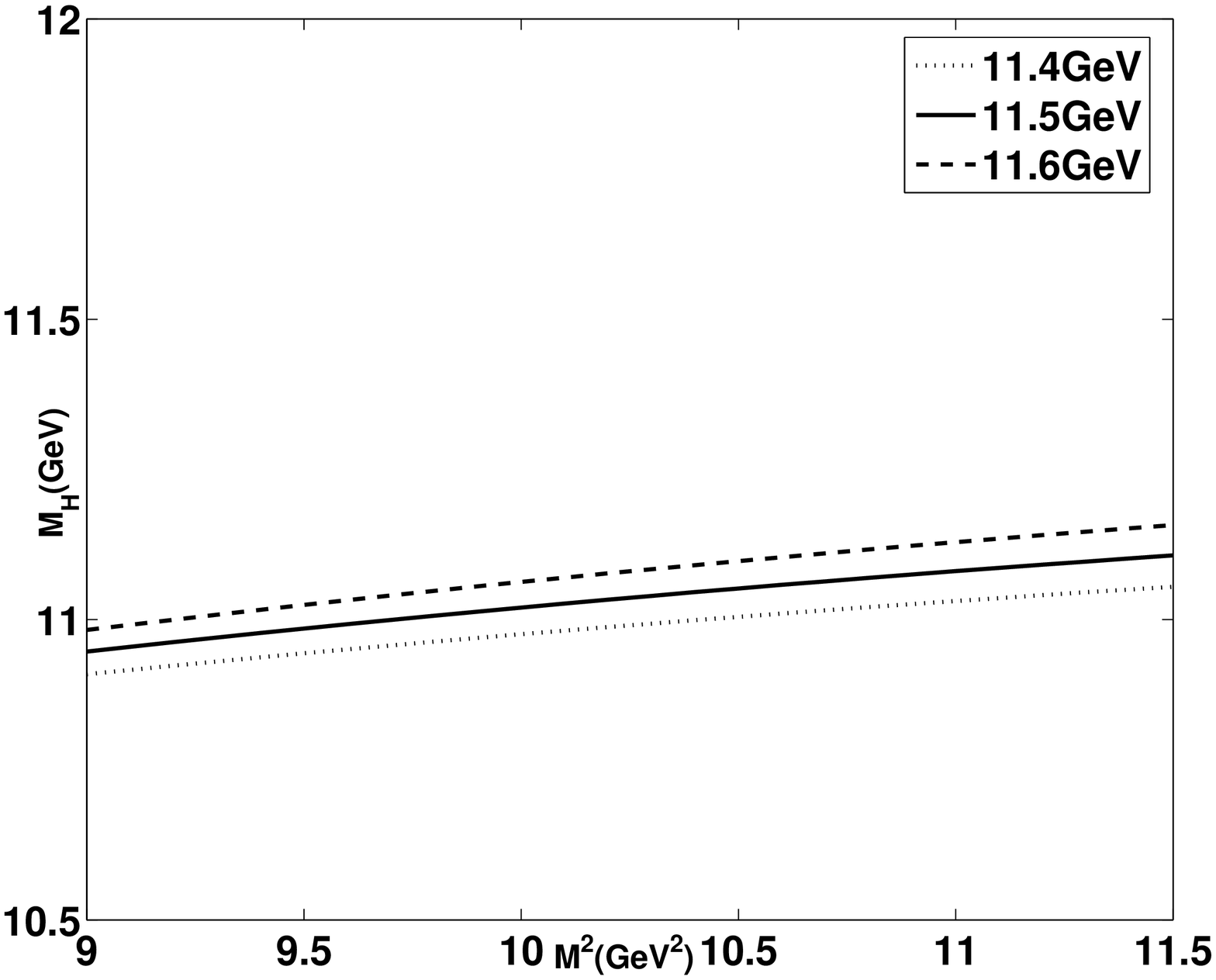}}\caption{The
dependence on $M^2$ for the masses of $D\bar{D}_{0}^{*}$ and
$B\bar{B}_{0}^{*}$ from sum rule (\ref{sum rule}). The continuum
thresholds are taken as $\sqrt{s_0}=4.5\sim4.7~\mbox{GeV}$ and
$\sqrt{s_0}=11.4\sim11.6~\mbox{GeV}$, respectively.} \label{fig:7}
\end{figure}

\begin{figure}
\centerline{\epsfysize=5.5truecm
\epsfbox{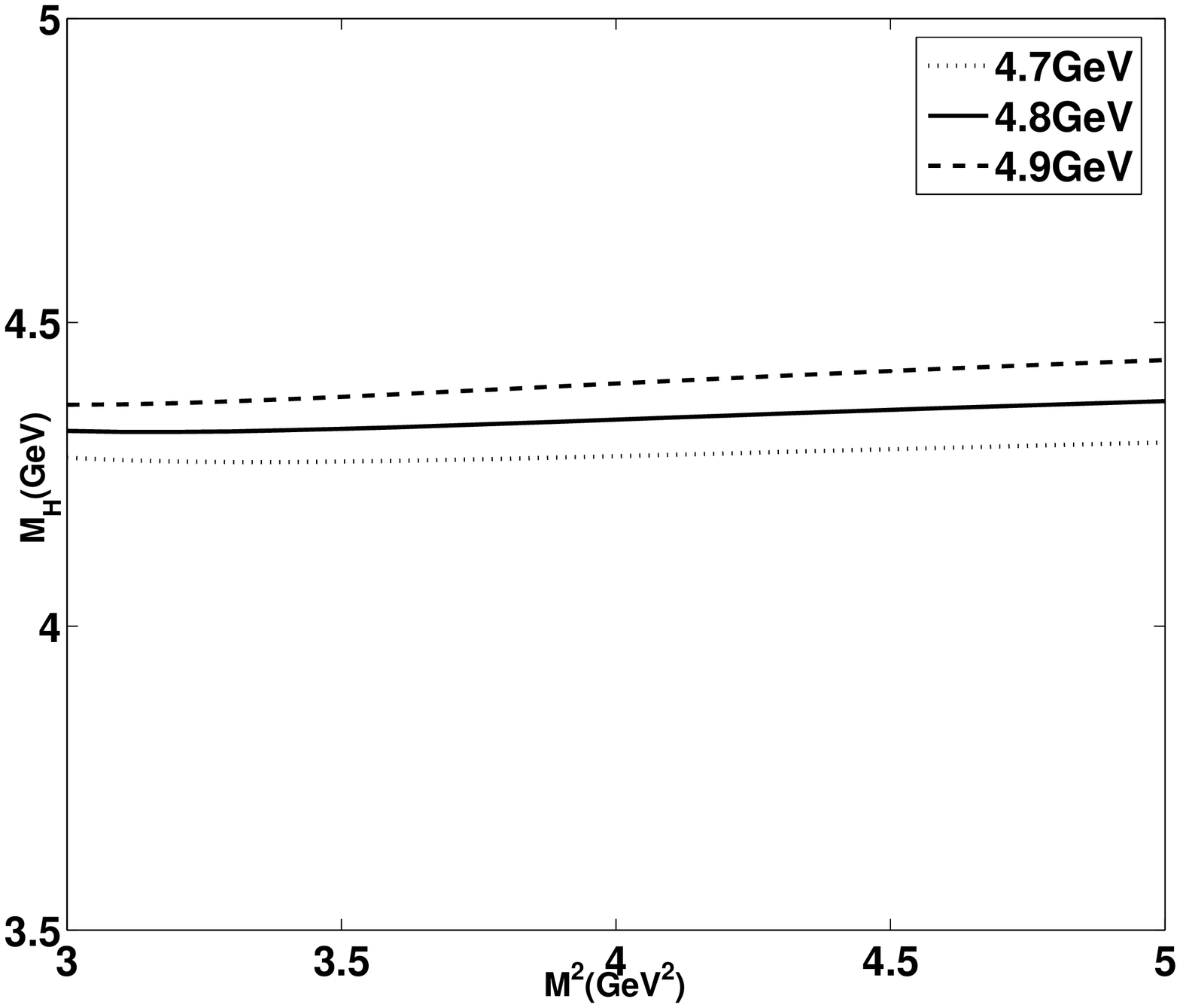}\epsfysize=5.5truecm\epsfbox{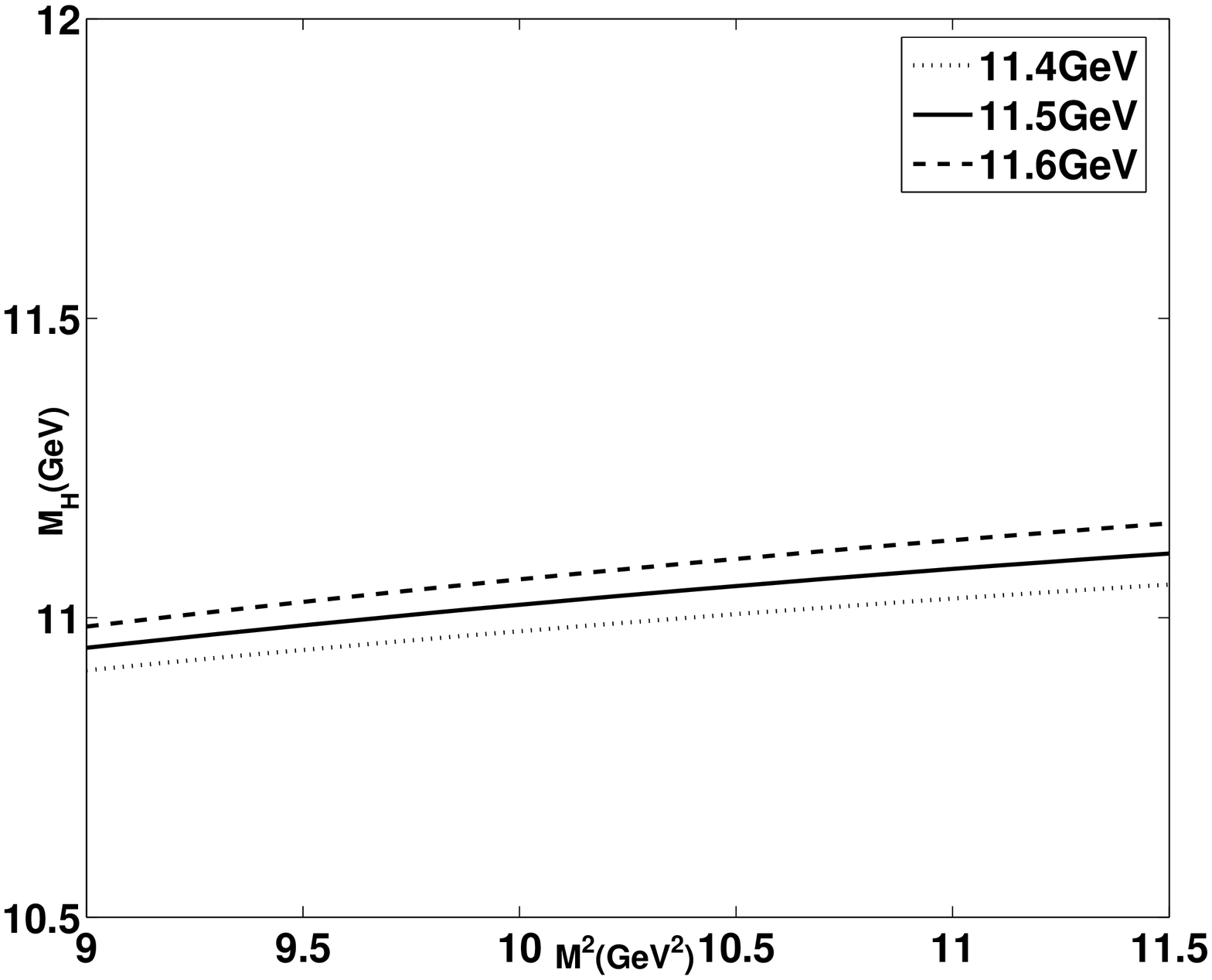}}\caption{The
dependence on $M^2$ for the masses of $D_{1}\bar{D}$ and
$B_{1}\bar{B}$ from sum rule (\ref{sum rule 1}). The continuum
thresholds are taken as $\sqrt{s_0}=4.7\sim4.9~\mbox{GeV}$ and
$\sqrt{s_0}=11.4\sim11.6~\mbox{GeV}$, respectively.} \label{fig:8}
\end{figure}

\begin{figure}
\centerline{\epsfysize=5.5truecm
\epsfbox{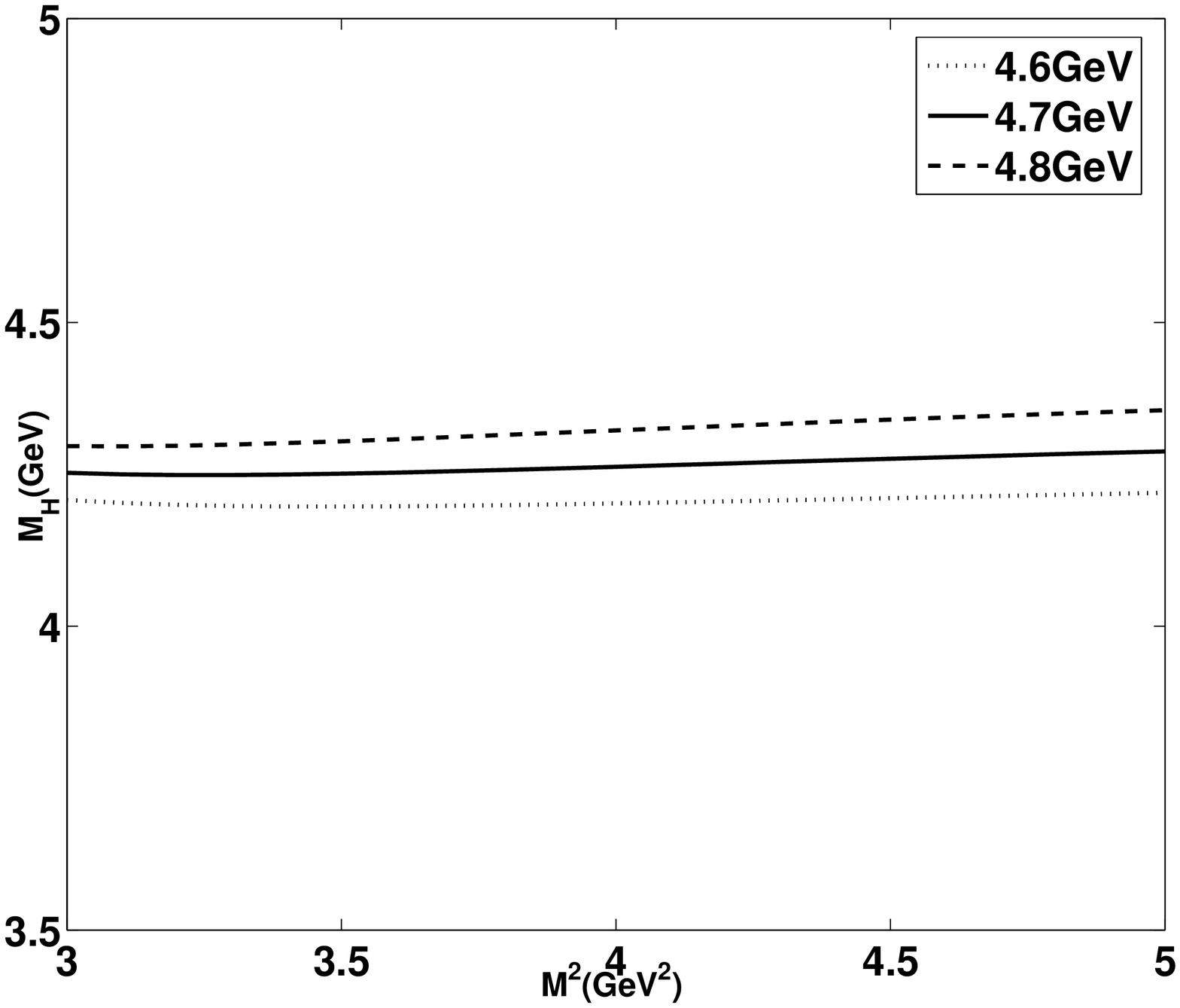}\epsfysize=5.5truecm\epsfbox{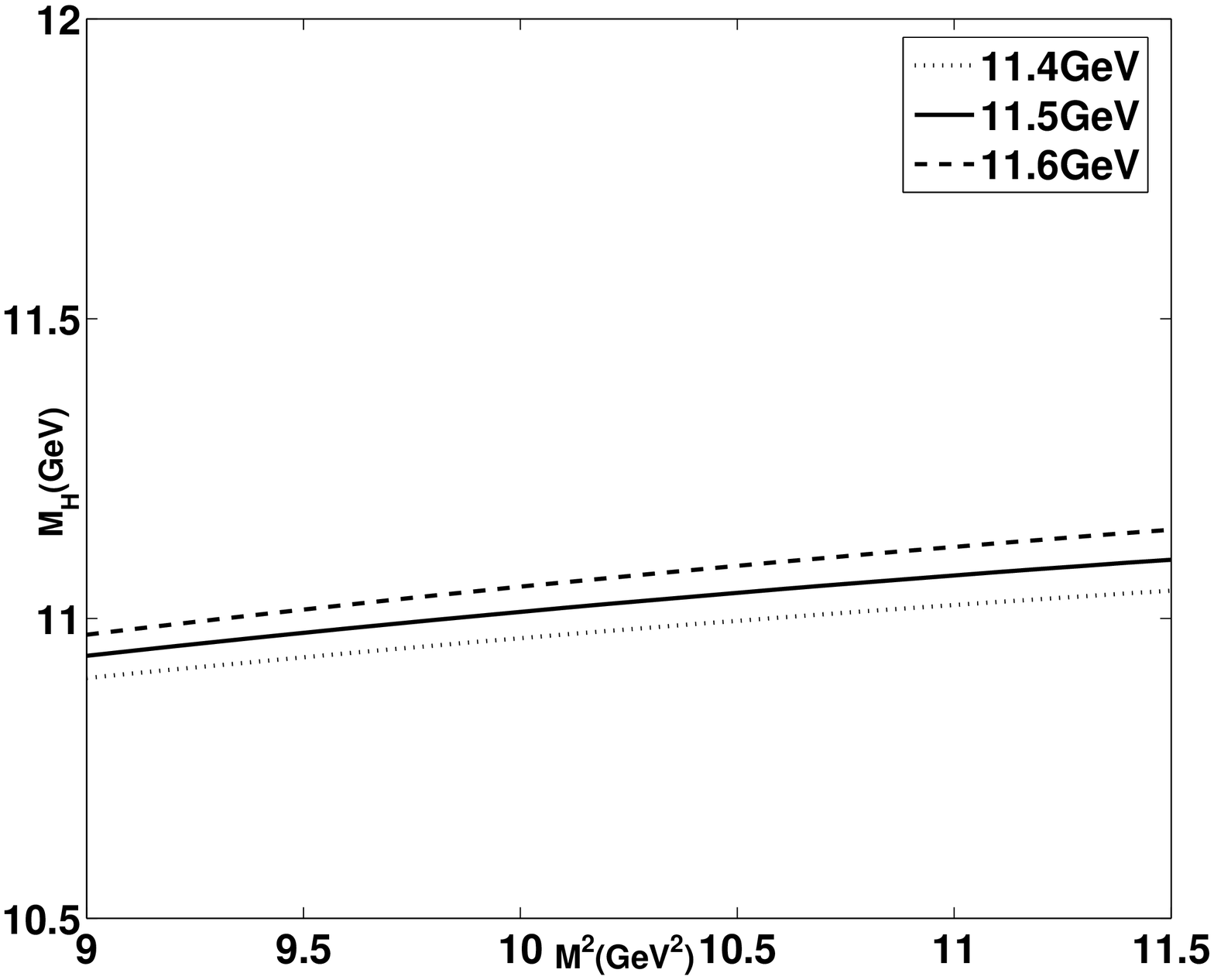}}\caption{The
dependence on $M^2$ for the masses of $D^{*}\bar{D}_{0}^{*}$ and
$B^{*}\bar{B}_{0}^{*}$ from sum rule (\ref{sum rule 1}). The
continuum thresholds are taken as $\sqrt{s_0}=4.6\sim4.8~\mbox{GeV}$
and $\sqrt{s_0}=11.4\sim11.6~\mbox{GeV}$, respectively.}
\label{fig:9}
\end{figure}

\begin{figure}
\centerline{\epsfysize=5.5truecm
\epsfbox{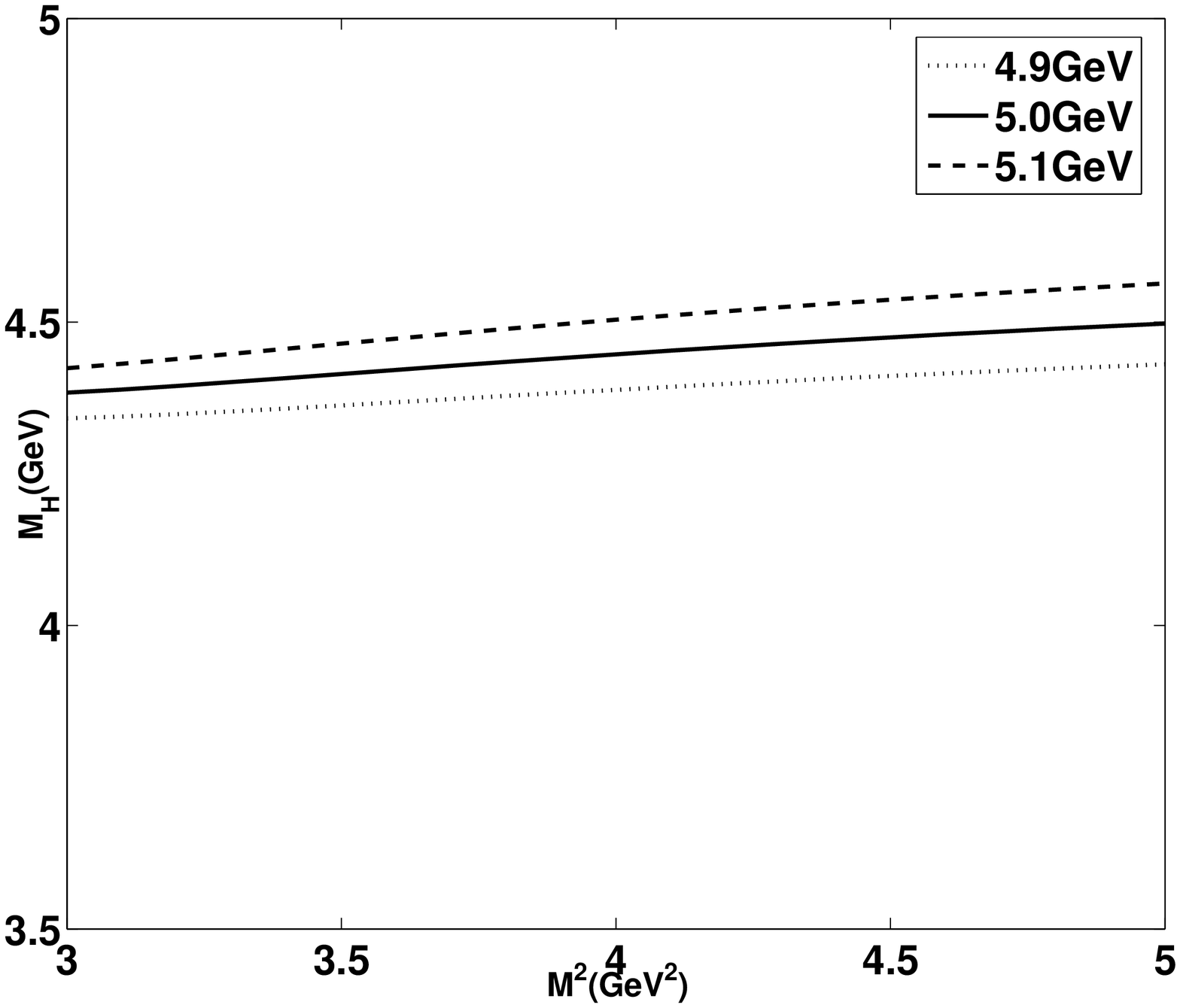}\epsfysize=5.5truecm\epsfbox{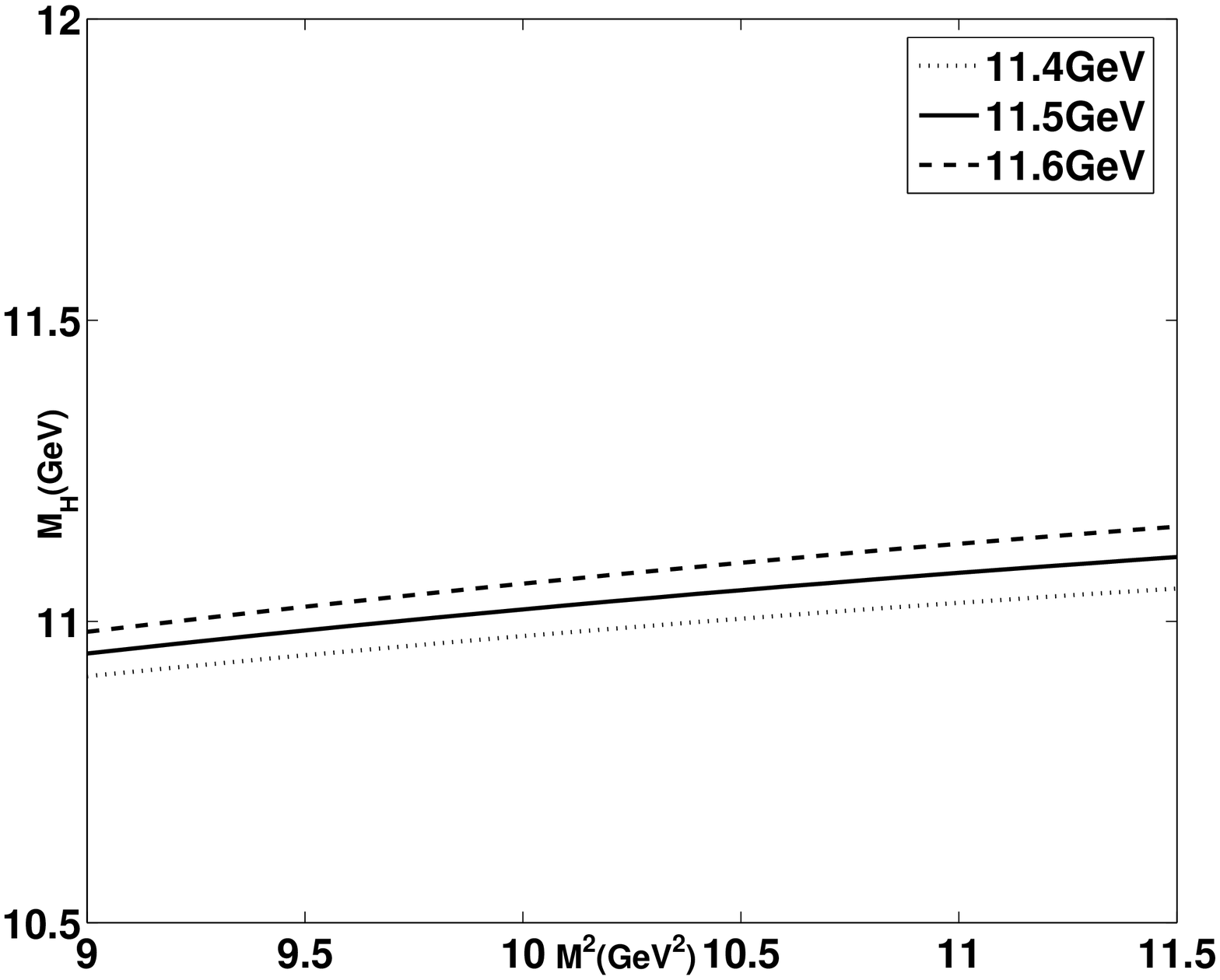}}\caption{The
dependence on $M^2$ for the masses of $D^{*}\bar{D}_{1}$ and
$B^{*}\bar{B}_{1}$ from sum rule (\ref{sum rule}). The continuum
thresholds are taken as $\sqrt{s_0}=4.9\sim5.1~\mbox{GeV}$ and
$\sqrt{s_0}=11.4\sim11.6~\mbox{GeV}$, respectively.} \label{fig:10}
\end{figure}

\begin{figure}
\centerline{\epsfysize=5.5truecm
\epsfbox{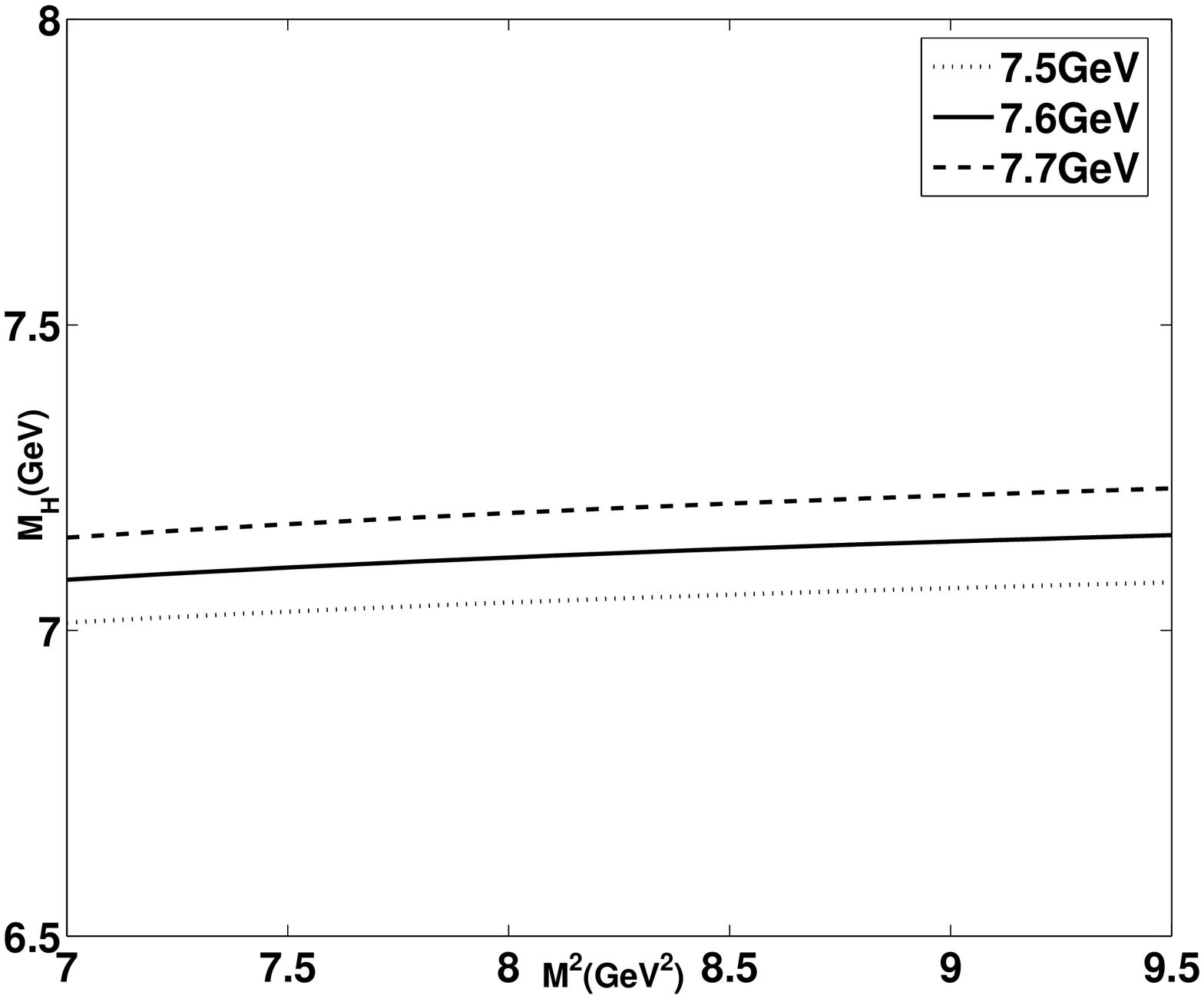}\epsfysize=5.5truecm\epsfbox{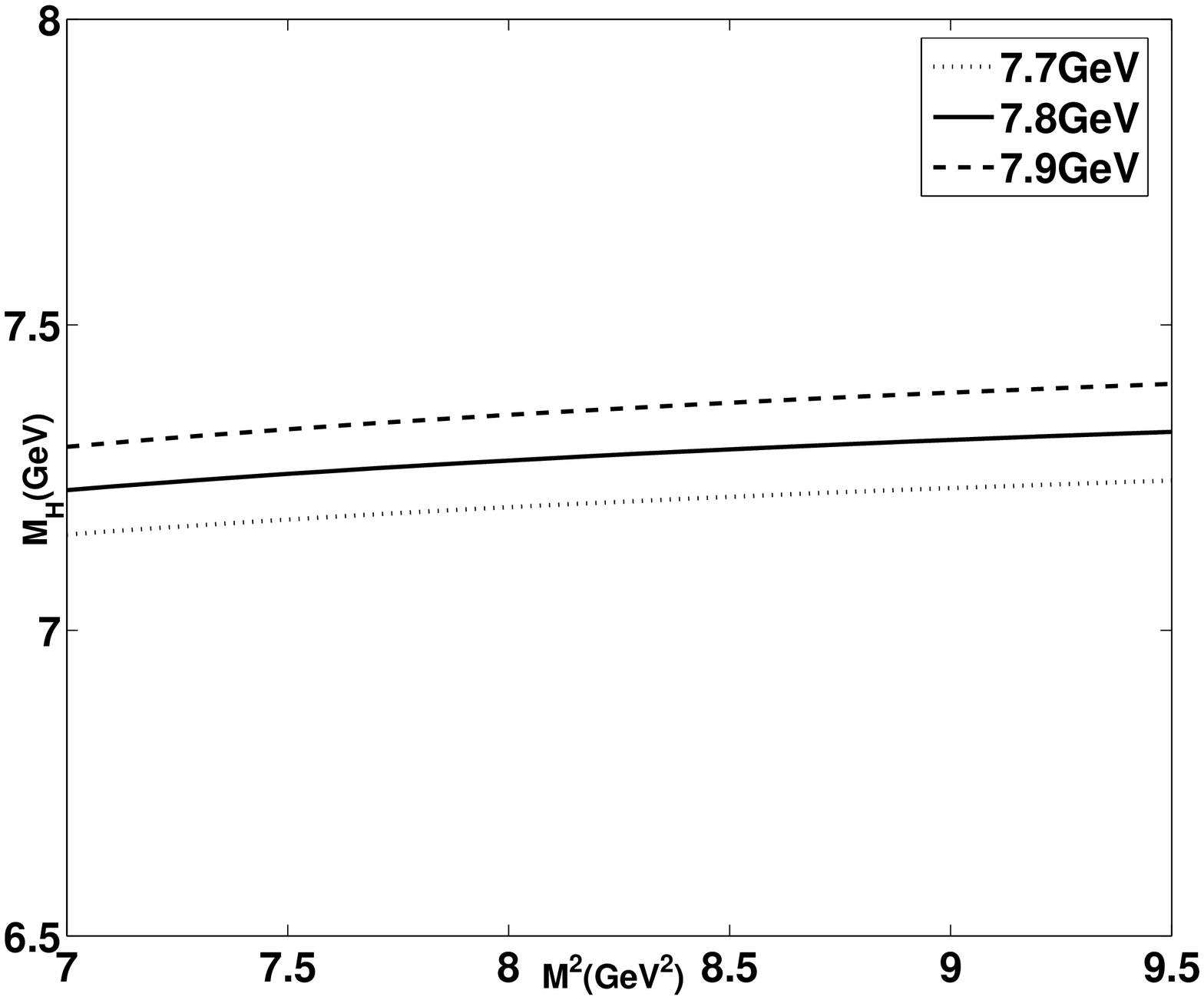}}\caption{The
dependence on $M^2$ for the masses of $B\bar{D}$ and
$B^{*}\bar{D}^{*}$ from sum rule (\ref{sum rule}). The continuum
thresholds are taken as $\sqrt{s_0}=7.5\sim7.7~\mbox{GeV}$ and
$\sqrt{s_0}=7.7\sim7.9~\mbox{GeV}$, respectively.} \label{fig:11}
\end{figure}

\begin{figure}
\centerline{\epsfysize=5.5truecm
\epsfbox{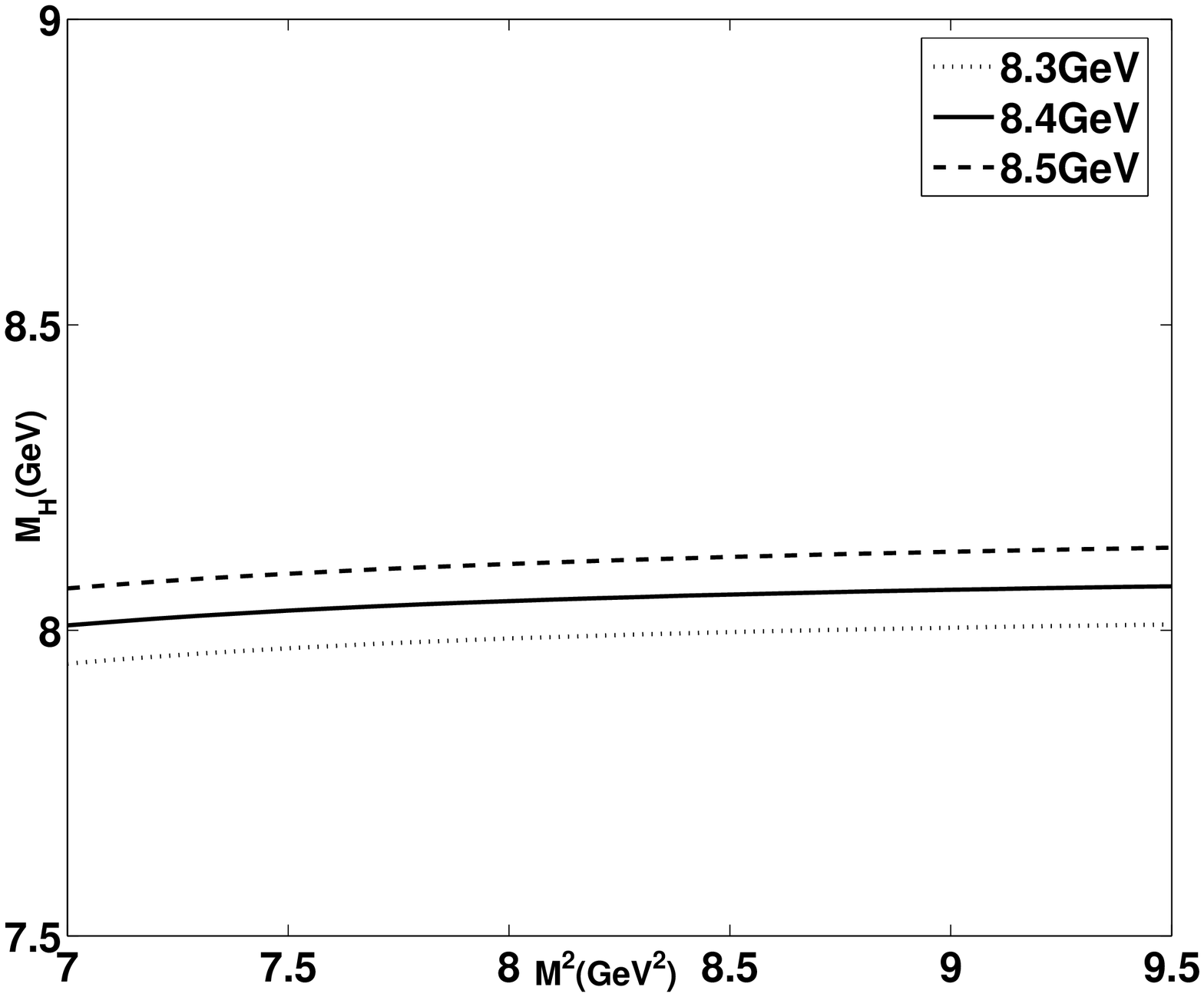}\epsfysize=5.5truecm\epsfbox{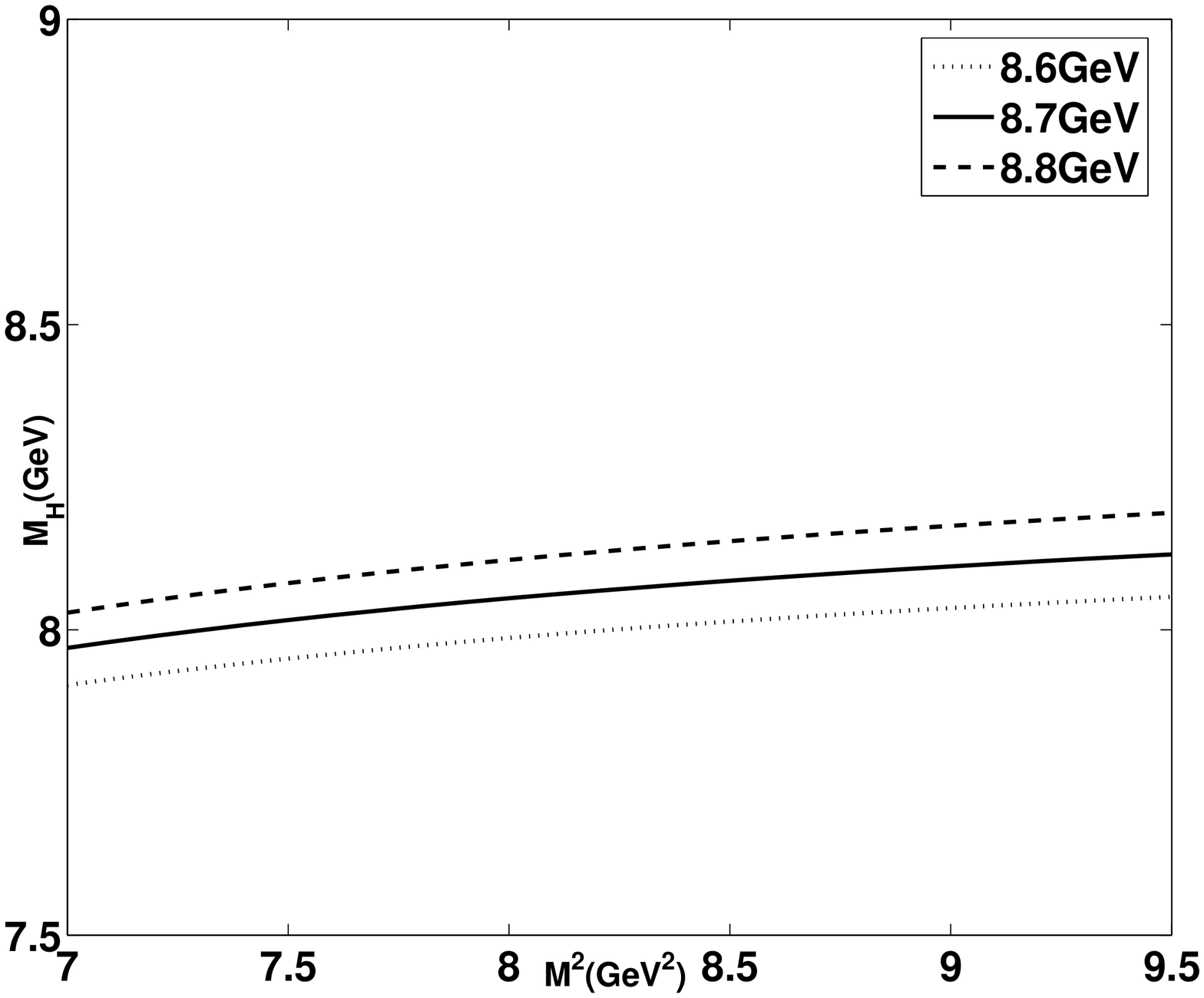}}\caption{The
dependence on $M^2$ for the masses of $B_{0}^{*}\bar{D}_{0}^{*}$ and
$B_{1}\bar{D}_{1}$ from sum rule (\ref{sum rule}). The continuum
thresholds are taken as $\sqrt{s_0}=8.3\sim8.5~\mbox{GeV}$ and
$\sqrt{s_0}=8.6\sim8.8~\mbox{GeV}$, respectively.} \label{fig:12}
\end{figure}

\begin{figure}
\centerline{\epsfysize=5.5truecm
\epsfbox{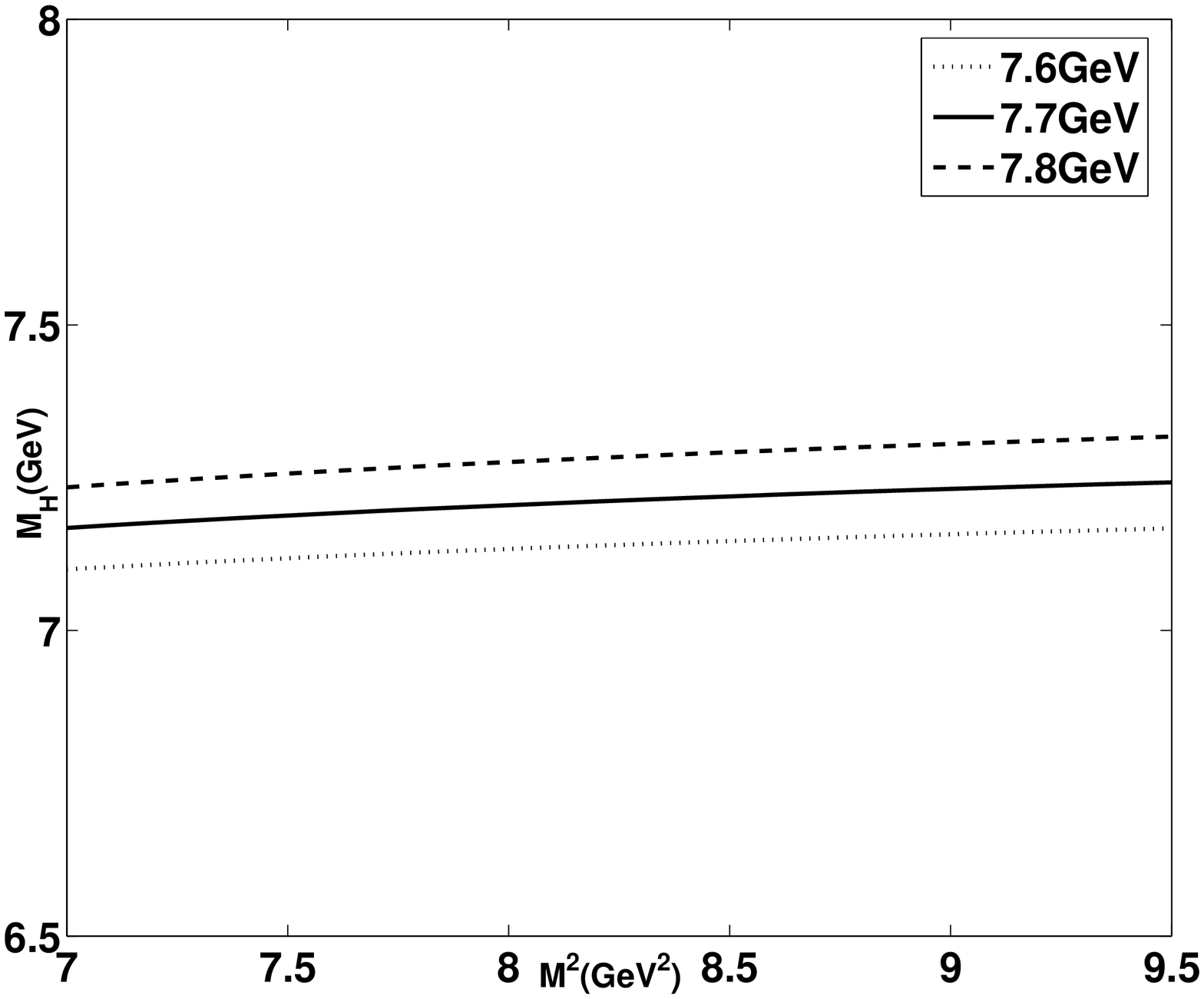}\epsfysize=5.5truecm\epsfbox{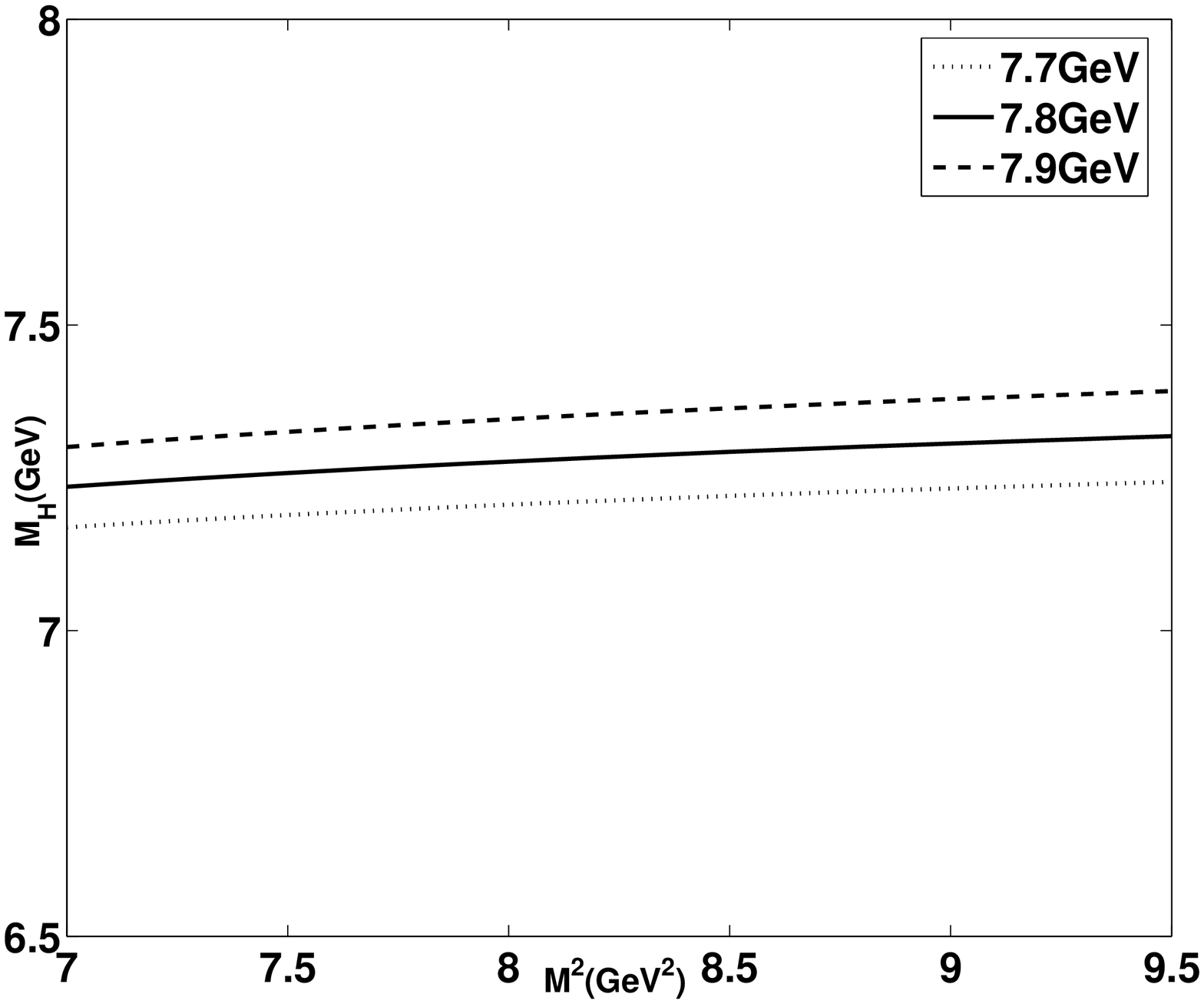}}\caption{The
dependence on $M^2$ for the masses of $D^{*}\bar{B}$ and
$B^{*}\bar{D}$ from sum rule (\ref{sum rule 1}). The continuum
thresholds are taken as $\sqrt{s_0}=7.6\sim7.8~\mbox{GeV}$ and
$\sqrt{s_0}=7.7\sim7.9~\mbox{GeV}$, respectively.} \label{fig:13}
\end{figure}

\begin{figure}
\centerline{\epsfysize=5.5truecm
\epsfbox{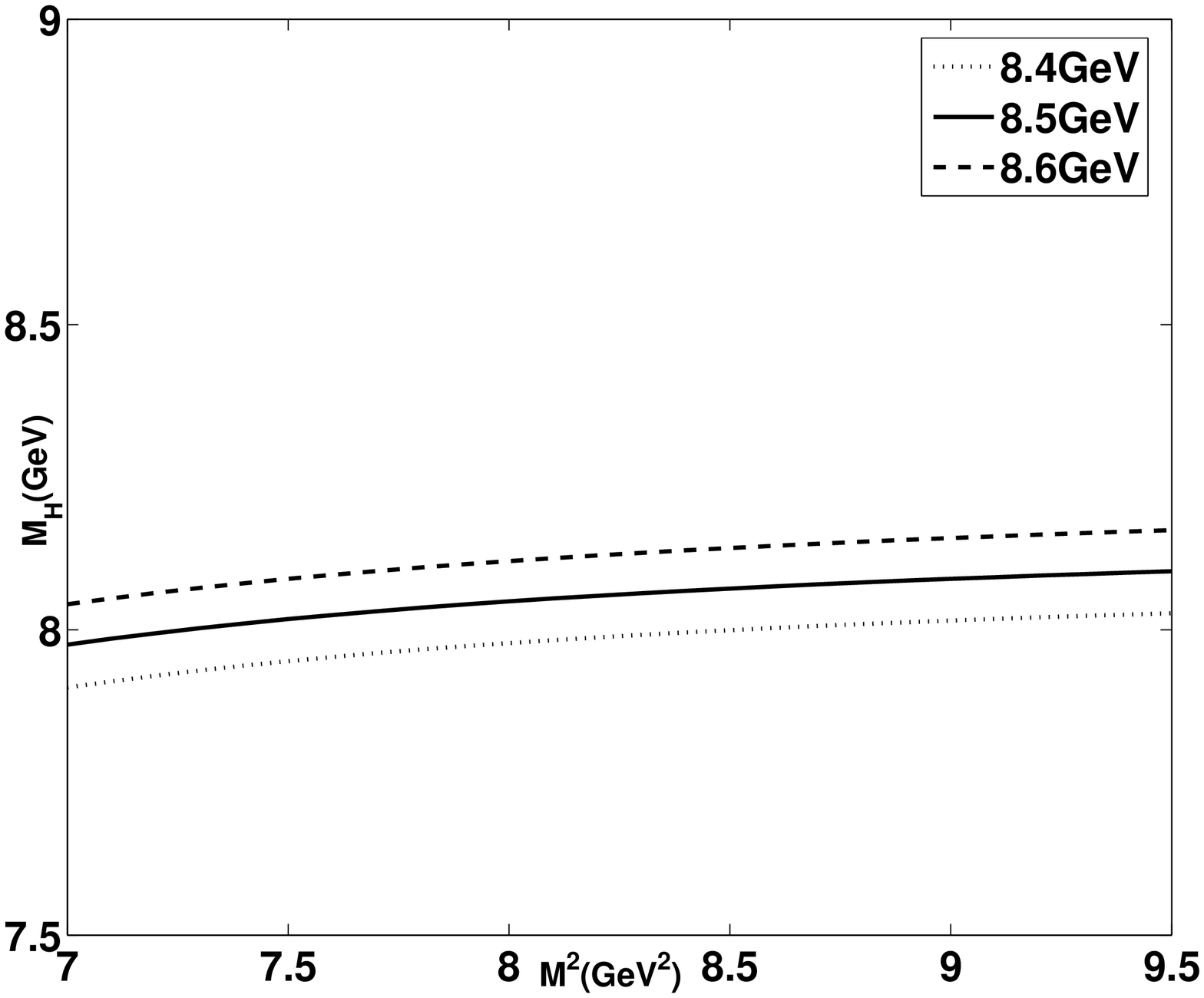}\epsfysize=5.5truecm\epsfbox{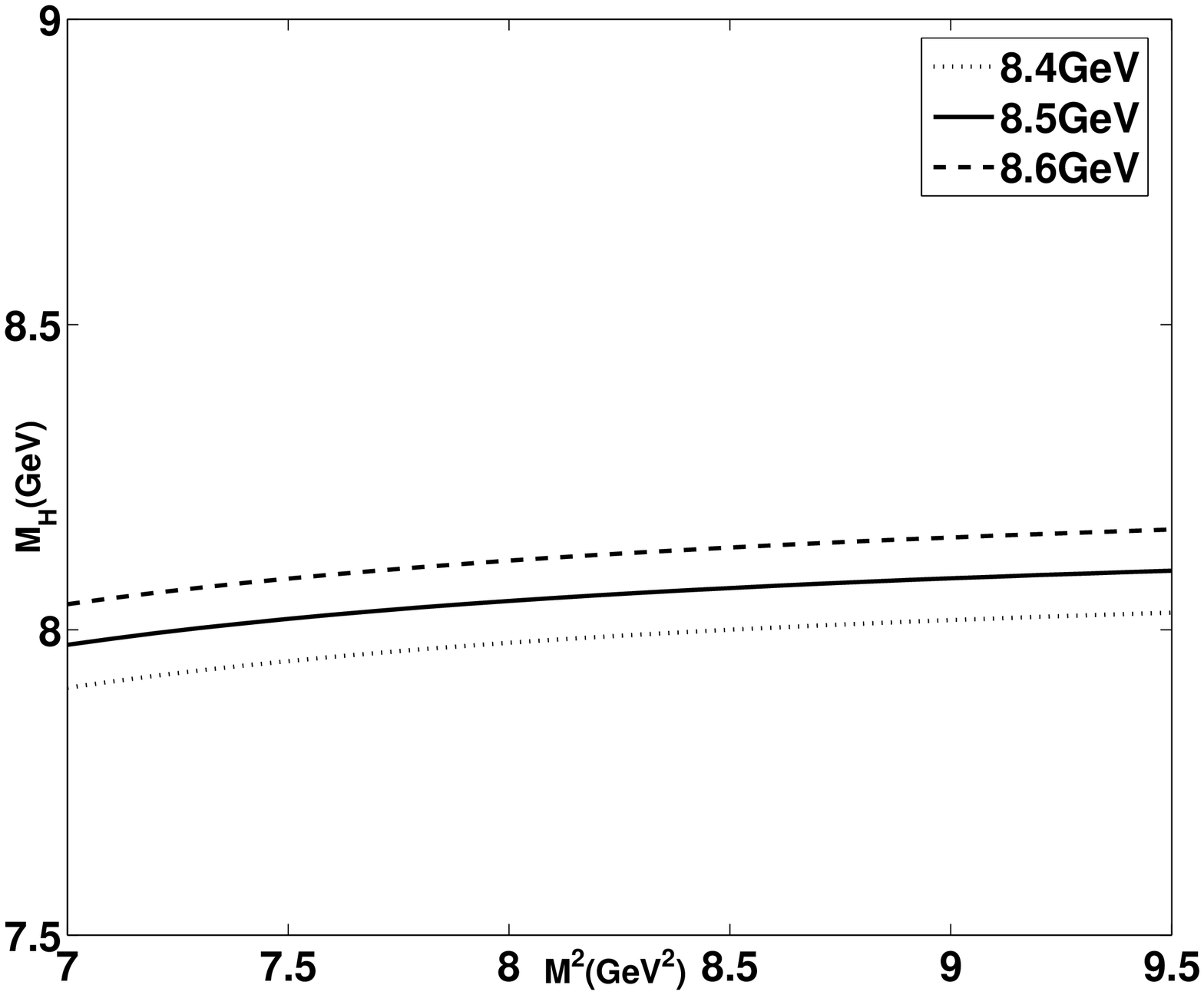}}\caption{The
dependence on $M^2$ for the masses of $D_{1}\bar{B}_{0}^{*}$ and
$B_{1}\bar{D}_{0}^{*}$ from sum rule (\ref{sum rule 1}). The
continuum thresholds are taken as $\sqrt{s_0}=8.4\sim8.6~\mbox{GeV}$
and $\sqrt{s_0}=8.4\sim8.6~\mbox{GeV}$, respectively.}
\label{fig:14}
\end{figure}

\begin{figure}
\centerline{\epsfysize=5.5truecm
\epsfbox{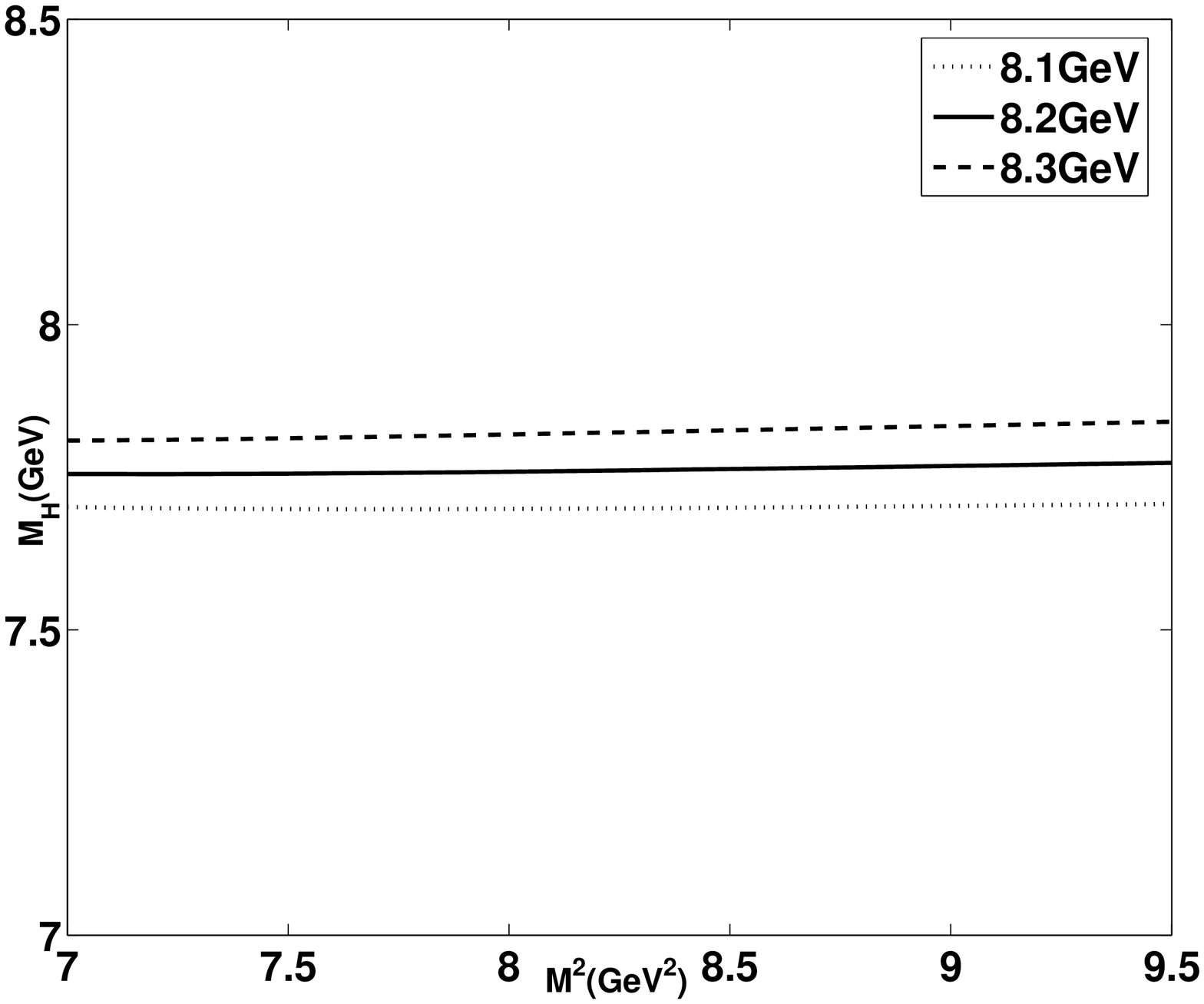}\epsfysize=5.5truecm\epsfbox{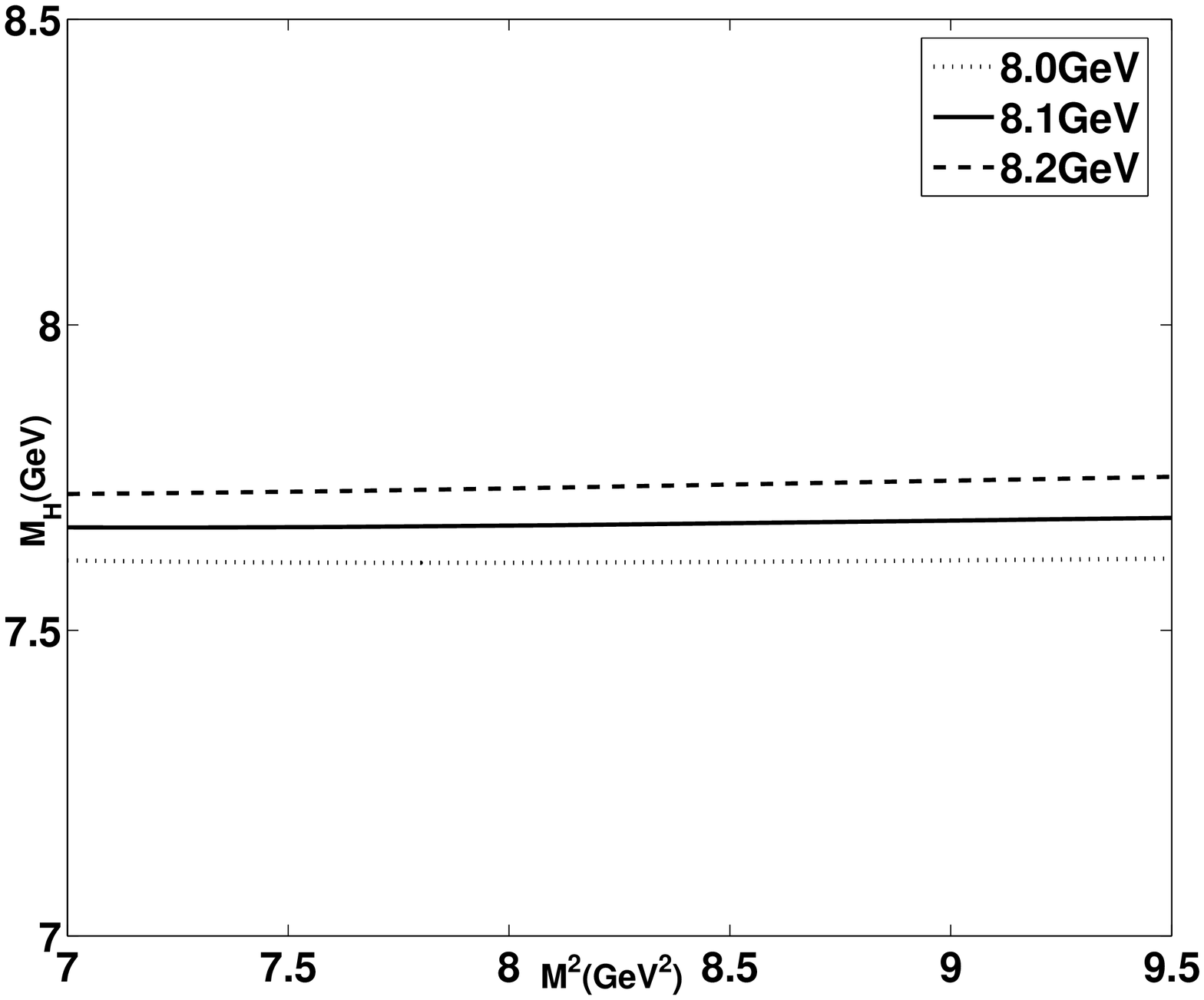}}\caption{The
dependence on $M^2$ for the masses of $D^{*}\bar{B}_{0}^{*}$ and
$B^{*}\bar{D}_{0}^{*}$ from sum rule (\ref{sum rule 1}). The
continuum thresholds are taken as $\sqrt{s_0}=8.1\sim8.3~\mbox{GeV}$
and $\sqrt{s_0}=8.0\sim8.2~\mbox{GeV}$, respectively.}
\label{fig:15}
\end{figure}

\begin{figure}
\centerline{\epsfysize=5.5truecm
\epsfbox{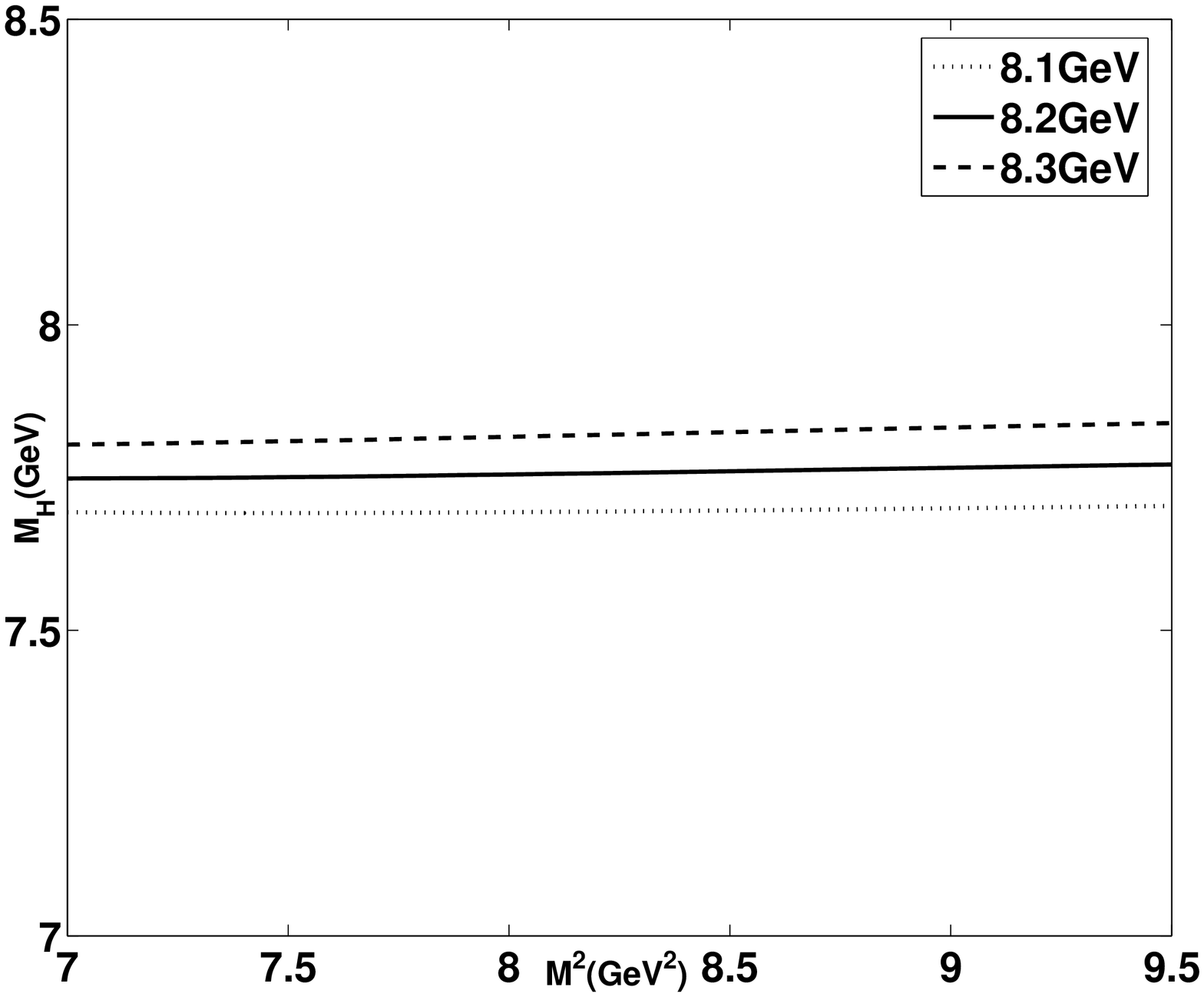}\epsfysize=5.5truecm\epsfbox{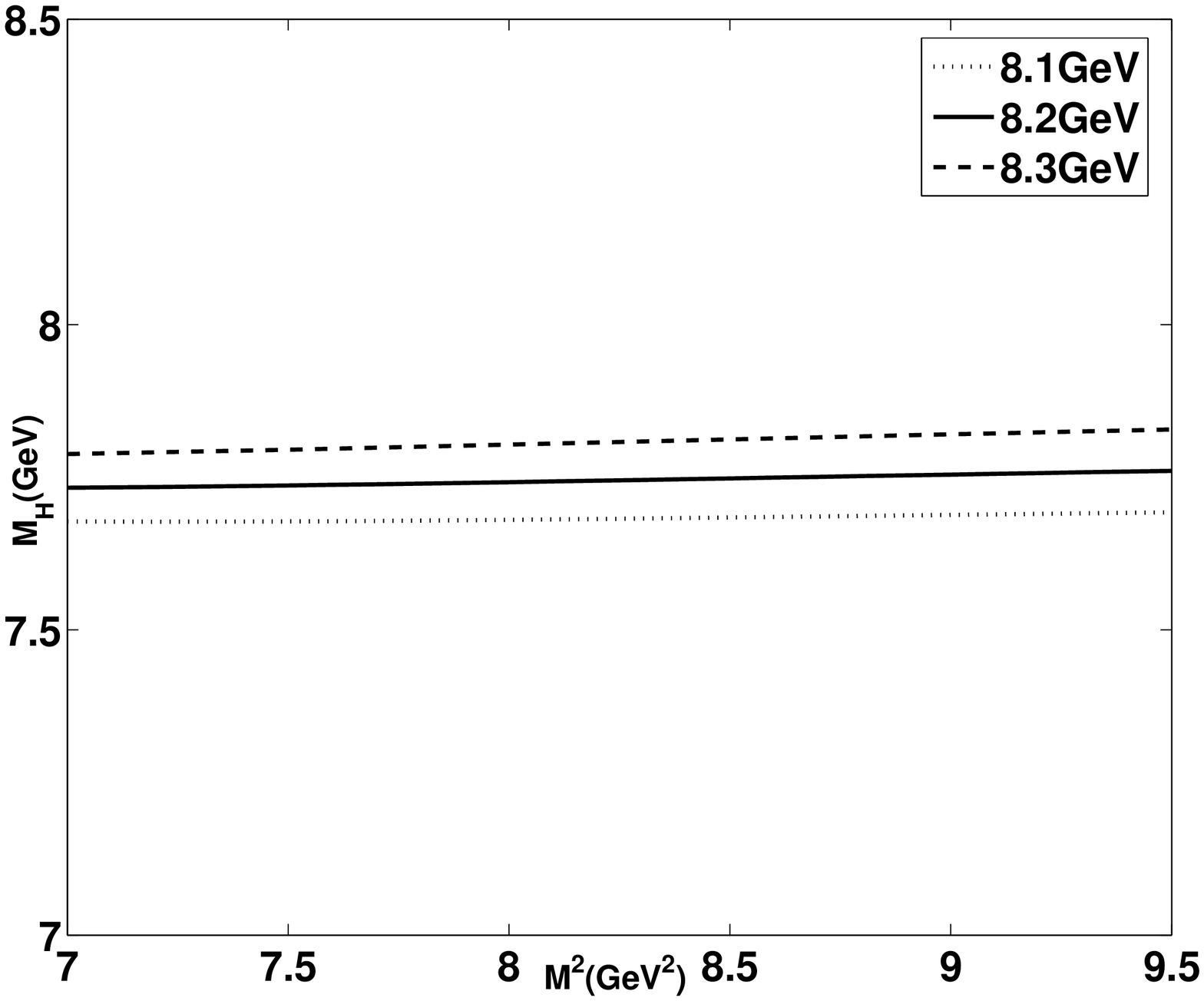}}\caption{The
dependence on $M^2$ for the masses of $D^{*}\bar{B}_{1}$ and
$B^{*}\bar{D}_{1}$ from sum rule (\ref{sum rule}). The continuum
thresholds are taken as $\sqrt{s_0}=8.1\sim8.3~\mbox{GeV}$ and
$\sqrt{s_0}=8.1\sim8.3~\mbox{GeV}$, respectively.} \label{fig:16}
\end{figure}

In summary, the QCD sum rules have been employed to calculate the
masses for $\{Q\bar{q}\}\{\bar{Q}^{(')}q\}$ molecular states,
including the contributions of the operators up to dimension six in
OPE. We have attained mass spectra for
$\{Q\bar{q}\}\{\bar{Q}^{(')}q\}$ molecular states in the end. In
molecular pictures, the numerical results for the masses of
$X(3872)$, $Z^{+}(4430)$, and $Y(3930)$ agree well with their
corresponding experimental values, which can support that $X(3872)$
could be a $D^{*}\bar{D}$ molecular state, $Z^{+}(4430)$ be a
$D^{*}\bar{D}_{1}$ molecular state, and $Y(3930)$ be a
$D^{*}\bar{D}^{*}$ one. On all accounts, all the numerical results
are expecting further experimental identification and it is looking
forward to more experimental evidence on molecular states.

\begin{figure}
\centerline{\epsfysize=5.5truecm
\epsfbox{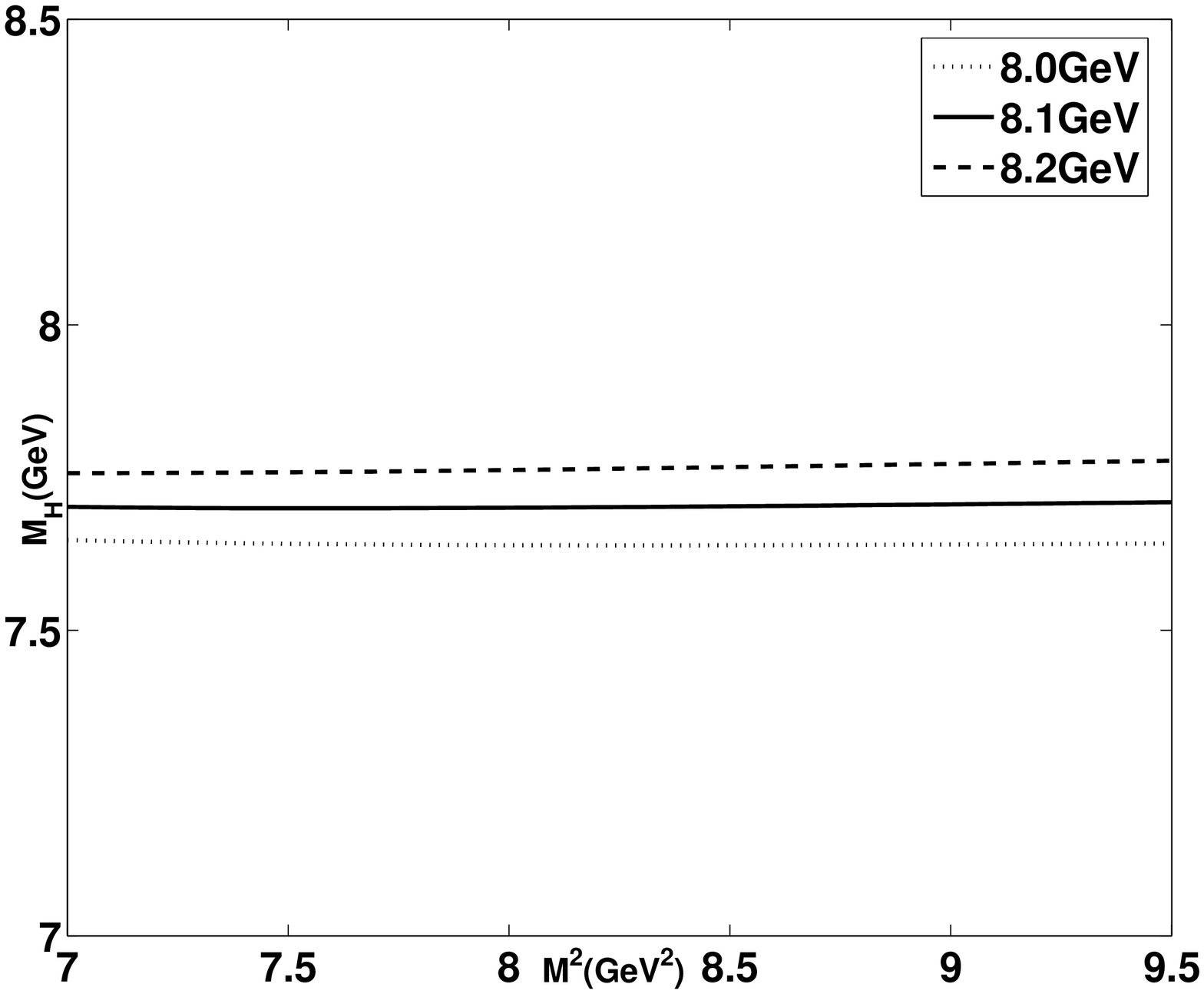}\epsfysize=5.5truecm\epsfbox{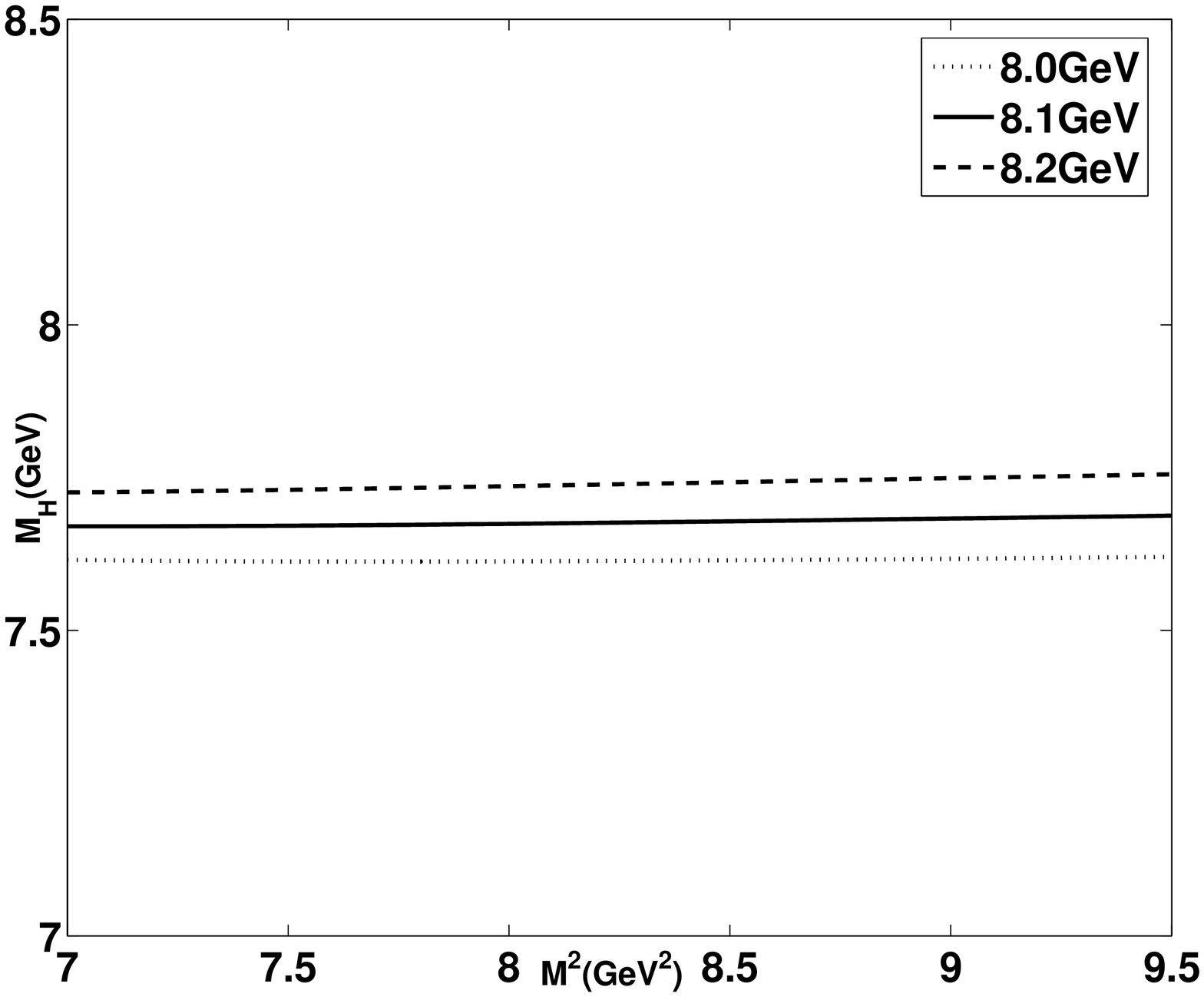}}\caption{The
dependence on $M^2$ for the masses of $D\bar{B}_{0}^{*}$ and
$B\bar{D}_{0}^{*}$ from sum rule (\ref{sum rule}). The continuum
thresholds are taken as $\sqrt{s_0}=8.0\sim8.2~\mbox{GeV}$ and
$\sqrt{s_0}=8.0\sim8.2~\mbox{GeV}$, respectively.} \label{fig:17}
\end{figure}

\begin{figure}
\centerline{\epsfysize=5.5truecm
\epsfbox{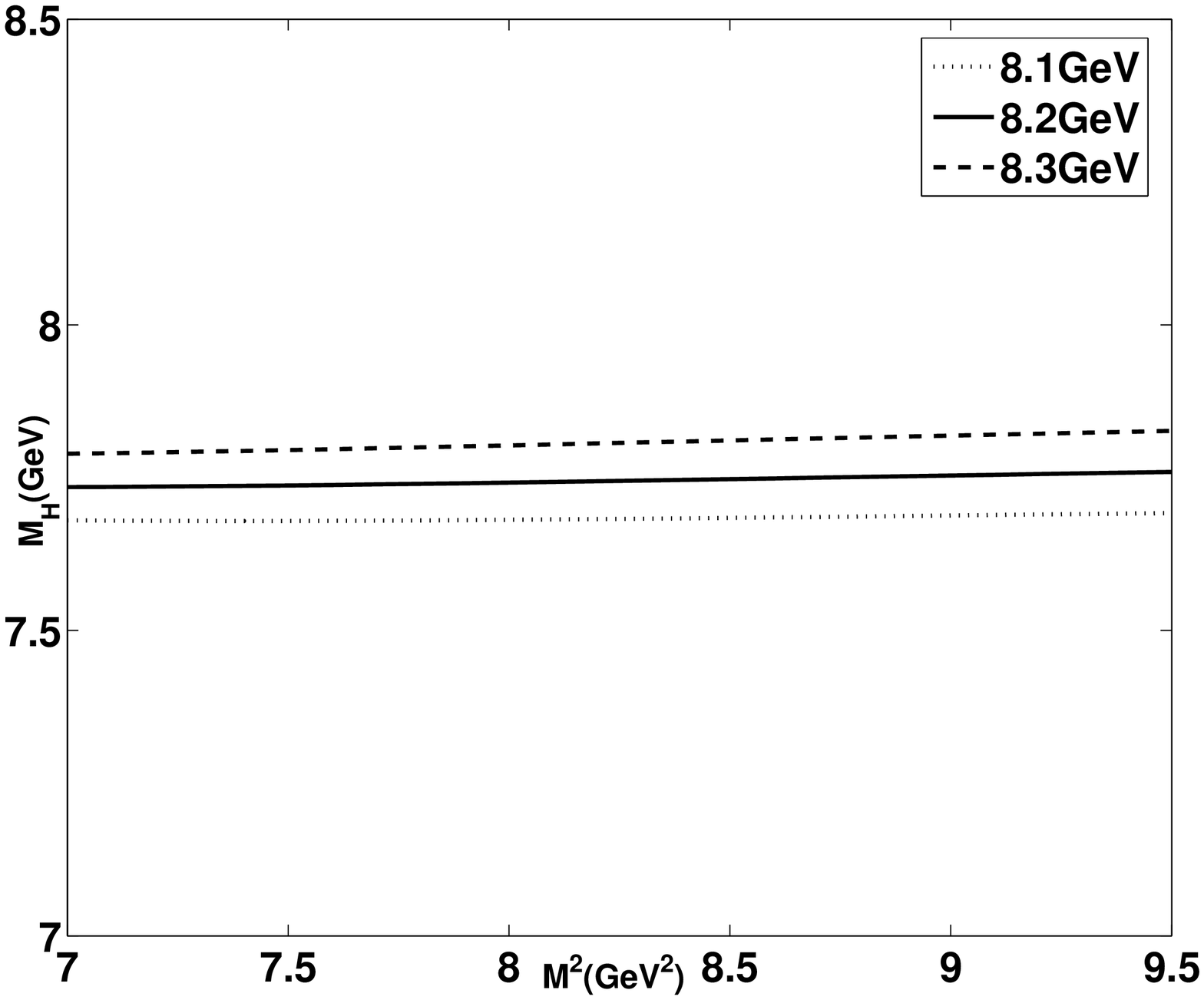}\epsfysize=5.5truecm\epsfbox{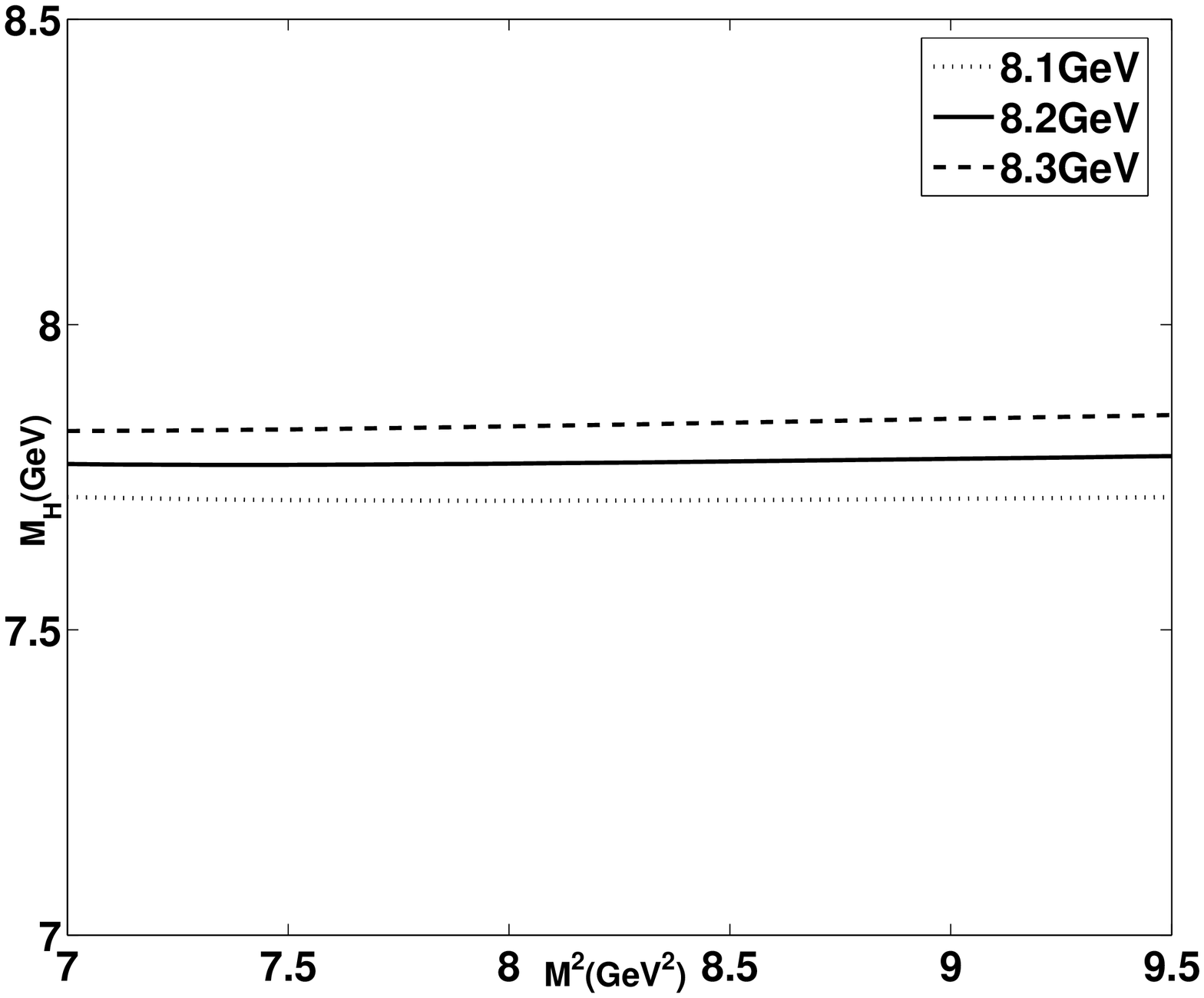}}\caption{The
dependence on $M^2$ for the masses of $D_{1}\bar{B}$ and
$B_{1}\bar{D}$ from sum rule (\ref{sum rule 1}). The continuum
thresholds are taken as $\sqrt{s_0}=8.1\sim8.3~\mbox{GeV}$ and
$\sqrt{s_0}=8.1\sim8.3~\mbox{GeV}$, respectively.} \label{fig:18}
\end{figure}

\appendix
\section*{appendix}
The spectral densities are distinguished for two kinds of doubly
heavy molecular states, namely, containing the same or differently
heavy quarks. It is defined that $r(m_{Q_1},m_{Q_2}) = \alpha
m_{Q_1}^2 + \beta m_{Q_2}^2 - \alpha \beta s$. Concretely,
$r(m_{Q},m_{Q}) = \alpha m_{Q}^2 + \beta m_{Q}^2 - \alpha \beta s$
and $r(m_{Q},m_{Q'}) = \alpha m_{Q}^2 + \beta m_{Q'}^2 - \alpha
\beta s$. First, with

\begin{eqnarray}
\rho^{\mbox{OPE}}(s)=\rho^{\mbox{pert}}(s)+\rho^{\langle\bar{q}q\rangle}(s)+\rho^{\langle\bar{q}q\rangle^{2}}(s)+\rho^{\langle
g\bar{q}\sigma\cdot G q\rangle}(s)+\rho^{\langle
g^{2}G^{2}\rangle}(s)+\rho^{\langle g^{3}G^{3}\rangle}(s),\nonumber
\end{eqnarray}

\begin{eqnarray}
\rho^{\mbox{pert}}(s)&=&\frac{3}{2^{11}\pi^{6}}\int_{\alpha_{min}}^{\alpha_{max}}\frac{d\alpha}{\alpha^{3}}\int_{\beta_{min}}^{1-\alpha}\frac{d\beta}{\beta^{3}}(1-\alpha-\beta)r(m_{Q},m_{Q})^{4},\nonumber
\end{eqnarray}

\begin{eqnarray}
\rho^{\langle\bar{q}q\rangle}(s)&=&-\frac{3\langle\bar{q}q\rangle}{2^{6}\pi^{4}}m_{Q}\int_{\alpha_{min}}^{\alpha_{max}}\frac{d\alpha}{\alpha^{2}}\int_{\beta_{min}}^{1-\alpha}\frac{d\beta}{\beta}r(m_{Q},m_{Q})^{2},\nonumber
\end{eqnarray}

\begin{eqnarray}
\rho^{\langle\bar{q}q\rangle^{2}}(s)&=&\frac{\langle\bar{q}q\rangle^{2}}{2^{4}\pi^{2}}m_{Q}^{2}\sqrt{1-4m_{Q}^{2}/s},\nonumber
\end{eqnarray}

\begin{eqnarray}
\rho^{\langle g\bar{q}\sigma\cdot G q\rangle}(s)&=&-\frac{3\langle
g\bar{q}\sigma\cdot G
q\rangle}{2^{7}\pi^{4}}m_{Q}\int_{\alpha_{min}}^{\alpha_{max}}\frac{d\alpha}{1-\alpha}[m_{Q}^{2}-\alpha(1-\alpha)
s],\nonumber
\end{eqnarray}

\begin{eqnarray}
\rho^{\langle g^{2}G^{2}\rangle}(s)&=&\frac{\langle
g^{2}G^{2}\rangle}{2^{10}\pi^{6}}m_{Q}^{2}\int_{\alpha_{min}}^{\alpha_{max}}\frac{d\alpha}{\alpha^{3}}\int_{\beta_{min}}^{1-\alpha}d\beta(1-\alpha-\beta)r(m_{Q},m_{Q}),\nonumber
\end{eqnarray}

\begin{eqnarray}
\rho^{\langle g^{3}G^{3}\rangle}(s)&=&\frac{\langle
g^{3}G^{3}\rangle}{2^{12}\pi^{6}}\int_{\alpha_{min}}^{\alpha_{max}}\frac{d\alpha}{\alpha^{3}}\int_{\beta_{min}}^{1-\alpha}d\beta(1-\alpha-\beta)[r(m_{Q},m_{Q})+2m_{Q}^{2}\beta],\nonumber
\end{eqnarray}
for $(Q\bar{q})(\bar{Q}q)$,

\begin{eqnarray}
\rho^{\mbox{OPE}}(s)=-\{\rho^{\mbox{pert}}(s)+\rho^{\langle\bar{q}q\rangle}(s)+\rho^{\langle\bar{q}q\rangle^{2}}(s)+\rho^{\langle
g\bar{q}\sigma\cdot G q\rangle}(s)+\rho^{\langle
g^{2}G^{2}\rangle}(s)+\rho^{\langle
g^{3}G^{3}\rangle}(s)\},\nonumber
\end{eqnarray}

\begin{eqnarray}
\rho^{\mbox{pert}}(s)&=&-\frac{3}{2^{12}\pi^{6}}\int_{\alpha_{min}}^{\alpha_{max}}\frac{d\alpha}{\alpha^{3}}\int_{\beta_{min}}^{1-\alpha}\frac{d\beta}{\beta^{3}}(1-\alpha-\beta)(1+\alpha+\beta)r(m_{Q},m_{Q})^{4},\nonumber
\end{eqnarray}

\begin{eqnarray}
\rho^{\langle\bar{q}q\rangle}(s)&=&\frac{3\langle\bar{q}q\rangle}{2^{7}\pi^{4}}m_{Q}\int_{\alpha_{min}}^{\alpha_{max}}\frac{d\alpha}{\alpha^{2}}\int_{\beta_{min}}^{1-\alpha}\frac{d\beta}{\beta}(1+\alpha+\beta)r(m_{Q},m_{Q})^{2},\nonumber
\end{eqnarray}

\begin{eqnarray}
\rho^{\langle\bar{q}q\rangle^{2}}(s)&=&-\frac{\langle\bar{q}q\rangle^{2}}{2^{4}\pi^{2}}m_{Q}^{2}\sqrt{1-4m_{Q}^{2}/s},\nonumber
\end{eqnarray}

\begin{eqnarray}
\rho^{\langle g\bar{q}\sigma\cdot G q\rangle}(s)&=&\frac{3\langle
g\bar{q}\sigma\cdot G
q\rangle}{2^{8}\pi^{4}}m_{Q}\int_{\alpha_{min}}^{\alpha_{max}}d\alpha\times\{-\int_{\beta_{min}}^{1-\alpha}\frac{d\beta}{\beta}r(m_{Q},m_{Q})+\frac{2}{1-\alpha}[m_{Q}^{2}-\alpha(1-\alpha)
s]\},\nonumber
\end{eqnarray}

\begin{eqnarray}
\rho^{\langle g^{2}G^{2}\rangle}(s)&=&-\frac{\langle
g^{2}G^{2}\rangle}{2^{11}\pi^{6}}m_{Q}^{2}\int_{\alpha_{min}}^{\alpha_{max}}\frac{d\alpha}{\alpha^{3}}\int_{\beta_{min}}^{1-\alpha}d\beta(1-\alpha-\beta)(1+\alpha+\beta)r(m_{Q},m_{Q}),\nonumber
\end{eqnarray}

\begin{eqnarray}
\rho^{\langle g^{3}G^{3}\rangle}(s)&=&-\frac{\langle
g^{3}G^{3}\rangle}{2^{13}\pi^{6}}\int_{\alpha_{min}}^{\alpha_{max}}\frac{d\alpha}{\alpha^{3}}\int_{\beta_{min}}^{1-\alpha}d\beta(1-\alpha-\beta)(1+\alpha+\beta)[r(m_{Q},m_{Q})+2
m_{Q}^{2}\beta],\nonumber
\end{eqnarray}
for $(Q\bar{q})^{*}(\bar{Q}q)$,

\begin{eqnarray}
\rho^{\mbox{OPE}}(s)=\rho^{\mbox{pert}}(s)+\rho^{\langle\bar{q}q\rangle}(s)+\rho^{\langle\bar{q}q\rangle^{2}}(s)+\rho^{\langle
g\bar{q}\sigma\cdot G q\rangle}(s)+\rho^{\langle
g^{2}G^{2}\rangle}(s)+\rho^{\langle g^{3}G^{3}\rangle}(s),\nonumber
\end{eqnarray}

\begin{eqnarray}
\rho^{\mbox{pert}}(s)&=&\frac{3}{2^{9}\pi^{6}}\int_{\alpha_{min}}^{\alpha_{max}}\frac{d\alpha}{\alpha^{3}}\int_{\beta_{min}}^{1-\alpha}\frac{d\beta}{\beta^{3}}(1-\alpha-\beta)r(m_{Q},m_{Q})^{4},\nonumber
\end{eqnarray}

\begin{eqnarray}
\rho^{\langle\bar{q}q\rangle}(s)&=&-\frac{3\langle\bar{q}q\rangle}{2^{5}\pi^{4}}m_{Q}\int_{\alpha_{min}}^{\alpha_{max}}\frac{d\alpha}{\alpha^{2}}\int_{\beta_{min}}^{1-\alpha}\frac{d\beta}{\beta}r(m_{Q},m_{Q})^{2},\nonumber
\end{eqnarray}

\begin{eqnarray}
\rho^{\langle\bar{q}q\rangle^{2}}(s)&=&\frac{\langle\bar{q}q\rangle^{2}}{2^{2}\pi^{2}}m_{Q}^{2}\sqrt{1-4m_{Q}^{2}/s},\nonumber
\end{eqnarray}

\begin{eqnarray}
\rho^{\langle g\bar{q}\sigma\cdot G q\rangle}(s)&=&-\frac{3\langle
g\bar{q}\sigma\cdot G
q\rangle}{2^{6}\pi^{4}}m_{Q}\int_{\alpha_{min}}^{\alpha_{max}}\frac{d\alpha}{1-\alpha}[m_{Q}^{2}-\alpha(1-\alpha)
s],\nonumber
\end{eqnarray}

\begin{eqnarray}
\rho^{\langle g^{2}G^{2}\rangle}(s)&=&\frac{\langle
g^{2}G^{2}\rangle}{2^{8}\pi^{6}}m_{Q}^{2}\int_{\alpha_{min}}^{\alpha_{max}}\frac{d\alpha}{\alpha^{3}}\int_{\beta_{min}}^{1-\alpha}d\beta(1-\alpha-\beta)r(m_{Q},m_{Q}),\nonumber
\end{eqnarray}

\begin{eqnarray}
\rho^{\langle g^{3}G^{3}\rangle}(s)&=&\frac{\langle
g^{3}G^{3}\rangle}{2^{10}\pi^{6}}\int_{\alpha_{min}}^{\alpha_{max}}\frac{d\alpha}{\alpha^{3}}\int_{\beta_{min}}^{1-\alpha}d\beta(1-\alpha-\beta)[r(m_{Q},m_{Q})+2
m_{Q}^{2}\beta],\nonumber
\end{eqnarray}
for $(Q\bar{q})^{*}(\bar{Q}q)^{*}$,

\begin{eqnarray}
\rho^{\mbox{OPE}}(s)=\rho^{\mbox{pert}}(s)+\rho^{\langle\bar{q}q\rangle}(s)+\rho^{\langle\bar{q}q\rangle^{2}}(s)+\rho^{\langle
g\bar{q}\sigma\cdot G q\rangle}(s)+\rho^{\langle
g^{2}G^{2}\rangle}(s)+\rho^{\langle g^{3}G^{3}\rangle}(s),\nonumber
\end{eqnarray}

\begin{eqnarray}
\rho^{\mbox{pert}}(s)&=&\frac{3}{2^{11}\pi^{6}}\int_{\alpha_{min}}^{\alpha_{max}}\frac{d\alpha}{\alpha^{3}}\int_{\beta_{min}}^{1-\alpha}\frac{d\beta}{\beta^{3}}(1-\alpha-\beta)r(m_{Q},m_{Q})^{4},\nonumber
\end{eqnarray}

\begin{eqnarray}
\rho^{\langle\bar{q}q\rangle}(s)&=&\frac{3\langle\bar{q}q\rangle}{2^{6}\pi^{4}}m_{Q}\int_{\alpha_{min}}^{\alpha_{max}}\frac{d\alpha}{\alpha^{2}}\int_{\beta_{min}}^{1-\alpha}\frac{d\beta}{\beta}r(m_{Q},m_{Q})^{2},\nonumber
\end{eqnarray}

\begin{eqnarray}
\rho^{\langle\bar{q}q\rangle^{2}}(s)&=&\frac{\langle\bar{q}q\rangle^{2}}{2^{4}\pi^{2}}m_{Q}^{2}\sqrt{1-4m_{Q}^{2}/s},\nonumber
\end{eqnarray}

\begin{eqnarray}
\rho^{\langle g\bar{q}\sigma\cdot G q\rangle}(s)&=&\frac{3\langle
g\bar{q}\sigma\cdot G
q\rangle}{2^{7}\pi^{4}}m_{Q}\int_{\alpha_{min}}^{\alpha_{max}}\frac{d\alpha}{1-\alpha}[m_{Q}^{2}-\alpha(1-\alpha)
s],\nonumber
\end{eqnarray}

\begin{eqnarray}
\rho^{\langle g^{2}G^{2}\rangle}(s)&=&\frac{\langle
g^{2}G^{2}\rangle}{2^{10}\pi^{6}}m_{Q}^{2}\int_{\alpha_{min}}^{\alpha_{max}}\frac{d\alpha}{\alpha^{3}}\int_{\beta_{min}}^{1-\alpha}d\beta(1-\alpha-\beta)r(m_{Q},m_{Q}),\nonumber
\end{eqnarray}

\begin{eqnarray}
\rho^{\langle g^{3}G^{3}\rangle}(s)&=&\frac{\langle
g^{3}G^{3}\rangle}{2^{12}\pi^{6}}\int_{\alpha_{min}}^{\alpha_{max}}\frac{d\alpha}{\alpha^{3}}\int_{\beta_{min}}^{1-\alpha}d\beta(1-\alpha-\beta)[r(m_{Q},m_{Q})+2
m_{Q}^{2}\beta],\nonumber
\end{eqnarray}
for $(Q\bar{q})_{0}^{*}(\bar{Q}q)_{0}^{*}$,

\begin{eqnarray}
\rho^{\mbox{OPE}}(s)=\rho^{\mbox{pert}}(s)+\rho^{\langle\bar{q}q\rangle}(s)+\rho^{\langle\bar{q}q\rangle^{2}}(s)+\rho^{\langle
g\bar{q}\sigma\cdot G q\rangle}(s)+\rho^{\langle
g^{2}G^{2}\rangle}(s)+\rho^{\langle g^{3}G^{3}\rangle}(s),\nonumber
\end{eqnarray}

\begin{eqnarray}
\rho^{\mbox{pert}}(s)&=&\frac{3}{2^{9}\pi^{6}}\int_{\alpha_{min}}^{\alpha_{max}}\frac{d\alpha}{\alpha^{3}}\int_{\beta_{min}}^{1-\alpha}\frac{d\beta}{\beta^{3}}(1-\alpha-\beta)r(m_{Q},m_{Q})^{4},\nonumber
\end{eqnarray}

\begin{eqnarray}
\rho^{\langle\bar{q}q\rangle}(s)&=&\frac{3\langle\bar{q}q\rangle}{2^{5}\pi^{4}}m_{Q}\int_{\alpha_{min}}^{\alpha_{max}}\frac{d\alpha}{\alpha^{2}}\int_{\beta_{min}}^{1-\alpha}\frac{d\beta}{\beta}r(m_{Q},m_{Q})^{2},\nonumber
\end{eqnarray}

\begin{eqnarray}
\rho^{\langle\bar{q}q\rangle^{2}}(s)&=&\frac{\langle\bar{q}q\rangle^{2}}{2^{2}\pi^{2}}m_{Q}^{2}\sqrt{1-4m_{Q}^{2}/s},\nonumber
\end{eqnarray}

\begin{eqnarray}
\rho^{\langle g\bar{q}\sigma\cdot G q\rangle}(s)&=&\frac{3\langle
g\bar{q}\sigma\cdot G
q\rangle}{2^{6}\pi^{4}}m_{Q}\int_{\alpha_{min}}^{\alpha_{max}}\frac{d\alpha}{1-\alpha}[m_{Q}^{2}-\alpha(1-\alpha)
s],\nonumber
\end{eqnarray}

\begin{eqnarray}
\rho^{\langle g^{2}G^{2}\rangle}(s)&=&\frac{\langle
g^{2}G^{2}\rangle}{2^{8}\pi^{6}}m_{Q}^{2}\int_{\alpha_{min}}^{\alpha_{max}}\frac{d\alpha}{\alpha^{3}}\int_{\beta_{min}}^{1-\alpha}d\beta(1-\alpha-\beta)r(m_{Q},m_{Q}),\nonumber
\end{eqnarray}

\begin{eqnarray}
\rho^{\langle g^{3}G^{3}\rangle}(s)&=&\frac{\langle
g^{3}G^{3}\rangle}{2^{10}\pi^{6}}\int_{\alpha_{min}}^{\alpha_{max}}\frac{d\alpha}{\alpha^{3}}\int_{\beta_{min}}^{1-\alpha}d\beta(1-\alpha-\beta)[r(m_{Q},m_{Q})+2
m_{Q}^{2}\beta],\nonumber
\end{eqnarray}
for $(Q\bar{q})_{1}(\bar{Q}q)_{1}$,

\begin{eqnarray}
\rho^{\mbox{OPE}}(s)=-\{\rho^{\mbox{pert}}(s)+\rho^{\langle\bar{q}q\rangle}(s)+\rho^{\langle\bar{q}q\rangle^{2}}(s)+\rho^{\langle
g\bar{q}\sigma\cdot G q\rangle}(s)+\rho^{\langle
g^{2}G^{2}\rangle}(s)+\rho^{\langle
g^{3}G^{3}\rangle}(s)\},\nonumber
\end{eqnarray}

\begin{eqnarray}
\rho^{\mbox{pert}}(s)&=&-\frac{3}{2^{12}\pi^{6}}\int_{\alpha_{min}}^{\alpha_{max}}\frac{d\alpha}{\alpha^{3}}\int_{\beta_{min}}^{1-\alpha}\frac{d\beta}{\beta^{3}}(1-\alpha-\beta)(1+\alpha+\beta)r(m_{Q},m_{Q})^{4},\nonumber
\end{eqnarray}

\begin{eqnarray}
\rho^{\langle\bar{q}q\rangle}(s)&=&-\frac{3\langle\bar{q}q\rangle}{2^{7}\pi^{4}}m_{Q}\int_{\alpha_{min}}^{\alpha_{max}}\frac{d\alpha}{\alpha^{2}}\int_{\beta_{min}}^{1-\alpha}\frac{d\beta}{\beta}(1+\alpha+\beta)r(m_{Q},m_{Q})^{2},\nonumber
\end{eqnarray}

\begin{eqnarray}
\rho^{\langle\bar{q}q\rangle^{2}}(s)&=&-\frac{\langle\bar{q}q\rangle^{2}}{2^{4}\pi^{2}}m_{Q}^{2}\sqrt{1-4m_{Q}^{2}/s},\nonumber
\end{eqnarray}

\begin{eqnarray}
\rho^{\langle g\bar{q}\sigma\cdot G q\rangle}(s)&=&\frac{3\langle
g\bar{q}\sigma\cdot G
q\rangle}{2^{8}\pi^{4}}m_{Q}\int_{\alpha_{min}}^{\alpha_{max}}d\alpha\{\int_{\beta_{min}}^{1-\alpha}\frac{d\beta}{\beta}r(m_{Q},m_{Q})-\frac{2}{1-\alpha}[m_{Q}^{2}-\alpha(1-\alpha)
s]\},\nonumber
\end{eqnarray}

\begin{eqnarray}
\rho^{\langle g^{2}G^{2}\rangle}(s)&=&-\frac{\langle
g^{2}G^{2}\rangle}{2^{11}\pi^{6}}m_{Q}^{2}\int_{\alpha_{min}}^{\alpha_{max}}\frac{d\alpha}{\alpha^{3}}\int_{\beta_{min}}^{1-\alpha}d\beta(1-\alpha-\beta)(1+\alpha+\beta)r(m_{Q},m_{Q}),\nonumber
\end{eqnarray}

\begin{eqnarray}
\rho^{\langle g^{3}G^{3}\rangle}(s)&=&-\frac{\langle
g^{3}G^{3}\rangle}{2^{13}\pi^{6}}\int_{\alpha_{min}}^{\alpha_{max}}\frac{d\alpha}{\alpha^{3}}\int_{\beta_{min}}^{1-\alpha}d\beta(1-\alpha-\beta)(1+\alpha+\beta)[r(m_{Q},m_{Q})+
2m_{Q}^{2}\beta],\nonumber
\end{eqnarray}
for $(Q\bar{q})_{1}(\bar{Q}q)_{0}^{*}$,

\begin{eqnarray}
\rho^{\mbox{OPE}}(s)=\rho^{\mbox{pert}}(s)+\rho^{\langle\bar{q}q\rangle}(s)+\rho^{\langle\bar{q}q\rangle^{2}}(s)+\rho^{\langle
g\bar{q}\sigma\cdot G q\rangle}(s)+\rho^{\langle
g^{2}G^{2}\rangle}(s)+\rho^{\langle g^{3}G^{3}\rangle}(s),\nonumber
\end{eqnarray}

\begin{eqnarray}
\rho^{\mbox{pert}}(s)&=&\frac{3}{2^{11}\pi^{6}}\int_{\alpha_{min}}^{\alpha_{max}}\frac{d\alpha}{\alpha^{3}}\int_{\beta_{min}}^{1-\alpha}\frac{d\beta}{\beta^{3}}(1-\alpha-\beta)r(m_{Q},m_{Q})^{4},\nonumber
\end{eqnarray}

\begin{eqnarray}
\rho^{\langle\bar{q}q\rangle^{2}}(s)&=&-\frac{\langle\bar{q}q\rangle^{2}}{2^{4}\pi^{2}}m_{Q}^{2}\sqrt{1-4m_{Q}^{2}/s},\nonumber
\end{eqnarray}

\begin{eqnarray}
\rho^{\langle g^{2}G^{2}\rangle}(s)&=&\frac{\langle
g^{2}G^{2}\rangle}{2^{10}\pi^{6}}m_{Q}^{2}\int_{\alpha_{min}}^{\alpha_{max}}\frac{d\alpha}{\alpha^{3}}\int_{\beta_{min}}^{1-\alpha}d\beta(1-\alpha-\beta)r(m_{Q},m_{Q}),\nonumber
\end{eqnarray}

\begin{eqnarray}
\rho^{\langle g^{3}G^{3}\rangle}(s)&=&\frac{\langle
g^{3}G^{3}\rangle}{2^{12}\pi^{6}}\int_{\alpha_{min}}^{\alpha_{max}}\frac{d\alpha}{\alpha^{3}}\int_{\beta_{min}}^{1-\alpha}d\beta(1-\alpha-\beta)[r(m_{Q},m_{Q})+2
m_{Q}^{2}\beta],\nonumber
\end{eqnarray}
for $(Q\bar{q})(\bar{Q}q)_{0}^{*}$,

\begin{eqnarray}
\rho^{\mbox{OPE}}(s)=-\{\rho^{\mbox{pert}}(s)+\rho^{\langle\bar{q}q\rangle}(s)+\rho^{\langle\bar{q}q\rangle^{2}}(s)+\rho^{\langle
g\bar{q}\sigma\cdot G q\rangle}(s)+\rho^{\langle
g^{2}G^{2}\rangle}(s)+\rho^{\langle
g^{3}G^{3}\rangle}(s)\},\nonumber
\end{eqnarray}

\begin{eqnarray}
\rho^{\mbox{pert}}(s)&=&-\frac{3}{2^{12}\pi^{6}}\int_{\alpha_{min}}^{\alpha_{max}}\frac{d\alpha}{\alpha^{3}}\int_{\beta_{min}}^{1-\alpha}\frac{d\beta}{\beta^{3}}(1-\alpha-\beta)(1+\alpha+\beta)r(m_{Q},m_{Q})^{4},\nonumber
\end{eqnarray}

\begin{eqnarray}
\rho^{\langle\bar{q}q\rangle}(s)&=&-\frac{3\langle\bar{q}q\rangle}{2^{7}\pi^{4}}m_{Q}\int_{\alpha_{min}}^{\alpha_{max}}\frac{d\alpha}{\alpha^{2}}\int_{\beta_{min}}^{1-\alpha}\frac{d\beta}{\beta}(1-\alpha-\beta)r(m_{Q},m_{Q})^{2},\nonumber
\end{eqnarray}

\begin{eqnarray}
\rho^{\langle\bar{q}q\rangle^{2}}(s)&=&\frac{\langle\bar{q}q\rangle^{2}}{2^{4}\pi^{2}}m_{Q}^{2}\sqrt{1-4m_{Q}^{2}/s},\nonumber
\end{eqnarray}

\begin{eqnarray}
\rho^{\langle g\bar{q}\sigma\cdot G q\rangle}(s)&=&-\frac{3\langle
g\bar{q}\sigma\cdot G
q\rangle}{2^{8}\pi^{4}}m_{Q}\int_{\alpha_{min}}^{\alpha_{max}}d\alpha\int_{\beta_{min}}^{1-\alpha}\frac{d\beta}{\beta}r(m_{Q},m_{Q}),\nonumber
\end{eqnarray}

\begin{eqnarray}
\rho^{\langle g^{2}G^{2}\rangle}(s)&=&-\frac{\langle
g^{2}G^{2}\rangle}{2^{11}\pi^{6}}m_{Q}^{2}\int_{\alpha_{min}}^{\alpha_{max}}\frac{d\alpha}{\alpha^{3}}\int_{\beta_{min}}^{1-\alpha}d\beta(1-\alpha-\beta)(1+\alpha+\beta)r(m_{Q},m_{Q}),\nonumber
\end{eqnarray}

\begin{eqnarray}
\rho^{\langle g^{3}G^{3}\rangle}(s)&=&-\frac{\langle
g^{3}G^{3}\rangle}{2^{13}\pi^{6}}\int_{\alpha_{min}}^{\alpha_{max}}\frac{d\alpha}{\alpha^{3}}\int_{\beta_{min}}^{1-\alpha}d\beta(1-\alpha-\beta)(1+\alpha+\beta)[r(m_{Q},m_{Q})+2
m_{Q}^{2}\beta],\nonumber
\end{eqnarray}
for $(Q\bar{q})_{1}(\bar{Q}q)$,

\begin{eqnarray}
\rho^{\mbox{OPE}}(s)=-\{\rho^{\mbox{pert}}(s)+\rho^{\langle\bar{q}q\rangle}(s)+\rho^{\langle\bar{q}q\rangle^{2}}(s)+\rho^{\langle
g\bar{q}\sigma\cdot G q\rangle}(s)+\rho^{\langle
g^{2}G^{2}\rangle}(s)+\rho^{\langle
g^{3}G^{3}\rangle}(s)\},\nonumber
\end{eqnarray}

\begin{eqnarray}
\rho^{\mbox{pert}}(s)&=&-\frac{3}{2^{12}\pi^{6}}\int_{\alpha_{min}}^{\alpha_{max}}\frac{d\alpha}{\alpha^{3}}\int_{\beta_{min}}^{1-\alpha}\frac{d\beta}{\beta^{3}}(1-\alpha-\beta)(1+\alpha+\beta)r(m_{Q},m_{Q})^{4},\nonumber
\end{eqnarray}

\begin{eqnarray}
\rho^{\langle\bar{q}q\rangle}(s)&=&\frac{3\langle\bar{q}q\rangle}{2^{7}\pi^{4}}m_{Q}\int_{\alpha_{min}}^{\alpha_{max}}\frac{d\alpha}{\alpha^{2}}\int_{\beta_{min}}^{1-\alpha}\frac{d\beta}{\beta}(1-\alpha-\beta)r(m_{Q},m_{Q})^{2},\nonumber
\end{eqnarray}

\begin{eqnarray}
\rho^{\langle\bar{q}q\rangle^{2}}(s)&=&\frac{\langle\bar{q}q\rangle^{2}}{2^{4}\pi^{2}}m_{Q}^{2}\sqrt{1-4m_{Q}^{2}/s},\nonumber
\end{eqnarray}

\begin{eqnarray}
\rho^{\langle g\bar{q}\sigma\cdot G q\rangle}(s)&=&\frac{3\langle
g\bar{q}\sigma\cdot G
q\rangle}{2^{8}\pi^{4}}m_{Q}\int_{\alpha_{min}}^{\alpha_{max}}d\alpha\int_{\beta_{min}}^{1-\alpha}\frac{d\beta}{\beta}r(m_{Q},m_{Q}),\nonumber
\end{eqnarray}

\begin{eqnarray}
\rho^{\langle g^{2}G^{2}\rangle}(s)&=&-\frac{\langle
g^{2}G^{2}\rangle}{2^{11}\pi^{6}}m_{Q}^{2}\int_{\alpha_{min}}^{\alpha_{max}}\frac{d\alpha}{\alpha^{3}}\int_{\beta_{min}}^{1-\alpha}d\beta(1-\alpha-\beta)(1+\alpha+\beta)r(m_{Q},m_{Q}),\nonumber
\end{eqnarray}

\begin{eqnarray}
\rho^{\langle g^{3}G^{3}\rangle}(s)&=&-\frac{\langle
g^{3}G^{3}\rangle}{2^{13}\pi^{6}}\int_{\alpha_{min}}^{\alpha_{max}}\frac{d\alpha}{\alpha^{3}}\int_{\beta_{min}}^{1-\alpha}d\beta(1-\alpha-\beta)(1+\alpha+\beta)[r(m_{Q},m_{Q})+2
m_{Q}^{2}\beta],\nonumber
\end{eqnarray}
for $(Q\bar{q})^{*}(\bar{Q}q)_{0}^{*}$, and

\begin{eqnarray}
\rho^{\mbox{OPE}}(s)=\rho^{\mbox{pert}}(s)+\rho^{\langle\bar{q}q\rangle}(s)+\rho^{\langle\bar{q}q\rangle^{2}}(s)+\rho^{\langle
g\bar{q}\sigma\cdot G q\rangle}(s)+\rho^{\langle
g^{2}G^{2}\rangle}(s)+\rho^{\langle g^{3}G^{3}\rangle}(s),\nonumber
\end{eqnarray}

\begin{eqnarray}
\rho^{\mbox{pert}}(s)&=&\frac{3}{2^{9}\pi^{6}}\int_{\alpha_{min}}^{\alpha_{max}}\frac{d\alpha}{\alpha^{3}}\int_{\beta_{min}}^{1-\alpha}\frac{d\beta}{\beta^{3}}(1-\alpha-\beta)r(m_{Q},m_{Q})^{4},\nonumber
\end{eqnarray}

\begin{eqnarray}
\rho^{\langle\bar{q}q\rangle^{2}}(s)&=&-\frac{\langle\bar{q}q\rangle^{2}}{2^{2}\pi^{2}}m_{Q}^{2}\sqrt{1-4m_{Q}^{2}/s},\nonumber
\end{eqnarray}

\begin{eqnarray}
\rho^{\langle g^{2}G^{2}\rangle}(s)&=&\frac{\langle
g^{2}G^{2}\rangle}{2^{8}\pi^{6}}m_{Q}^{2}\int_{\alpha_{min}}^{\alpha_{max}}\frac{d\alpha}{\alpha^{3}}\int_{\beta_{min}}^{1-\alpha}d\beta(1-\alpha-\beta)r(m_{Q},m_{Q}),\nonumber
\end{eqnarray}

\begin{eqnarray}
\rho^{\langle g^{3}G^{3}\rangle}(s)&=&\frac{\langle
g^{3}G^{3}\rangle}{2^{10}\pi^{6}}\int_{\alpha_{min}}^{\alpha_{max}}\frac{d\alpha}{\alpha^{3}}\int_{\beta_{min}}^{1-\alpha}d\beta(1-\alpha-\beta)[r(m_{Q},m_{Q})+2
m_{Q}^{2}\beta],\nonumber
\end{eqnarray}
for $(Q\bar{q})^{*}(\bar{Q}q)_{1}$. The integration limits are given
by $\alpha_{min}=(1-\sqrt{1-4m_{Q}^{2}/s})/2$,
$\alpha_{max}=(1+\sqrt{1-4m_{Q}^{2}/s})/2$, and $\beta_{min}=\alpha
m_{Q}^{2}/(s\alpha-m_{Q}^{2})$.

Second, with

\begin{eqnarray}
\rho^{\mbox{OPE}}(s)=\rho^{\mbox{pert}}(s)+\rho^{\langle\bar{q}q\rangle}(s)+\rho^{\langle\bar{q}q\rangle^{2}}(s)+\rho^{\langle
g\bar{q}\sigma\cdot G q\rangle}(s)+\rho^{\langle
g^{2}G^{2}\rangle}(s)+\rho^{\langle g^{3}G^{3}\rangle}(s),\nonumber
\end{eqnarray}

\begin{eqnarray}
\rho^{\mbox{pert}}(s)&=&\frac{3}{2^{11}\pi^{6}}\int_{\alpha_{min}}^{\alpha_{max}}\frac{d\alpha}{\alpha^{3}}\int_{\beta_{min}}^{1-\alpha}\frac{d\beta}{\beta^{3}}(1-\alpha-\beta)r(m_{Q},m_{Q'})^{4},\nonumber
\end{eqnarray}

\begin{eqnarray}
\rho^{\langle\bar{q}q\rangle}(s)&=&-\frac{3\langle\bar{q}q\rangle}{2^{7}\pi^{4}}\int_{\alpha_{min}}^{\alpha_{max}}d\alpha\int_{\beta_{min}}^{1-\alpha}d\beta(
\frac{m_{Q'}}{\alpha^{2}\beta}+\frac{m_{Q}}{\alpha\beta^{2}})r(m_{Q},m_{Q'})^{2},\nonumber
\end{eqnarray}

\begin{eqnarray}
\rho^{\langle\bar{q}q\rangle^{2}}(s)&=&\frac{\langle\bar{q}q\rangle^{2}}{2^{4}\pi^{2}}m_{Q}m_{Q'}\sqrt{(s-m_{Q}^{2}+m_{Q'}^{2})^{2}-4m_{Q'}^{2}s}/s,\nonumber
\end{eqnarray}

\begin{eqnarray}
\rho^{\langle g\bar{q}\sigma\cdot G q\rangle}(s)&=&-\frac{3\langle
g\bar{q}\sigma\cdot G
q\rangle}{2^{8}\pi^{4}}\int_{\alpha_{min}}^{\alpha_{max}}d\alpha(\frac{m_{Q'}}{\alpha}+\frac{m_{Q}}{1-\alpha})[\alpha
m_{Q}^{2}+(1-\alpha) m_{Q'}^{2}-\alpha(1-\alpha) s],\nonumber
\end{eqnarray}

\begin{eqnarray}
\rho^{\langle g^{2}G^{2}\rangle}(s)&=&\frac{\langle
g^{2}G^{2}\rangle}{2^{11}\pi^{6}}\int_{\alpha_{min}}^{\alpha_{max}}d\alpha\int_{\beta_{min}}^{1-\alpha}d\beta(1-\alpha-\beta)(\frac{m_{Q'}^{2}}{\alpha^{3}}+\frac{m_{Q}^{2}}{\beta^{3}})r(m_{Q},m_{Q'}),\nonumber
\end{eqnarray}

\begin{eqnarray}
\rho^{\langle g^{3}G^{3}\rangle}(s)&=&\frac{\langle
g^{3}G^{3}\rangle}{2^{13}\pi^{6}}\int_{\alpha_{min}}^{\alpha_{max}}d\alpha\int_{\beta_{min}}^{1-\alpha}d\beta(1-\alpha-\beta)[(\frac{1}{\alpha^{3}}+\frac{1}{\beta^{3}})r(m_{Q},m_{Q'})+2(\frac{m_{Q'}^{2}\beta}{\alpha^{3}}+\frac{m_{Q}^{2}\alpha}{\beta^{3}})],\nonumber
\end{eqnarray}
for $(Q\bar{q})(\bar{Q'}q)$,

\begin{eqnarray}
\rho^{\mbox{OPE}}(s)=-\{\rho^{\mbox{pert}}(s)+\rho^{\langle\bar{q}q\rangle}(s)+\rho^{\langle\bar{q}q\rangle^{2}}(s)+\rho^{\langle
g\bar{q}\sigma\cdot G q\rangle}(s)+\rho^{\langle
g^{2}G^{2}\rangle}(s)+\rho^{\langle
g^{3}G^{3}\rangle}(s)\},\nonumber
\end{eqnarray}

\begin{eqnarray}
\rho^{\mbox{pert}}(s)&=&-\frac{3}{2^{12}\pi^{6}}\int_{\alpha_{min}}^{\alpha_{max}}\frac{d\alpha}{\alpha^{3}}\int_{\beta_{min}}^{1-\alpha}\frac{d\beta}{\beta^{3}}(1-\alpha-\beta)(1+\alpha+\beta)r(m_{Q},m_{Q'})^{4},\nonumber
\end{eqnarray}

\begin{eqnarray}
\rho^{\langle\bar{q}q\rangle}(s)&=&\frac{3\langle\bar{q}q\rangle}{2^{7}\pi^{4}}\int_{\alpha_{min}}^{\alpha_{max}}d\alpha\int_{\beta_{min}}^{1-\alpha}d\beta[\frac{m_{Q'}(\alpha+\beta)}{\alpha^{2}\beta}+\frac{m_{Q}}{\alpha\beta^{2}}]r(m_{Q},m_{Q'})^{2},\nonumber
\end{eqnarray}

\begin{eqnarray}
\rho^{\langle\bar{q}q\rangle^{2}}(s)&=&-\frac{\langle\bar{q}q\rangle^{2}}{2^{4}\pi^{2}}m_{Q}m_{Q'}\sqrt{(s-m_{Q}^{2}+m_{Q'}^{2})^{2}-4m_{Q'}^{2}s}/s,\nonumber
\end{eqnarray}

\begin{eqnarray}
\rho^{\langle g\bar{q}\sigma\cdot G q\rangle}(s)&=&\frac{3\langle
g\bar{q}\sigma\cdot G
q\rangle}{2^{8}\pi^{4}}\int_{\alpha_{min}}^{\alpha_{max}}d\alpha\times\{-\frac{m_{Q'}}{\alpha}\int_{\beta_{min}}^{1-\alpha}d\beta
r(m_{Q},m_{Q'})+
(\frac{m_{Q'}}{\alpha}+\frac{m_{Q}}{1-\alpha})[\alpha
m_{Q}^{2}+(1-\alpha) m_{Q'}^{2}-\alpha(1-\alpha) s]\},\nonumber
\end{eqnarray}

\begin{eqnarray}
\rho^{\langle g^{2}G^{2}\rangle}(s)&=&-\frac{\langle
g^{2}G^{2}\rangle}{2^{12}\pi^{6}}\int_{\alpha_{min}}^{\alpha_{max}}d\alpha\int_{\beta_{min}}^{1-\alpha}d\beta(1-\alpha-\beta)(1+\alpha+\beta)(\frac{m_{Q'}^{2}}{\alpha^{3}}+\frac{m_{Q}^{2}}{\beta^{3}})r(m_{Q},m_{Q'}),\nonumber
\end{eqnarray}

\begin{eqnarray}
\rho^{\langle g^{3}G^{3}\rangle}(s)&=&-\frac{\langle
g^{3}G^{3}\rangle}{2^{14}\pi^{6}}\int_{\alpha_{min}}^{\alpha_{max}}d\alpha\int_{\beta_{min}}^{1-\alpha}d\beta(1-\alpha-\beta)(1+\alpha+\beta)[(\frac{1}{\alpha^{3}}+\frac{1}{\beta^{3}})r(m_{Q},m_{Q'})+2(\frac{m_{Q'}^{2}\beta}{\alpha^{3}}+\frac{m_{Q}^{2}\alpha}{\beta^{3}})],\nonumber
\end{eqnarray}
for $(Q\bar{q})^{*}(\bar{Q'}q)$,

\begin{eqnarray}
\rho^{\mbox{OPE}}(s)=\rho^{\mbox{pert}}(s)+\rho^{\langle\bar{q}q\rangle}(s)+\rho^{\langle\bar{q}q\rangle^{2}}(s)+\rho^{\langle
g\bar{q}\sigma\cdot G q\rangle}(s)+\rho^{\langle
g^{2}G^{2}\rangle}(s)+\rho^{\langle g^{3}G^{3}\rangle}(s),\nonumber
\end{eqnarray}

\begin{eqnarray}
\rho^{\mbox{pert}}(s)&=&\frac{3}{2^{9}\pi^{6}}\int_{\alpha_{min}}^{\alpha_{max}}\frac{d\alpha}{\alpha^{3}}\int_{\beta_{min}}^{1-\alpha}\frac{d\beta}{\beta^{3}}(1-\alpha-\beta)r(m_{Q},m_{Q'})^{4},\nonumber
\end{eqnarray}

\begin{eqnarray}
\rho^{\langle\bar{q}q\rangle}(s)&=&-\frac{3\langle\bar{q}q\rangle}{2^{6}\pi^{4}}\int_{\alpha_{min}}^{\alpha_{max}}d\alpha\int_{\beta_{min}}^{1-\alpha}d\beta(\frac{m_{Q'}}{\alpha^{2}\beta}+\frac{m_{Q}}{\alpha\beta^{2}})r(m_{Q},m_{Q'})^{2},\nonumber
\end{eqnarray}

\begin{eqnarray}
\rho^{\langle\bar{q}q\rangle^{2}}(s)&=&\frac{\langle\bar{q}q\rangle^{2}}{2^{2}\pi^{2}}m_{Q}m_{Q'}\sqrt{(s-m_{Q}^{2}+m_{Q'}^{2})^{2}-4m_{Q'}^{2}s}/s,\nonumber
\end{eqnarray}

\begin{eqnarray}
\rho^{\langle g\bar{q}\sigma\cdot G q\rangle}(s)&=&-\frac{3\langle
g\bar{q}\sigma\cdot G
q\rangle}{2^{7}\pi^{4}}\int_{\alpha_{min}}^{\alpha_{max}}d\alpha(\frac{m_{Q'}}{\alpha}+\frac{m_{Q}}{1-\alpha})[\alpha
m_{Q}^{2}+(1-\alpha) m_{Q'}^{2}-\alpha(1-\alpha) s],\nonumber
\end{eqnarray}

\begin{eqnarray}
\rho^{\langle g^{2}G^{2}\rangle}(s)&=&\frac{\langle
g^{2}G^{2}\rangle}{2^{9}\pi^{6}}\int_{\alpha_{min}}^{\alpha_{max}}d\alpha\int_{\beta_{min}}^{1-\alpha}d\beta(1-\alpha-\beta)(\frac{m_{Q'}^{2}}{\alpha^{3}}+\frac{m_{Q}^{2}}{\beta^{3}})r(m_{Q},m_{Q'}),\nonumber
\end{eqnarray}

\begin{eqnarray}
\rho^{\langle g^{3}G^{3}\rangle}(s)&=&\frac{\langle
g^{3}G^{3}\rangle}{2^{11}\pi^{6}}\int_{\alpha_{min}}^{\alpha_{max}}d\alpha\int_{\beta_{min}}^{1-\alpha}d\beta(1-\alpha-\beta)[(\frac{1}{\alpha^{3}}+\frac{1}{\beta^{3}})r(m_{Q},m_{Q'})+2(\frac{m_{Q'}^{2}\beta}{\alpha^{3}}+\frac{m_{Q}^{2}\alpha}{\beta^{3}})]
,\nonumber
\end{eqnarray}
for $(Q\bar{q})^{*}(\bar{Q'}q)^{*}$,

\begin{eqnarray}
\rho^{\mbox{OPE}}(s)=\rho^{\mbox{pert}}(s)+\rho^{\langle\bar{q}q\rangle}(s)+\rho^{\langle\bar{q}q\rangle^{2}}(s)+\rho^{\langle
g\bar{q}\sigma\cdot G q\rangle}(s)+\rho^{\langle
g^{2}G^{2}\rangle}(s)+\rho^{\langle g^{3}G^{3}\rangle}(s),\nonumber
\end{eqnarray}

\begin{eqnarray}
\rho^{\mbox{pert}}(s)&=&\frac{3}{2^{11}\pi^{6}}\int_{\alpha_{min}}^{\alpha_{max}}\frac{d\alpha}{\alpha^{3}}\int_{\beta_{min}}^{1-\alpha}\frac{d\beta}{\beta^{3}}(1-\alpha-\beta)r(m_{Q},m_{Q'})^{4},\nonumber
\end{eqnarray}

\begin{eqnarray}
\rho^{\langle\bar{q}q\rangle}(s)&=&\frac{3\langle\bar{q}q\rangle}{2^{7}\pi^{4}}\int_{\alpha_{min}}^{\alpha_{max}}d\alpha\int_{\beta_{min}}^{1-\alpha}d\beta(\frac{m_{Q'}}{\alpha^{2}\beta}+\frac{m_{Q}}{\alpha\beta^{2}})r(m_{Q},m_{Q'})^{2},\nonumber
\end{eqnarray}

\begin{eqnarray}
\rho^{\langle\bar{q}q\rangle^{2}}(s)&=&\frac{\langle\bar{q}q\rangle^{2}}{2^{4}\pi^{2}}m_{Q}m_{Q'}\sqrt{(s-m_{Q}^{2}+m_{Q'}^{2})^{2}-4m_{Q'}^{2}s}/s,\nonumber
\end{eqnarray}

\begin{eqnarray}
\rho^{\langle g\bar{q}\sigma\cdot G q\rangle}(s)&=&\frac{3\langle
g\bar{q}\sigma\cdot G
q\rangle}{2^{8}\pi^{4}}\int_{\alpha_{min}}^{\alpha_{max}}d\alpha(\frac{m_{Q'}}{\alpha}+\frac{m_{Q}}{1-\alpha})[\alpha
m_{Q}^{2}+(1-\alpha) m_{Q'}^{2}-\alpha(1-\alpha) s],\nonumber
\end{eqnarray}

\begin{eqnarray}
\rho^{\langle g^{2}G^{2}\rangle}(s)&=&\frac{\langle
g^{2}G^{2}\rangle}{2^{11}\pi^{6}}\int_{\alpha_{min}}^{\alpha_{max}}d\alpha\int_{\beta_{min}}^{1-\alpha}d\beta(1-\alpha-\beta)(\frac{m_{Q'}^{2}}{\alpha^{3}}+\frac{m_{Q}^{2}}{\beta^{3}})r(m_{Q},m_{Q'}),\nonumber
\end{eqnarray}

\begin{eqnarray}
\rho^{\langle g^{3}G^{3}\rangle}(s)&=&\frac{\langle
g^{3}G^{3}\rangle}{2^{13}\pi^{6}}\int_{\alpha_{min}}^{\alpha_{max}}d\alpha\int_{\beta_{min}}^{1-\alpha}d\beta(1-\alpha-\beta)[(\frac{1}{\alpha^{3}}+\frac{1}{\beta^{3}})r(m_{Q},m_{Q'})+2(\frac{m_{Q'}^{2}\beta}{\alpha^{3}}+\frac{m_{Q}^{2}\alpha}{\beta^{3}})]
,\nonumber
\end{eqnarray}
for $(Q\bar{q})_{0}^{*}(\bar{Q'}q)_{0}^{*}$,

\begin{eqnarray}
\rho^{\mbox{OPE}}(s)=\rho^{\mbox{pert}}(s)+\rho^{\langle\bar{q}q\rangle}(s)+\rho^{\langle\bar{q}q\rangle^{2}}(s)+\rho^{\langle
g\bar{q}\sigma\cdot G q\rangle}(s)+\rho^{\langle
g^{2}G^{2}\rangle}(s)+\rho^{\langle g^{3}G^{3}\rangle}(s),\nonumber
\end{eqnarray}

\begin{eqnarray}
\rho^{\mbox{pert}}(s)&=&\frac{3}{2^{9}\pi^{6}}\int_{\alpha_{min}}^{\alpha_{max}}\frac{d\alpha}{\alpha^{3}}\int_{\beta_{min}}^{1-\alpha}\frac{d\beta}{\beta^{3}}(1-\alpha-\beta)r(m_{Q},m_{Q'})^{4},\nonumber
\end{eqnarray}

\begin{eqnarray}
\rho^{\langle\bar{q}q\rangle}(s)&=&\frac{3\langle\bar{q}q\rangle}{2^{6}\pi^{4}}\int_{\alpha_{min}}^{\alpha_{max}}d\alpha\int_{\beta_{min}}^{1-\alpha}d\beta(\frac{m_{Q'}}{\alpha^{2}\beta}+\frac{m_{Q}}{\alpha\beta^{2}})r(m_{Q},m_{Q'})^{2},\nonumber
\end{eqnarray}

\begin{eqnarray}
\rho^{\langle\bar{q}q\rangle^{2}}(s)&=&\frac{\langle\bar{q}q\rangle^{2}}{2^{2}\pi^{2}}m_{Q}m_{Q'}\sqrt{(s-m_{Q}^{2}+m_{Q'}^{2})^{2}-4m_{Q'}^{2}s}/s,\nonumber
\end{eqnarray}

\begin{eqnarray}
\rho^{\langle g\bar{q}\sigma\cdot G q\rangle}(s)&=&\frac{3\langle
g\bar{q}\sigma\cdot G
q\rangle}{2^{7}\pi^{4}}\int_{\alpha_{min}}^{\alpha_{max}}d\alpha(\frac{m_{Q'}}{\alpha}+\frac{m_{Q}}{1-\alpha})[\alpha
m_{Q}^{2}+(1-\alpha) m_{Q'}^{2}-\alpha(1-\alpha) s],\nonumber
\end{eqnarray}

\begin{eqnarray}
\rho^{\langle g^{2}G^{2}\rangle}(s)&=&\frac{\langle
g^{2}G^{2}\rangle}{2^{9}\pi^{6}}\int_{\alpha_{min}}^{\alpha_{max}}d\alpha\int_{\beta_{min}}^{1-\alpha}d\beta(1-\alpha-\beta)(\frac{m_{Q'}^{2}}{\alpha^{3}}+\frac{m_{Q}^{2}}{\beta^{3}})r(m_{Q},m_{Q'}),\nonumber
\end{eqnarray}

\begin{eqnarray}
\rho^{\langle g^{3}G^{3}\rangle}(s)&=&\frac{\langle
g^{3}G^{3}\rangle}{2^{11}\pi^{6}}\int_{\alpha_{min}}^{\alpha_{max}}d\alpha\int_{\beta_{min}}^{1-\alpha}d\beta(1-\alpha-\beta)[(\frac{1}{\alpha^{3}}+\frac{1}{\beta^{3}})r(m_{Q},m_{Q'})+2(\frac{m_{Q'}^{2}\beta}{\alpha^{3}}+\frac{m_{Q}^{2}\alpha}{\beta^{3}})],\nonumber
\end{eqnarray}
for $(Q\bar{q})_{1}(\bar{Q'}q)_{1}$,

\begin{eqnarray}
\rho^{\mbox{OPE}}(s)=-\{\rho^{\mbox{pert}}(s)+\rho^{\langle\bar{q}q\rangle}(s)+\rho^{\langle\bar{q}q\rangle^{2}}(s)+\rho^{\langle
g\bar{q}\sigma\cdot G q\rangle}(s)+\rho^{\langle
g^{2}G^{2}\rangle}(s)+\rho^{\langle
g^{3}G^{3}\rangle}(s)\},\nonumber
\end{eqnarray}

\begin{eqnarray}
\rho^{\mbox{pert}}(s)&=&-\frac{3}{2^{12}\pi^{6}}\int_{\alpha_{min}}^{\alpha_{max}}\frac{d\alpha}{\alpha^{3}}\int_{\beta_{min}}^{1-\alpha}\frac{d\beta}{\beta^{3}}(1-\alpha-\beta)(1+\alpha+\beta)r(m_{Q},m_{Q'})^{4},\nonumber
\end{eqnarray}

\begin{eqnarray}
\rho^{\langle\bar{q}q\rangle}(s)&=&-\frac{3\langle\bar{q}q\rangle}{2^{7}\pi^{4}}\int_{\alpha_{min}}^{\alpha_{max}}d\alpha\int_{\beta_{min}}^{1-\alpha}d\beta[\frac{m_{Q'}(\alpha+\beta)}{\alpha^{2}\beta}+\frac{m_{Q}}{\alpha\beta^{2}}]r(m_{Q},m_{Q'})^{2},\nonumber
\end{eqnarray}

\begin{eqnarray}
\rho^{\langle\bar{q}q\rangle^{2}}(s)&=&-\frac{\langle\bar{q}q\rangle^{2}}{2^{4}\pi^{2}}m_{Q}m_{Q'}\sqrt{(s-m_{Q}^{2}+m_{Q'}^{2})^{2}-4m_{Q'}^{2}s}/s,\nonumber
\end{eqnarray}

\begin{eqnarray}
\rho^{\langle g\bar{q}\sigma\cdot G q\rangle}(s)&=&\frac{3\langle
g\bar{q}\sigma\cdot G
q\rangle}{2^{8}\pi^{4}}\int_{\alpha_{min}}^{\alpha_{max}}d\alpha\{\frac{m_{Q'}}{\alpha}\int_{\beta_{min}}^{1-\alpha}d\beta
r(m_{Q},m_{Q'})-(\frac{m_{Q'}}{\alpha}+\frac{m_{Q}}{1-\alpha})[\alpha
m_{Q}^{2}+(1-\alpha) m_{Q'}^{2}-\alpha(1-\alpha) s]\},\nonumber
\end{eqnarray}

\begin{eqnarray}
\rho^{\langle g^{2}G^{2}\rangle}(s)&=&-\frac{\langle
g^{2}G^{2}\rangle}{2^{12}\pi^{6}}\int_{\alpha_{min}}^{\alpha_{max}}d\alpha\int_{\beta_{min}}^{1-\alpha}d\beta(1-\alpha-\beta)(1+\alpha+\beta)(\frac{m_{Q'}^{2}}{\alpha^{3}}+\frac{m_{Q}^{2}}{\beta^{3}})r(m_{Q},m_{Q'}),\nonumber
\end{eqnarray}

\begin{eqnarray}
\rho^{\langle g^{3}G^{3}\rangle}(s)&=&-\frac{\langle
g^{3}G^{3}\rangle}{2^{14}\pi^{6}}\int_{\alpha_{min}}^{\alpha_{max}}d\alpha\int_{\beta_{min}}^{1-\alpha}d\beta(1-\alpha-\beta)(1+\alpha+\beta)[(\frac{1}{\alpha^{3}}+\frac{1}{\beta^{3}})r(m_{Q},m_{Q'})+2(\frac{m_{Q'}^{2}\beta}{\alpha^{3}}+\frac{m_{Q}^{2}\alpha}{\beta^{3}})],\nonumber
\end{eqnarray}
for $(Q\bar{q})_{1}(\bar{Q'}q)_{0}^{*}$,

\begin{eqnarray}
\rho^{\mbox{OPE}}(s)=\rho^{\mbox{pert}}(s)+\rho^{\langle\bar{q}q\rangle}(s)+\rho^{\langle\bar{q}q\rangle^{2}}(s)+\rho^{\langle
g\bar{q}\sigma\cdot G q\rangle}(s)+\rho^{\langle
g^{2}G^{2}\rangle}(s)+\rho^{\langle g^{3}G^{3}\rangle}(s),\nonumber
\end{eqnarray}

\begin{eqnarray}
\rho^{\mbox{pert}}(s)&=&\frac{3}{2^{11}\pi^{6}}\int_{\alpha_{min}}^{\alpha_{max}}\frac{d\alpha}{\alpha^{3}}\int_{\beta_{min}}^{1-\alpha}\frac{d\beta}{\beta^{3}}(1-\alpha-\beta)r(m_{Q},m_{Q'})^{4},\nonumber
\end{eqnarray}

\begin{eqnarray}
\rho^{\langle\bar{q}q\rangle}(s)&=&\frac{3\langle\bar{q}q\rangle}{2^{7}\pi^{4}}\int_{\alpha_{min}}^{\alpha_{max}}d\alpha\int_{\beta_{min}}^{1-\alpha}d\beta(\frac{m_{Q'}}{\alpha^{2}\beta}-\frac{m_{Q}}{\alpha\beta^{2}})r(m_{Q},m_{Q'})^{2},\nonumber
\end{eqnarray}

\begin{eqnarray}
\rho^{\langle\bar{q}q\rangle^{2}}(s)&=&-\frac{\langle\bar{q}q\rangle^{2}}{2^{4}\pi^{2}}m_{Q}m_{Q'}\sqrt{(s-m_{Q}^{2}+m_{Q'}^{2})^{2}-4m_{Q'}^{2}s}/s,\nonumber
\end{eqnarray}

\begin{eqnarray}
\rho^{\langle g\bar{q}\sigma\cdot G q\rangle}(s)&=&\frac{3\langle
g\bar{q}\sigma\cdot G
q\rangle}{2^{8}\pi^{4}}\int_{\alpha_{min}}^{\alpha_{max}}d\alpha(\frac{m_{Q'}}{\alpha}-\frac{m_{Q}}{1-\alpha})[\alpha
m_{Q}^{2}+(1-\alpha) m_{Q'}^{2}-\alpha(1-\alpha) s],\nonumber
\end{eqnarray}

\begin{eqnarray}
\rho^{\langle g^{2}G^{2}\rangle}(s)&=&\frac{\langle
g^{2}G^{2}\rangle}{2^{11}\pi^{6}}\int_{\alpha_{min}}^{\alpha_{max}}d\alpha\int_{\beta_{min}}^{1-\alpha}d\beta(1-\alpha-\beta)(\frac{m_{Q'}^{2}}{\alpha^{3}}+\frac{m_{Q}^{2}}{\beta^{3}})r(m_{Q},m_{Q'}),\nonumber
\end{eqnarray}

\begin{eqnarray}
\rho^{\langle g^{3}G^{3}\rangle}(s)&=&\frac{\langle
g^{3}G^{3}\rangle}{2^{13}\pi^{6}}\int_{\alpha_{min}}^{\alpha_{max}}d\alpha\int_{\beta_{min}}^{1-\alpha}d\beta(1-\alpha-\beta)[(\frac{1}{\alpha^{3}}+\frac{1}{\beta^{3}})r(m_{Q},m_{Q'})+2(\frac{m_{Q'}^{2}\beta}{\alpha^{3}}+\frac{m_{Q}^{2}\alpha}{\beta^{3}})]
,\nonumber
\end{eqnarray}
for $(Q\bar{q})(\bar{Q'}q)_{0}^{*}$,

\begin{eqnarray}
\rho^{\mbox{OPE}}(s)=-\{\rho^{\mbox{pert}}(s)+\rho^{\langle\bar{q}q\rangle}(s)+\rho^{\langle\bar{q}q\rangle^{2}}(s)+\rho^{\langle
g\bar{q}\sigma\cdot G q\rangle}(s)+\rho^{\langle
g^{2}G^{2}\rangle}(s)+\rho^{\langle
g^{3}G^{3}\rangle}(s)\},\nonumber
\end{eqnarray}

\begin{eqnarray}
\rho^{\mbox{pert}}(s)&=&-\frac{3}{2^{12}\pi^{6}}\int_{\alpha_{min}}^{\alpha_{max}}\frac{d\alpha}{\alpha^{3}}\int_{\beta_{min}}^{1-\alpha}\frac{d\beta}{\beta^{3}}(1-\alpha-\beta)(1+\alpha+\beta)r(m_{Q},m_{Q'})^{4},\nonumber
\end{eqnarray}

\begin{eqnarray}
\rho^{\langle\bar{q}q\rangle}(s)&=&\frac{3\langle\bar{q}q\rangle}{2^{7}\pi^{4}}\int_{\alpha_{min}}^{\alpha_{max}}d\alpha\int_{\beta_{min}}^{1-\alpha}d\beta[\frac{m_{Q'}(\alpha+\beta)}{\alpha^{2}\beta}-\frac{m_{Q}}{\alpha\beta^{2}}]r(m_{Q},m_{Q'})^{2},\nonumber
\end{eqnarray}

\begin{eqnarray}
\rho^{\langle\bar{q}q\rangle^{2}}(s)&=&\frac{\langle\bar{q}q\rangle^{2}}{2^{4}\pi^{2}}m_{Q}m_{Q'}\sqrt{(s-m_{Q}^{2}+m_{Q'}^{2})^{2}-4m_{Q'}^{2}s}/s,\nonumber
\end{eqnarray}

\begin{eqnarray}
\rho^{\langle g\bar{q}\sigma\cdot G q\rangle}(s)&=&\frac{3\langle
g\bar{q}\sigma\cdot G
q\rangle}{2^{8}\pi^{4}}\int_{\alpha_{min}}^{\alpha_{max}}d\alpha\times\{-\frac{m_{Q'}}{\alpha}\int_{\beta_{min}}^{1-\alpha}d\beta
r(m_{Q},m_{Q'})+(\frac{m_{Q'}}{\alpha}-\frac{m_{Q}}{1-\alpha})[\alpha
m_{Q}^{2}+(1-\alpha) m_{Q'}^{2}-\alpha(1-\alpha) s]\},\nonumber
\end{eqnarray}

\begin{eqnarray}
\rho^{\langle g^{2}G^{2}\rangle}(s)&=&-\frac{\langle
g^{2}G^{2}\rangle}{2^{12}\pi^{6}}\int_{\alpha_{min}}^{\alpha_{max}}d\alpha\int_{\beta_{min}}^{1-\alpha}d\beta(1-\alpha-\beta)(1+\alpha+\beta)(\frac{m_{Q'}^{2}}{\alpha^{3}}+\frac{m_{Q}^{2}}{\beta^{3}})r(m_{Q},m_{Q'}),\nonumber
\end{eqnarray}

\begin{eqnarray}
\rho^{\langle g^{3}G^{3}\rangle}(s)&=&-\frac{\langle
g^{3}G^{3}\rangle}{2^{14}\pi^{6}}\int_{\alpha_{min}}^{\alpha_{max}}d\alpha\int_{\beta_{min}}^{1-\alpha}d\beta(1-\alpha-\beta)(1+\alpha+\beta)[(\frac{1}{\alpha^{3}}+\frac{1}{\beta^{3}})r(m_{Q},m_{Q'})+2(\frac{m_{Q'}^{2}\beta}{\alpha^{3}}+\frac{m_{Q}^{2}\alpha}{\beta^{3}})]
,\nonumber
\end{eqnarray}
for $(Q\bar{q})_{1}(\bar{Q'}q)$,

\begin{eqnarray}
\rho^{\mbox{OPE}}(s)=-\{\rho^{\mbox{pert}}(s)+\rho^{\langle\bar{q}q\rangle}(s)+\rho^{\langle\bar{q}q\rangle^{2}}(s)+\rho^{\langle
g\bar{q}\sigma\cdot G q\rangle}(s)+\rho^{\langle
g^{2}G^{2}\rangle}(s)+\rho^{\langle
g^{3}G^{3}\rangle}(s)\},\nonumber
\end{eqnarray}

\begin{eqnarray}
\rho^{\mbox{pert}}(s)&=&-\frac{3}{2^{12}\pi^{6}}\int_{\alpha_{min}}^{\alpha_{max}}\frac{d\alpha}{\alpha^{3}}\int_{\beta_{min}}^{1-\alpha}\frac{d\beta}{\beta^{3}}(1-\alpha-\beta)(1+\alpha+\beta)r(m_{Q},m_{Q'})^{4},\nonumber
\end{eqnarray}

\begin{eqnarray}
\rho^{\langle\bar{q}q\rangle}(s)&=&\frac{3\langle\bar{q}q\rangle}{2^{7}\pi^{4}}\int_{\alpha_{min}}^{\alpha_{max}}d\alpha\int_{\beta_{min}}^{1-\alpha}d\beta[-\frac{m_{Q'}(\alpha+\beta)}{\alpha^{2}\beta}+\frac{m_{Q}}{\alpha\beta^{2}}]r(m_{Q},m_{Q'})^{2},\nonumber
\end{eqnarray}

\begin{eqnarray}
\rho^{\langle\bar{q}q\rangle^{2}}(s)&=&\frac{\langle\bar{q}q\rangle^{2}}{2^{4}\pi^{2}}m_{Q}m_{Q'}\sqrt{(s-m_{Q}^{2}+m_{Q'}^{2})^{2}-4m_{Q'}^{2}s}/s,\nonumber
\end{eqnarray}

\begin{eqnarray}
\rho^{\langle g\bar{q}\sigma\cdot G q\rangle}(s)&=&\frac{3\langle
g\bar{q}\sigma\cdot G
q\rangle}{2^{8}\pi^{4}}\int_{\alpha_{min}}^{\alpha_{max}}d\alpha\{\frac{m_{Q'}}{\alpha}\int_{\beta_{min}}^{1-\alpha}d\beta
r(m_{Q},m_{Q'})+(-\frac{m_{Q'}}{\alpha}+\frac{m_{Q}}{1-\alpha})[\alpha
m_{Q}^{2}+(1-\alpha) m_{Q'}^{2}-\alpha(1-\alpha) s]\},\nonumber
\end{eqnarray}

\begin{eqnarray}
\rho^{\langle g^{2}G^{2}\rangle}(s)&=&-\frac{\langle
g^{2}G^{2}\rangle}{2^{12}\pi^{6}}\int_{\alpha_{min}}^{\alpha_{max}}d\alpha\int_{\beta_{min}}^{1-\alpha}d\beta(1-\alpha-\beta)(1+\alpha+\beta)(\frac{m_{Q'}^{2}}{\alpha^{3}}+\frac{m_{Q}^{2}}{\beta^{3}})r(m_{Q},m_{Q'}),\nonumber
\end{eqnarray}

\begin{eqnarray}
\rho^{\langle g^{3}G^{3}\rangle}(s)&=&-\frac{\langle
g^{3}G^{3}\rangle}{2^{14}\pi^{6}}\int_{\alpha_{min}}^{\alpha_{max}}d\alpha\int_{\beta_{min}}^{1-\alpha}d\beta(1-\alpha-\beta)(1+\alpha+\beta)[(\frac{1}{\alpha^{3}}+\frac{1}{\beta^{3}})r(m_{Q},m_{Q'})+2(\frac{m_{Q'}^{2}\beta}{\alpha^{3}}+\frac{m_{Q}^{2}\alpha}{\beta^{3}})]
,\nonumber
\end{eqnarray}
for $(Q\bar{q})^{*}(\bar{Q'}q)_{0}^{*}$, and

\begin{eqnarray}
\rho^{\mbox{OPE}}(s)=\rho^{\mbox{pert}}(s)+\rho^{\langle\bar{q}q\rangle}(s)+\rho^{\langle\bar{q}q\rangle^{2}}(s)+\rho^{\langle
g\bar{q}\sigma\cdot G q\rangle}(s)+\rho^{\langle
g^{2}G^{2}\rangle}(s)+\rho^{\langle g^{3}G^{3}\rangle}(s),\nonumber
\end{eqnarray}

\begin{eqnarray}
\rho^{\mbox{pert}}(s)&=&\frac{3}{2^{9}\pi^{6}}\int_{\alpha_{min}}^{\alpha_{max}}\frac{d\alpha}{\alpha^{3}}\int_{\beta_{min}}^{1-\alpha}\frac{d\beta}{\beta^{3}}(1-\alpha-\beta)r(m_{Q},m_{Q'})^{4},\nonumber
\end{eqnarray}

\begin{eqnarray}
\rho^{\langle\bar{q}q\rangle}(s)&=&\frac{3\langle\bar{q}q\rangle}{2^{6}\pi^{4}}\int_{\alpha_{min}}^{\alpha_{max}}d\alpha\int_{\beta_{min}}^{1-\alpha}d\beta(\frac{m_{Q'}}{\alpha^{2}\beta}-\frac{m_{Q}}{\alpha\beta^{2}})r(m_{Q},m_{Q'})^{2},\nonumber
\end{eqnarray}

\begin{eqnarray}
\rho^{\langle\bar{q}q\rangle^{2}}(s)&=&-\frac{\langle\bar{q}q\rangle^{2}}{2^{2}\pi^{2}}m_{Q}m_{Q'}\sqrt{(s-m_{Q}^{2}+m_{Q'}^{2})^{2}-4m_{Q'}^{2}s}/s,\nonumber
\end{eqnarray}

\begin{eqnarray}
\rho^{\langle g\bar{q}\sigma\cdot G q\rangle}(s)&=&\frac{3\langle
g\bar{q}\sigma\cdot G
q\rangle}{2^{7}\pi^{4}}\int_{\alpha_{min}}^{\alpha_{max}}d\alpha(\frac{m_{Q'}}{\alpha}-\frac{m_{Q}}{1-\alpha})[\alpha
m_{Q}^{2}+(1-\alpha) m_{Q'}^{2}-\alpha(1-\alpha) s],\nonumber
\end{eqnarray}

\begin{eqnarray}
\rho^{\langle g^{2}G^{2}\rangle}(s)&=&\frac{\langle
g^{2}G^{2}\rangle}{2^{9}\pi^{6}}\int_{\alpha_{min}}^{\alpha_{max}}d\alpha\int_{\beta_{min}}^{1-\alpha}d\beta(1-\alpha-\beta)(\frac{m_{Q'}^{2}}{\alpha^{3}}+\frac{m_{Q}^{2}}{\beta^{3}})r(m_{Q},m_{Q'}),\nonumber
\end{eqnarray}

\begin{eqnarray}
\rho^{\langle g^{3}G^{3}\rangle}(s)&=&\frac{\langle
g^{3}G^{3}\rangle}{2^{11}\pi^{6}}\int_{\alpha_{min}}^{\alpha_{max}}d\alpha\int_{\beta_{min}}^{1-\alpha}d\beta(1-\alpha-\beta)[(\frac{1}{\alpha^{3}}+\frac{1}{\beta^{3}})r(m_{Q},m_{Q'})+2(\frac{m_{Q'}^{2}\beta}{\alpha^{3}}+\frac{m_{Q}^{2}\alpha}{\beta^{3}})]
,\nonumber
\end{eqnarray}

for $(Q\bar{q})^{*}(\bar{Q'}q)_{1}$. The integration limits are
given by
$\alpha_{min}=[s-m_{Q}^{2}+m_{Q'}^{2}-\sqrt{(s-m_{Q}^{2}+m_{Q'}^{2})^{2}-4m_{Q'}^{2}s}]/(2s)$,
$\alpha_{max}=[s-m_{Q}^{2}+m_{Q'}^{2}+\sqrt{(s-m_{Q}^{2}+m_{Q'}^{2})^{2}-4m_{Q'}^{2}s}]/(2s)$,
and $\beta_{min}=\alpha m_{Q}^{2}/(s\alpha-m_{Q'}^{2})$.

%%%%%%%%%%%%%%%%%%%%%%%%%%%%%%%%%%%%%%
\begin{acknowledgments}
J. R. Zhang is very indebted to Ming Zhong for helpful discussions.
This work was supported in part by the National Natural Science
Foundation of China under Contract No.10675167.
\end{acknowledgments}
%%%%%%%%%%%%%%%%%%%%%%%%%%%%%%%%%%%%%%%%%%%%%%%%%%%%


\begin{thebibliography}{99}
\bibitem{X3872}S.~K.~Choi {\it et al.}, (Belle Collaboration), Phys.
Rev. Lett. {\bf91}, 262001 (2003); V.~M.~Abazov {\it et al.}, (D0
Collaboration), Phys. Rev. Lett. {\bf 93}, 162002 (2004); D.~Acosta
{\it et al.}, (CDF Collaboration), Phys. Rev. Lett. {\bf 93}, 072001
(2004); B.~Aubert {\it et al.}, (BaBar Collaboration), Phys. Rev. D
{\bf 71}, 071103 (2005).


\bibitem{Y3930}S.~K.~Choi {\it et al.}, (Belle Collaboration), Phys.
Rev. Lett. {\bf94}, 182002 (2005); B.~Aubert {\it et al.}, (BaBar
Collaboration), Phys. Rev. Lett. {\bf101}, 082001 (2008).

\bibitem{Y4260}B.~Aubert {\it et al.}, (BaBar Collaboration), Phys.
Rev. Lett. {\bf95}, 142001 (2005); Q.~He {\it et al.}, (CLEO
Collaboration), Phys. Rev. D {\bf 74}, 091104(R) (2006); C.~Z.~Yuan
{\it et al.}, (Belle Collaboration), Phys. Rev. Lett. {\bf99},
182004 (2007).


\bibitem{Z3930}S.~Uehara {\it et al.}, (Belle Collaboration),
Phys. Rev. Lett. {\bf96}, 082003 (2006).


\bibitem{X3940}K.~Abe {\it et al.}, (Belle Collaboration), Phys. Rev. Lett. {\bf98}, 082001 (2007).



\bibitem{Z4430}K.~Abe {\it et al.}, (Belle Collaboration), Phys. Rev. Lett.
{\bf100}, 142001 (2008).

\bibitem{Z4050}R.~Mizuk {\it et al.}, (Belle Collaboration),
Phys. Rev. D {\bf78}, 072004 (2008).

\bibitem{Y4140}T.~Aaltonen {\it et al.},
(CDF Collaboration), arXiv:0903.2229.

\bibitem{Swanson}E.~S.~Swanson, Phys. Rep. {\bf 429}, 243 (2006).


\bibitem{PDG}C.~Amsler {\it et al.}, (Particle Data Group), Phys. Lett. B {\bf 667}, 1 (2008).



\bibitem{theory}X.~Liu, X.~Q.~Zeng,
and X.~Q.~Li, Phys. Rev. D {\bf 72}, 054023 (2005); Y.~J.~Zhang,
H.~C.~Chiang, P.~N.~Shen, and B.~S.~Zou, Phys. Rev. D {\bf74},
014013 (2006); J.~L.~Rosner, Phys. Rev. D {\bf76}, 114002 (2007);
G.~J.~Ding, Phys. Rev. D {\bf79}, 014001 (2009); G.~J.~Ding,
J.~F.~Liu, and M.~L.~Yan, Phys. Rev. D {\bf79}, 054005 (2009);
G.~J.~Ding, arXiv:0905.1188.


\bibitem{theory-X3872}F.~E.~Close and P.~R.~Page, Phys. Lett. B {\bf578}, 119
(2004); M.~B.~Voloshin, Phys. Lett. B {\bf579}, 316 (2004);
C.~Y.~Wong, Phys. Rev. C {\bf69}, 055202 (2004); E.~S.~Swanson,
Phys. Lett. B {\bf588}, 189 (2004); N.~A.~T$\ddot{\mbox{o}}$rnqvist,
Phys. Lett. B {\bf590}, 209 (2004); E.~S.~Swanson, Phys. Lett. B
{\bf598}, 197 (2004); C.~E.~Thomas and F.~E.~Close, Phys. Rev. D
{\bf78}, 034007 (2008); Y.~R.~Liu, X.~Liu, W.~Z.~Deng, and
S.~L.~Zhu, Euro. Phys. J. C {\bf56}, 63 (2008); Y.~R.~Liu and
Z.~Y.~Zhang, Phys. Rev. C {\bf79}, 035206 (2009).

\bibitem{theory-Z4430}C.~Meng and K.~T.~Chao, arXiv:0708.4222; X.~ Liu, Y.~R.~Liu, W.~Z.~Deng, and
S.~L.~Zhu, Phys. Rev. D {\bf77}, 034003 (2008); X.~Liu, Y.~R.~Liu,
W.~Z.~Deng, and S.~L.~Zhu, Phys. Rev. D {\bf77}, 094015 (2008).


\bibitem{theory-Y3930}X.~Liu, Z.~G.~Luo, Y.~R.~Liu, and S.~L.~Zhu,
arXiv:0808.0073.

\bibitem{Liu}X.~Liu and S.~L.~Zhu, arXiv:0903.2529.

\bibitem{theory-Y4140}N.~Mahajan, arXiv:0903.3107; T.~Branz, T.~Gutsche, and
V.~E.~Lyubovitskij, arXiv:0903.5424; G.~J.~Ding, arXiv:0904.1782.



\bibitem{long ago}M.~B.~Voloshin and L.~B.~Okun, JETP Lett. {\bf23}, 333 (1976).


\bibitem{cc}A.~D.~Rujula, H.~Georgi, and S.~L.~Glashow, Phys. Rev. Lett. {\bf38}, 317
(1977).


\bibitem{svzsum}M.~A.~Shifman, A.~I.~Vainshtein, and V.~I.~Zakharov, Nucl. Phys. {\bf B147}, 385 (1979); {\bf B147}, 448 (1979);
 V.~A.~Novikov, M.~A.~Shifman, A.~I.~Vainshtein, and V.~I.~Zakharov, Fortschr. Phys. {\bf 32}, 585 (1984).

\bibitem{overview}M.~A.~Shifman, Vacuum Structure and QCD Sum Rules,
North-Holland, Amsterdam 1992.

\bibitem{overview1}B.~L.~Ioffe, in
``The spin structure of the nucleon", edited by B.~Frois,
V.~W.~Hughes, N.~de Groot, World Scientific (1997), arXiv:9511401.

\bibitem{overview2}S.~Narison, QCD Spectral Sum Rules, World Scientific, Singapore,
1989.

\bibitem{overview3}P.~Colangelo and
A.~Khodjamirian, in: M.~Shifman (Ed.), At the Frontier of Particle
Physics: Handbook of QCD, vol. 3, Boris Ioffe Festschrift, World
Scientific, Sigapore, 2001, pp. 1495-1576, arXiv:0010175;
A.~Khodjamirian, talk given at Continuous Advances in QCD
2002/ARKADYFEST, arXiv:0209166.




\bibitem{zhang}J.~R.~Zhang and M.~Q.~Huang,
arXiv:0905.4178; J.~R.~Zhang and M.~Q.~Huang, arXiv:0905.4672.



\bibitem{SR-study}S.~H.~Lee, M.~Nielsen, and U.~Wiedner, arXiv:0803.1168; S.~H.~Lee, A.~Mihara,
F.~S.~Navarra, and M.~Nielsen, Phys. Lett. B {\bf661}, 28 (2008);
S.~H.~Lee, K.~Morita, and M.~Nielsen, Nucl. Phys. A {\bf815}, 29
(2009); R.~M.~Albuquerque and M.~Nielsen, Nucl. Phys. A {\bf815}, 53
(2009); R.~M.~Albuquerque, M.~E.~Bracco, and M.~Nielsen,
arXiv:0903.5540.

\bibitem{zgwang}Z.~G. Wang, arXiv:0903.5200.


\bibitem{reinders}L.~J.~Reinders, H.~R.~Rubinstein, and S.~Yazaki, Phys. Rep. {\bf 127}, 1 (1985).

\bibitem{reinders1}L.~J.~Reinders, H.~R.~Rubinstein, and S.~Yazaki, Nucl. Phys. {\bf
B186}, 109 (1981).





\bibitem{technique}R.~D.~Matheus, S.~Narison, M.~Nielsen, and J.~M.~Richard, Phys. Rev. D {\bf75}, 014005 (2007);
S.~H.~Lee, K.~Morita, and M.~Nielsen, Phys. Rev. D {\bf78}, 076001
(2008); M.~E.~Bracco, S.~H.~Lee, M.~Nielsen, and R.~R.~daSilva,
Phys. Lett. B {\bf671}, 240 (2009).

\bibitem{technique1}J.~R.~Zhang and
M.~Q.~Huang, Phys. Rev. D {\bf 77}, 094002 (2008); Phys. Rev. D {\bf
78}, 094007 (2008); Phys. Rev. D {\bf 78}, 094015 (2008); Phys.
Lett. B {\bf 674}, 28 (2009).

\end{thebibliography}
\end{document}